\documentclass[pdflatex,sn-mathphys-num]{sn-jnl}

\usepackage{graphicx}%
\usepackage{multirow}%
\usepackage{amsmath,amssymb,amsfonts}%
\usepackage{amsthm}%
\usepackage{mathrsfs}%
\usepackage[title]{appendix}%
\usepackage{textcomp}%
\usepackage{manyfoot}%
\usepackage{booktabs}%
\usepackage{algorithm}%
\usepackage{algorithmicx}%
\usepackage{algpseudocode}%
\usepackage{listings}%
\usepackage{comment}%
\usepackage[table]{xcolor}
\usepackage{bm}
\usepackage{ulem}
\usepackage{tikz}
\usepackage{hyperref}
\usepackage{orcidlink}

\theoremstyle{thmstyleone}%
\theoremstyle{thmstyletwo}%

\theoremstyle{thmstylethree}%

\newcommand{\snn} {\sqrt{s_{_{\rm NN}}}}

\let\OLDthebibliography\thebibliography
\renewcommand\thebibliography[1]{
  \OLDthebibliography{#1}
  \setlength{\parskip}{0pt}
  \setlength{\itemsep}{0pt plus 0.3ex}
}

\begin{document}

\title[Article Title]{Toward a Unified Understanding of the Dense Matter Equation of State}


\author[1,2]{\fnm{Kshitij} \sur{Agarwal}~\orcidlink{0000-0001-5781-3393}}\email{kshitij.agarwal@ts.infn.it}
\author[3]{\fnm{Johannes} \sur{Jahan}~\orcidlink{0000-0002-4557-4652}}\email{jjahan@central.uh.edu}
\author[4,9]{\fnm{Behruz} \sur{Kardan}~\orcidlink{0000-0002-8981-6051}}\email{bkardan@ikf.uni-frankfurt.de}
\author[5,6]{\fnm{Peter T. H.} \sur{Pang}~\orcidlink{0000-0001-7041-3239}}\email{thopang@nikhef.nl}
\author[7]{\fnm{Tom} \sur{Reichert}~\orcidlink{0000-0003-4353-5024}}\email{treichert@itp.uni-frankfurt.de}
\author[8]{\fnm{Alexandra C.} \sur{Semposki}~\orcidlink{0000-0003-2354-1523}}\email{semposki.1@osu.edu}


\affil[1]{\orgdiv{Department of Physics}, \orgname{University of Trieste}, \orgaddress{\city{Trieste}, \postcode{34127}, \country{Italy}}}

\affil[2]{\orgname{INFN - Section of Trieste}, \orgaddress{\city{Trieste}, \postcode{34127}, \country{Italy}}}

\affil[3]{\orgdiv{Department of Physics}, \orgname{University of Houston}, \orgaddress{\city{Houston}, \postcode{TX 77204}, \country{USA}}}

\affil[4]{\orgdiv{Institute for Nuclear Physics}, \orgname{Goethe University}, \orgaddress{\city{Frankfurt am Main}, \postcode{60438}, \country{Germany}}}

\affil[5]{\orgname{Nikhef}, \orgaddress{Science Park 105, \city{Amsterdam}, \postcode{1098 XG}, \country{Netherlands}}}

\affil[6]{\orgdiv{Institute for Gravitational and Subatomic Physics (GRASP)}, \orgname{Department of Physics, Utrecht University},  \orgaddress{\city{Utrecht}, \postcode{3584 CC}, \country{Netherlands}}}

\affil[7]{\orgdiv{Department of Physics}, \orgname{Duke University}, \orgaddress{\city{Durham}, \postcode{NC 27708}, \country{USA}}}

\affil[8]{\orgdiv{Department of Physics and Astronomy and Institute for Nuclear and Particle Physics}, \orgname{Ohio University}, \orgaddress{\city{Athens}, \postcode{OH 45701}, \country{USA}}}

\affil[9]{\orgname{Helmholtz Research Academy Hesse for FAIR (HFHF)}, \orgaddress{\city{Darmstadt}, \postcode{64291}, \country{Germany}}}

\abstract{Efforts to understand the equation of state (EOS) of dense nuclear matter at supra-saturation densities have grown more sophisticated over the past decade, driven by a surge in high-precision data from both terrestrial experiments and astrophysical observations. While for the former, heavy-ion collisions (HIC) represent a unique opportunity to constrain the EOS in a controlled laboratory setting, the latter can be precisely probed thanks to the advent of multi-messenger astronomy (MMA). However, as we move away from understanding drawn from individual sources and limited statistics to the era of precision physics with improved datasets, the need for a systematic way to combine them becomes clear. In this article, we trace the individual methods for extracting the EOS both for HIC and MMA. 
\textcolor{black}{We then review the current state-of-the-art collaborative efforts to combine these individual sources of information, focusing on: the Nuclear Physics and Multi-Messenger Astrophysics (NMMA) framework, which relies on Bayesian inference methods; the Modular Unified Solver for the Equation of State (MUSES) calculation engine, which integrates EOS priors with HIC data and produces predictions for key neutron star properties; and the Bayesian Analysis of Nuclear Dynamics (BAND) framework, which uses cutting-edge Bayesian methods to produce reliable and trustworthy predictions for nuclear and astrophysical problems.} We highlight the scientific advances \textcolor{black}{with respect to the EOS and neutron star properties} made possible by each framework and outline the remaining challenges that must be addressed to build a coherent, predictive picture of dense nuclear matter across all relevant regimes. \textcolor{black}{We conclude with a detailed discussion of how these frameworks might be integrated with each other to form a unified workflow for future EOS predictions.}} 




\maketitle
\tableofcontents

\section{Introduction}
\label{sec:intro}

\begin{figure}[!b]
    \centering
    \includegraphics[width=0.7\linewidth]{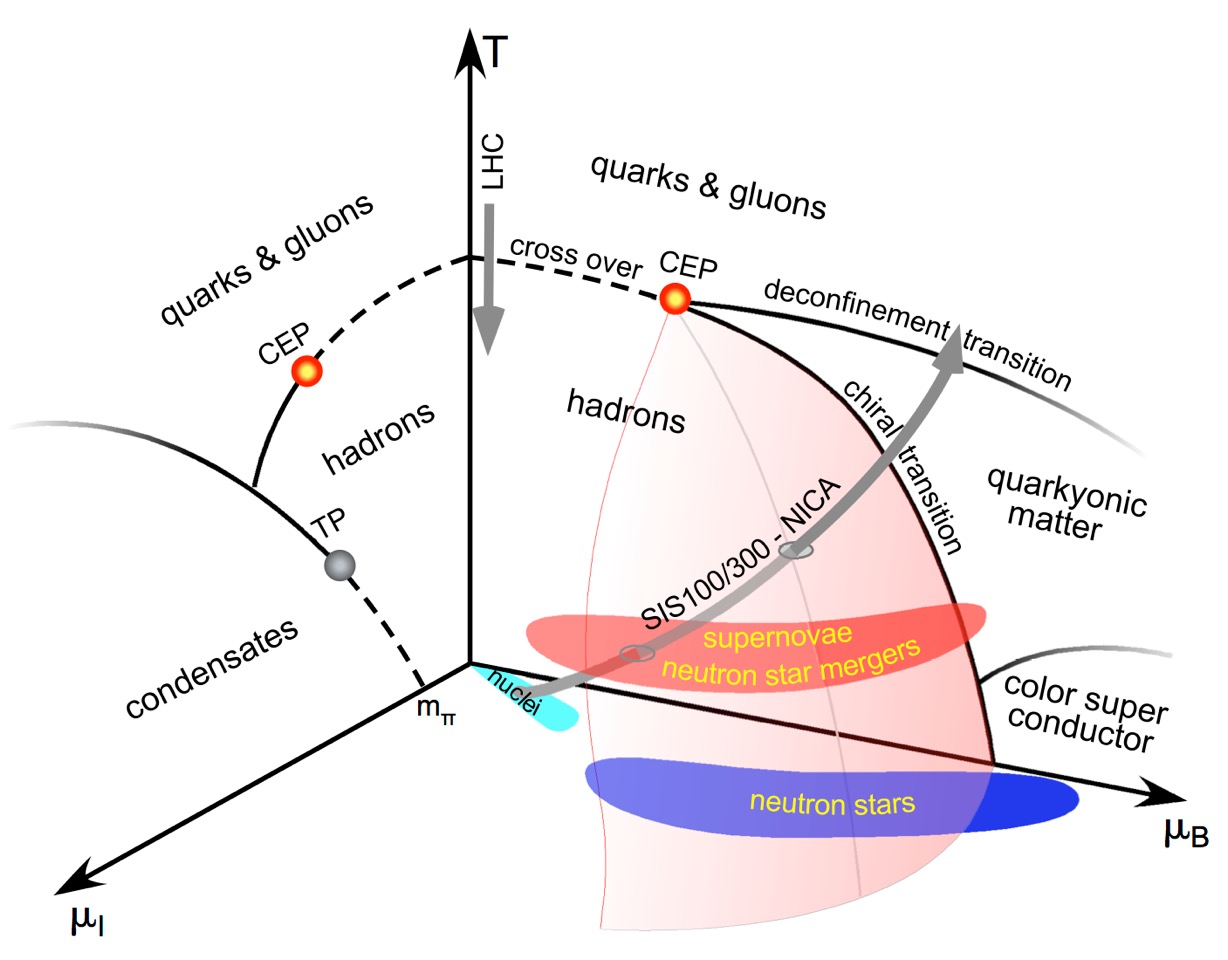}
    \caption{Schematic of the three-dimensional QCD phase diagram, represented as temperature ($T$), baryonic ($\mu_B$) and isospin ($\mu_I$) densities. A critical endpoint is conjectured to separate a smooth cross-over from a first-order phase transition, accompanied by a chiral phase transition at high densities. Exotic states such as quarkyonic matter or colour-superconducting phases could emerge at high baryon densities. Proton fractions for astrophysical systems evolve from 0.4 for supernovae down to 0.1 for cold neutron stars, where complementary measurements can be performed with heavy-ion collisions at FAIR, RHIC and NICA energies. Figure from H.R.~Schmidt (EKU Tübingen and GSI Darmstadt) and taken from~\cite{NupeccLRP2017}.
    }
    \label{fig:nupeccLRP2017PhaseDiagram}
\end{figure}

The study of Quantum Chromodynamics (QCD) matter and its properties across its phase diagram has been the central focus of nuclear physics and astrophysics for the past several decades~\cite{Gross:2022hyw, Lattimer:2021emm}. In particular, QCD matter at densities above nuclear saturation density, $n_{0} = 0.16~\textrm{fm}^{-3}$ (corresponding to the mass density of $2.17 \times 10^{14}~\textrm{g/cm}^{-3}$), is of special interest, as it is expected to host a rich structure, with evidence suggesting the emergence of deconfined quark matter phases at densities $\gtrsim 3n_{0}$~\cite{Annala:2023cwx} (see Fig.~\ref{fig:nupeccLRP2017PhaseDiagram}). The behaviour of dense nuclear matter at such extreme densities is described by the equation of state (EOS). 

\begin{figure}[!t]
    \centering
    \includegraphics[width=1\linewidth]{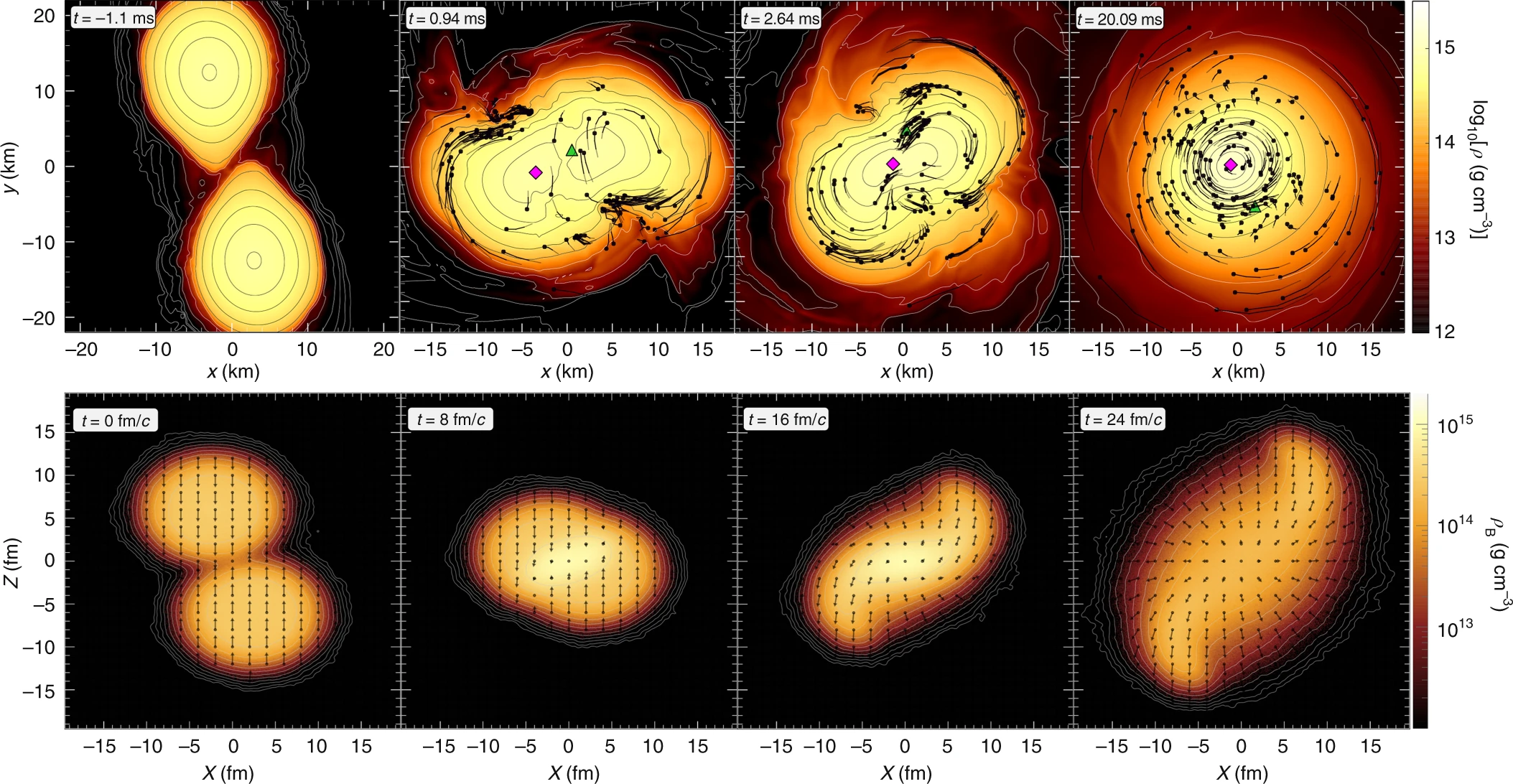}
    \caption{Snapshots of simulations showing the density evolution of a binary neutron star merger (top row) and a heavy ion-collision (bottom row).  While the former is shown for two neutron stars, each $1.35M_{\odot}$, merging into a compact object, the latter is of a non-central Au+Au collision at 2.42~GeV per nucleon. Despite dramatically different spatial and temporal scales, both events show similar densities and temperatures. Figure from~\cite{HADES:2019auv}.
    }
    \label{fig:hades_hic_ns_comp}
\end{figure}

\underline{Microscopically}, the dense matter EOS can be probed in heavy-ion collision experiments at relativistic incident energies ($\sim$ 5~GeV/nucleon in the fixed-target frame), thus creating an expanding fireball in the collision region with densities of about $5n_{0}$ (see Fig.~\ref{fig:hades_hic_ns_comp} - bottom row). Information about this hot and dense phase is obtained either from direct observables, such as rare or penetrating probes, or by comparing the measured final-state phase-space distribution of particle emission with the results of dynamic transport simulations, effectively tracing back the system’s evolution to the high-density phase.~\cite{Sorensen:2023zkk}. Experiments at facilities like the Facility for Antiproton and Ion Research (FAIR)~\cite{Durante:2019hzd, Aumann:2024unk}, Relativistic Heavy Ion Collider (RHIC)~\cite{Du:2024wjm}, Nuclotron-based Ion Collider fAcility (NICA)~\cite{Blaschke:2016NICA,Kekelidze:2017ghu}, Facility for Rare Isotope Beams (FRIB)~\cite{FRIB400, Brown:2024rml} and Radioactive Isotope Beam Factory (RIBF)~\cite{SpiRIT:2014yhq} enable the probing of the EOS of symmetric nuclear matter along with its isospin dependence under laboratory-controlled conditions, depending on the beam energy, the choice of target and projectile nuclei, and the varying dynamics due reaction centrality and orientation.

\underline{Macroscopically}, the dense matter EOS is accessible from the multi-messenger astrophysics analysis of neutron stars (NS) and their mergers, where densities of about $5n_{0}$ are reached (see Fig.~\ref{fig:hades_hic_ns_comp} - top row). This is inferred by the observation of the neutron star inspiral via gravitational waves and the electromagnetic signals of both the subsequent counterpart and the NS mass-radii measurements from radio and X-ray observations. This is enabled by precision measurements from the LIGO-Virgo-KAGRA (LVK) network~\cite{LIGOScientific:2014pky, VIRGO:2014yos, KAGRA:2021duu, Kiendrebeogo:2023hzf}, the Neutron Star Interior Composition ExploreR (NICER) mission~\cite{Rutherford:2024srk}, and the future enhanced X-ray Timing and Polarimetry (eXTP) mission~\cite{Zhang:2025iae,Li:2025uaw}.

Given the comparable conditions created in heavy-ion collisions and binary neutron star (BNS) mergers, the EOS serves as the link between the two information sources~\cite{Yao:2023yda}, potentially allowing a complementary and unified approach to help reduce uncertainties and to explore the EOS in a more complete and self-consistent way. In this review, we explore recent efforts to address this challenge. Sect.~\ref{sec:methods} reviews how the EOS information is extracted from both heavy-ion collisions and astrophysical sources. Sect.~\ref{sec:frameworks} covers the development of integrative frameworks, such as NMMA~\cite{Pang:2022rzc}, MUSES~\cite{ReinkePelicer:2025vuh}, and BAND~\cite{Phillips:2020dmw, bandframework}, that aim to merge EOS information from individual sources in statistically robust ways toward a unified EOS model. Finally, Sect.~\ref{sec:outlook} looks ahead at future challenges and the potential for a fully unified understanding of dense nuclear matter, as the field enters the ``precision era".

\section{Methods to Extract the Supra-Saturation Nuclear Matter EOS}
\label{sec:methods}

\begin{figure*}[!b]
    \centering
    \includegraphics[width=0.9\linewidth]{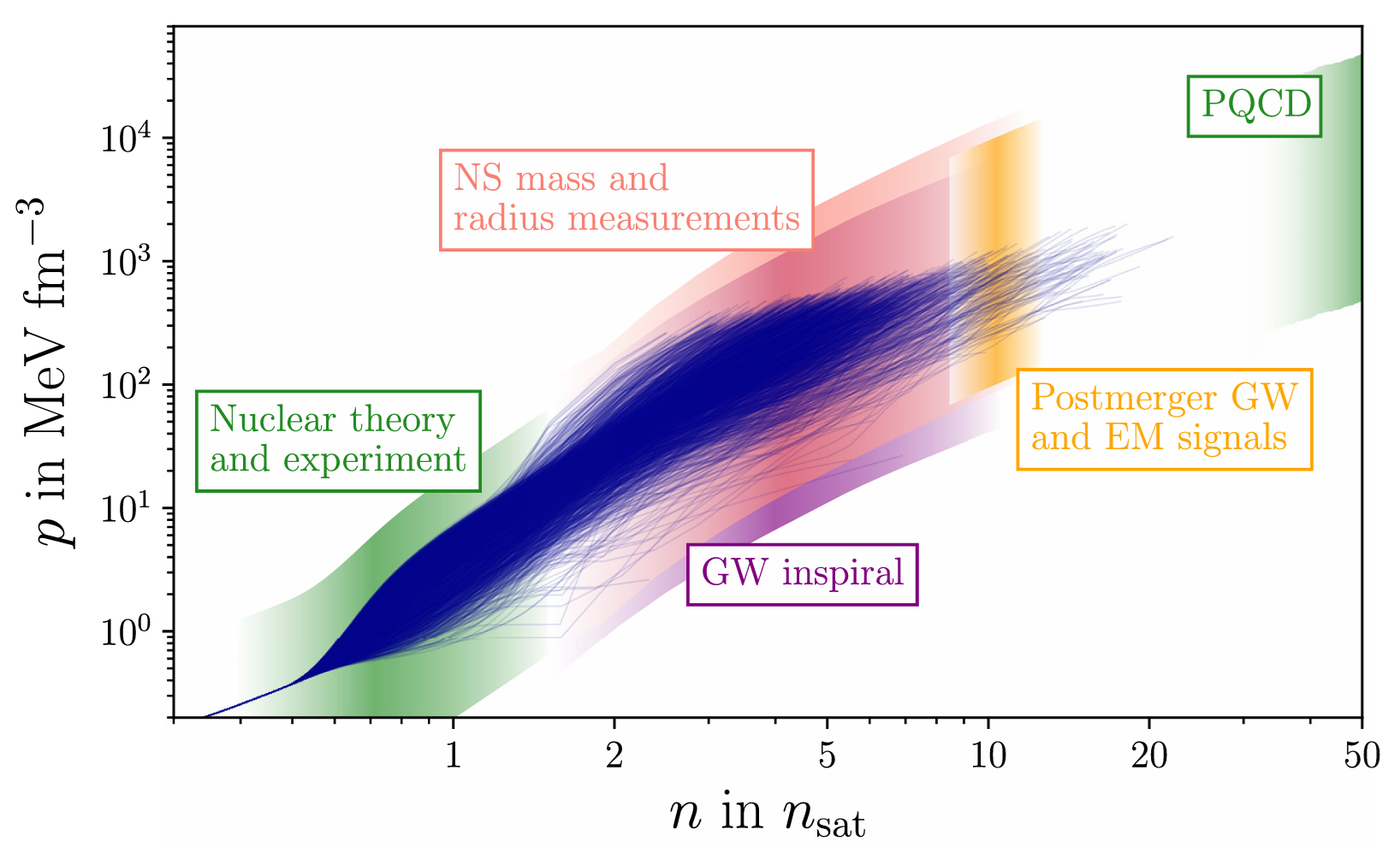}
    \caption{Schematic overview of information sources of the dense matter EOS. Dark blue lines show candidate EOS up to their maximum mass configurations, while coloured bands mark the density ranges constrained by different inputs. Figure from~\cite{Koehn:2024set}.
    } 
    \label{fig:eos_ns_obs}
\end{figure*}

\textcolor{black}{The EOS encapsulates the behaviour of strongly interacting matter under varying state parameters, such as pressure (or energy per baryon; $E$), baryon density $n_B$, isospin asymmetry $\delta$, and temperature $T$, thus providing insight into the phase structure and degrees of freedom of dense nuclear matter. For isospin-asymmetric matter, the energy per nucleon is shown as the following relation, where the energy per baryon can be expanded around symmetric nuclear matter ($\delta = 0$; $E_{\mathrm{SNM}}(n_B)$), while the remaining expansion terms correspond to the isospin dependence:
\begin{equation}
    \label{eq:eosdefinition}
    E(n_B,\delta) = E_{\mathrm{SNM}}(n_B) + E_{\mathrm{sym}}(n_B)\,\delta^2 + \mathcal{O}(\delta^4).
\end{equation}
Therein, the isospin asymmetry is given by $\delta = (n_n - n_p)/n_B$, where $n_n$ and $n_p$ are the densities of neutrons and protons, respectively. The second-order expansion coefficient $E_{\mathrm{sym}}(n_B)$ is referred to as the nuclear symmetry energy and is expressed as,
\begin{equation}
    \label{eq:esymdef1}
    E_{\mathrm{sym}}(n_B) = \frac{1}{2}\left.\frac{\partial^2 E(n_B,\delta)}{\partial \delta^2}\right|_{\delta=0}.
\end{equation}
If the higher-order expansion terms in $\delta$ ($\mathcal{O}(\delta^4)$) are negligibly small, the symmetry energy is described as the difference between pure neutron matter and symmetric nuclear matter,
\begin{equation}
    \label{eq:esymdef2}
    E_{\mathrm{sym}}(n_B) \approx E(n_B,\delta=1) - E(n_B,\delta=0).
\end{equation}
Notably, both definitions of $E_{\mathrm{sym}}(n_B)$ agree well around nuclear saturation density ($n_{\rm{sat}}$) \cite{Guven:2026fjb,Reed:2021nqk}, but deviate at larger densities. A detailed review on high-density nuclear symmetry energy is also planned to be published in this special issue and addresses this aspect in detail ~\cite{Li:2025xio}.}

\textcolor{black}{Both the symmetric and isospin-dependent components of the EOS can be inferred} from the experimental study of the emergent observable macroscopic properties that are sensitive to the underlying dynamics and require precise theoretical modelling for interpretation. The various information sources are based on different physical processes, and thus are sensitive to the EOS at different densities (see Fig.~\ref{fig:eos_ns_obs}). This section discusses the two complementary methods to extract nuclear matter EOS at supranuclear densities and reviews their current status: Sect.~\ref{subsec:methods_hic} describes relativistic heavy-ion collisions, which probe hot and dense matter under controlled laboratory conditions (green band in Fig.~\ref{fig:eos_ns_obs}; $\approx1-5n_{0}$); and Sect.~\ref{subsec:methods_mma} describes  multi-messenger astrophysical observations of neutron stars and their mergers (purple, pink, and orange bands in Fig.~\ref{fig:eos_ns_obs}; $\approx2-10n_{0}$).

\subsection{Relativistic Heavy-Ion Collisions}
\label{subsec:methods_hic}

\begin{figure*}[!b]
    \centering
    \includegraphics[width=\textwidth]{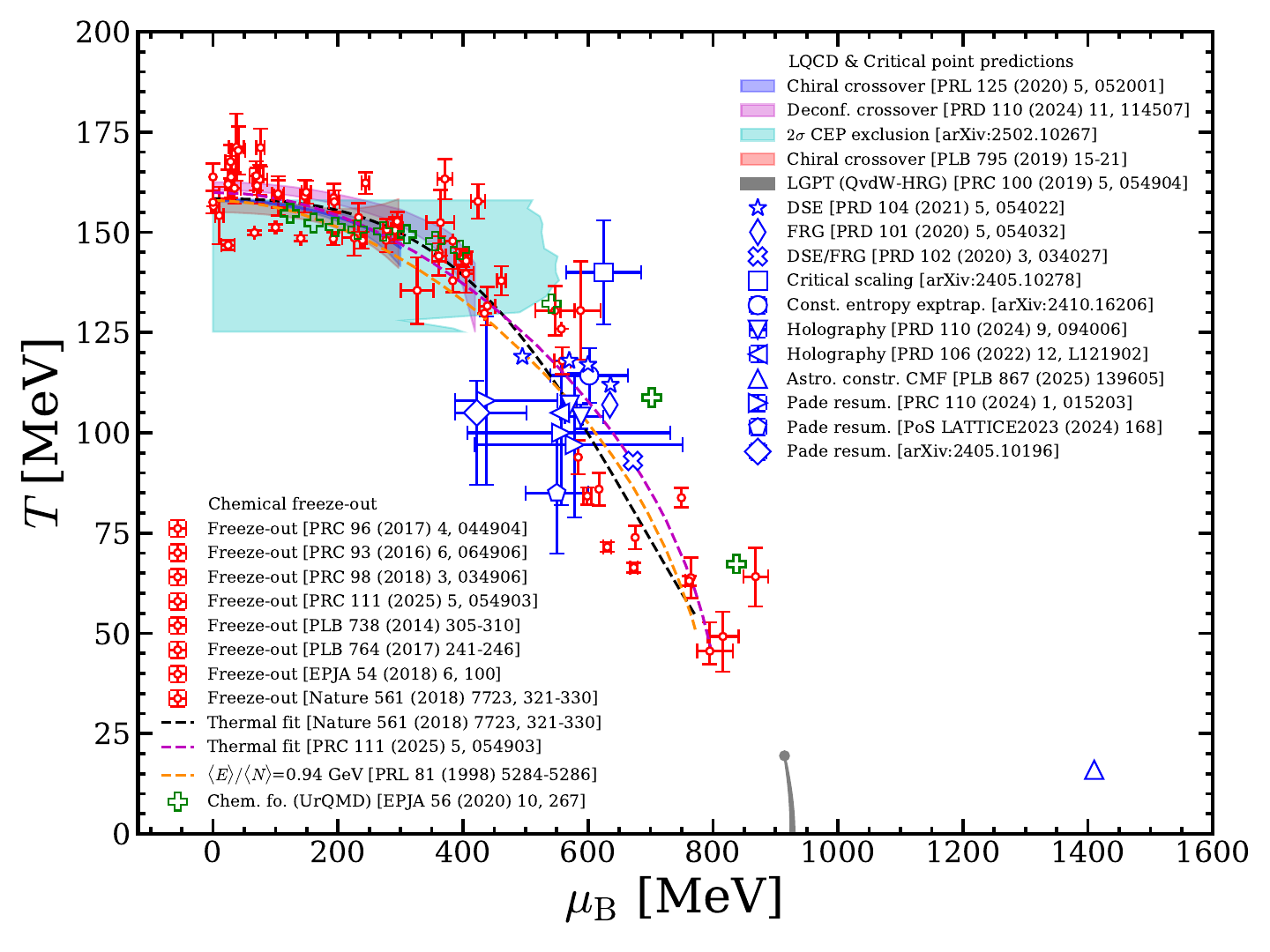}
    \caption{QCD Phase Diagram shown as a function of temperature ($T$) and baryon chemical potential ($\mu_B$). It highlights lattice QCD results, critical point predictions, freeze-out data and freeze-out calculations. The shaded areas show lattice QCD results for the chiral crossover \cite{Borsanyi:2020fev} (dark blue shaded area), the deconfinement crossover \cite{Borsanyi:2024xrx} (magenta shaded area), a constant entropy guided $2\sigma$ exclusion for the CEP \cite{Borsanyi:2025dyp} (light blue shaded area) and the chiral crossover \cite{HotQCD:2018pds} (red shaded area). The grey shaded area and the grey dot show the liquid-gas phase transition and its critical point calculated with the QvdW-HRG \cite{Poberezhnyuk:2019pxs}. The critical point predictions are all shown as blue symbols. The CEP predictions are from: lQCD guided truncated Dyson-Schwinger approach with backcoupled mesons \cite{Gunkel:2021oya} (star), lQCD guided functional Renormalization Group approach \cite{Fu:2019hdw} (diamond), lQCD guided generalized DSE/fRG approach \cite{Gao:2020qsj} (cross), finite size scaling of measured proton cumulants \cite{Sorensen:2024mry} (square), lQCD guided extrapolations along constant entropy density contours \cite{Shah:2024img} (circle), lQCD guided Bayesian inference in a holographic model \cite{Hippert:2023bel} (triangle-down), lQCD guided non-perturbative holographic model \cite{Cai:2022omk} (triangle-left), Chiral-Mean-Field model prediction constrained by combined neutron star $M-R$ and heavy-ion data \cite{Steinheimer:2025hsr} (triangle-up), lQCD guided Pad\'e resummation using a conformal map to track Lee-Yang edge singularities \cite{Basar:2023nkp} (triangle-right), lQCD guided and Ising-, Thirring- and Roberge-Weiss consistent multi-point Pad\'e expansion \cite{Clarke:2024seq} (pentagon) and lQCD guided multi-point pade expansion to locate Lee-Yang singularities of pressure \cite{Clarke:2024ugt} (thick diamond). Also shown are ``freeze-out data'' (various thermal model fits to measured yields, ratios or cumulants) as red dots. The data points are from \cite{STAR:2017sal,Vovchenko:2015idt,Vovchenko:2018fmh,Lysenko:2024hqp,Alba:2014eba,Andronic:2017pug,Becattini:2016xct,Sagun:2017eye} and fitted to data measured at SIS18, AGS, SPS, RHIC and LHC. The parametrized chemical freeze-out lines are taken from \cite{Andronic:2017pug} (black dashed line), \cite{Lysenko:2024hqp} (magenta dashed line) and using $\langle E\rangle /\langle N\rangle=0.94$ GeV \cite{Cleymans:1998fq} (orange dashed line) and the hadronic transport model based calculations of the chemical freeze-out is from \cite{Reichert:2020yhx} (green plusses). }
    \label{fig:QCD-PD}
\end{figure*}

Over the years, heavy-ion collision experiments have probed different regions of the QCD phase diagram by varying the collision energies and system sizes~\cite{Gross:2022hyw} (see Fig.~\ref{fig:QCD-PD} for a recent compilation of data related to the QCD phase diagram). The ultra-relativistic collisions ($\sqrt{s_{NN}} \gtrsim 100~\textrm{GeV}$) at the Large Hadron Collider (LHC) and RHIC have confirmed the existence of the Quark-Gluon Plasma (QGP) at nearly zero net-baryon density~\cite{Heinz:2000bk,Jacak:2012dx,Muller:2012zq,Braun-Munzinger:2015hba,ALICE:2022wpn,CMS:2024krd}, with the chemical freeze-out temperature consistent with the Lattice QCD (LQCD) pseudo-critical point~\cite{Andronic:2017pug,HotQCD:2018pds,Borsanyi:2020fev}. In contrast, intermediately lower-energy collisions ($\sqrt{s_{NN}} \lesssim 10~\textrm{GeV}$) have long been studied, yet the EOS of symmetric nuclear matter along with its isospin dependence in this regime ($\gtrsim 2n_{0}$) remains insufficiently constrained, owing on the one hand to the breakdown of first-principles methods available in LQCD, and on the other to the limited precision of experimental data together with the inherent challenges of modeling the detailed evolution of the medium. This subsection reviews the experimental (Sect.~\ref{subsubsec:methods_hic_exp}) and theoretical approaches (Sect.~\ref{subsubsec:methods_hic_theory}) to EOS extraction at these energies.

\subsubsection{Experiment}
\label{subsubsec:methods_hic_exp}

\underline{Symmetric Nuclear Matter:} The expansion dynamics of the hot and dense fireball and the subsequent particle production in heavy-ion collisions are governed by the compressibility of nuclear matter controlled by its EOS. Several key observables carry direct imprints of the supra-saturation phase. Among these observables, collective flow is particularly notable, as it describes the correlated motion of a large ensemble of emitted particles within a common velocity field or along similar directions, reflecting their shared dynamical origin~\cite{Herrmann:1999wu, Danielewicz:2000tr}. The azimuthal distribution of particles relative to a common reaction plane, characterized by Fourier coefficients ($v_n$) that quantify momentum-space anisotropy, encodes both the thermal and collective motion of particle emission arising from the underlying non-uniformity of the pressure gradients.~\cite{Danielewicz:2002pu}. Another sensitive probe is the yield of sub-threshold $K^+$ mesons which are produced through multiple sequential hadron–hadron collisions~\cite{Aichelin:1985rbt,Fuchs:2005zg}. Their production probability increases with medium density and provides an unperturbed measure of the maximum compression achieved due to their low in-medium cross-section. Together, these observables serve as quantitative probes of the stiffness and incompressibility ($K_0$) of nuclear matter at supra-saturation densities (the theoretical framework to extract EOS is described in Sect.~\ref{subsubsec:methods_hic_theory}). They have been extensively studied by several experiments, spanning densities $1-5n_{0}$ at the Schwerionensynchrotron-18 (SIS-18), Bevalac, Alternating Gradient Synchrotron (AGS) and RHIC. The first experimental observation of collective flow at Bevalac in 1984 was reported by the Plastic Ball collaboration~\cite{Gustafsson:1984ka, Gutbrod:1989wd, Gutbrod:2016ptc}, followed by the Streamer Chamber~\cite{Renfordt:1984et, Danielewicz:1985hn} and DIOGENE at Saturne~\cite{Gosset:1990gm, Demoulins:1990ac}. For the density regime $1-2.5n_{0}$, the Kaon Spectrometer (KaoS) Experiment at SIS-18 measured the sub-threshold $K^+$ production in Au+Au and C+C collisions at $\sqrt{s_{NN}}=2.16-2.70~\textrm{GeV}$~\cite{KAOS:2000ekm,Senger:2022ihi}. Moreover, the Four Pi (FOPI) Experiment at SIS-18 studied the directed ($v_1$) and elliptic flow ($v_2$) of protons and light fragments in Au+Au collisions at $\sqrt{s_{NN}}=1.92-2.52~\textrm{GeV}$~\cite{FOPI:2004bfz,FOPI:2011aa}, while the High-Acceptance DiElectron Spectrometer (HADES) at SIS-18 studied flow coefficients $v_n$ of the orders $n=1-6$ in Au+Au collisions at  $\sqrt{s_{NN}}=2.42~\textrm{GeV}$~\cite{HADES:2020lob,HADES:2022osk}. For densities above $2.5n_{0}$, collective flow has been studied at the AGS (E895~\cite{E895:2000maf, E895:1999ldn} and E877~\cite{E877:1997zjw}; $\sqrt{s_{NN}}=2.70-4.72~\textrm{GeV}$) and by the STAR Experiment at RHIC in its Fixed-Target (FXT) configuration ($\sqrt{s_{NN}}=3.0-4.5~\textrm{GeV}$)~\cite{STAR:2020dav,STAR:2021yiu}. 

\vspace{6pt}
\underline{Symmetry Energy:} The isospin dependence of the EOS is encoded in the symmetry energy, which becomes increasingly important for neutron-rich systems, such as neutron stars with isospin asymmetry $\delta \gtrsim 0.8$. Although heavy-ion collisions typically involve nuclei with relatively small asymmetry, this can be experimentally probed in heavy-ion collisions of highly asymmetric isotopes ($\delta \lesssim 0.25$)~\cite{Li:2014oda,Lynch:2021xkq}. This is reflected in both the differences in collective behaviour and yields of emitted neutrons and protons~\cite{Li:2000bj}, or their isospin partners such as charged pions~\cite{Li:2002qx} to constrain the symmetry energy $\gtrsim 1.5n_{0}$. The elliptic flow ($v_2$) ratio of neutrons and hydrogen has been studied in Au+Au collisions at SIS-18 by the FOPI-LAND~\cite{Russotto:2011hq} and Asymmetric-Matter EOS (ASY-EOS)~\cite{Russotto:2016ucm} experiments at $\sqrt{s_{NN}}=2.07~\textrm{GeV}$, both employing the Land Area Neutron Detector (LAND). Additionally, the SAMURAI Pion Reconstruction and Ion-Tracker (S$\pi$RIT) experiment at RIKEN has systematically explored the charged pion ratio ($\pi^-/\pi^+$) at sub-threshold energy ($\sqrt{s_{NN}}=2.01~\textrm{GeV}$) with radioactive tin collision systems of varying asymmetry (beams of $^{108, 112, 124, 132}\rm{Sn}$ on $^{112, 124}\rm{Sn}$ target)~\cite{SpiRIT:2020sfn,SpiRIT:2021gtq}.

\vspace{6pt}
\underline{Hyperon-Nucleon Interactions:} For dense nuclear matter, such as in the core of a neutron star, hyperons become energetically favourable at $2-3n_{0}$~\cite{1960AZh....37..193A,Vidana:2018bdi}. This reduces the Fermi pressure and the maximum mass of a neutron star below the observed maxima of $\approx 2M_\odot$ from pulsar mass measurements, resulting in the \textit{Hyperon Puzzle}~\cite{Weber:2000xd,Weber:2006da,Schaffner-Bielich:2020psc}. This softening in a hypernuclear EOS can be compensated by additional repulsion caused by two- and three-body hyperon-nucleon interactions~\cite{Weissenborn:2011ut,Bednarek:2011gd,Yamamoto:2014jga,Maslov:2015msa,Gerstung:2020ktv}. This can be experimentally probed in heavy-ion collisions by studying hypernuclei production and femtoscopic correlations~\cite{Tolos:2020aln}, thus providing more statistics data than the earlier nuclear emulsion and scattering experiments. Relativistic heavy-ion collisions both at A Large Ion Collider Experiment (ALICE) at LHC ($\rm{Pb}+\rm{Pb}$ $\sqrt{s_{NN}}=2.76,5.02~\textrm{TeV}$)~\cite{ALICE:2015oer,ALICE:2019vlx,ALICE:2022sco,ALICE:2024koa} and STAR at RHIC ($\rm{Au}+\rm{Au}$ $\sqrt{s_{NN}}=3,200~\textrm{GeV}$)~\cite{STAR:2021orx,STAR:2019wjm} have provided precise constraints on the lifetime and binding energy of the $A=3$ hypernuclei $({}^{3}_{\Lambda}\text{H})$~\cite{Chen:2023mel}. At the same time have harmonic flow measurements of hypernuclei become available \cite{STAR:2022fnj}. In parallel, the availability of large statistics and the development of femtoscopic methods, analysing particle momentum correlations have led to their use to study the strong interaction among the produced hadrons. The STAR experiment at RHIC has paved the way for the feasibility of these measurements in larger collision systems ($\rm{Au+Au}$; source size $5-6\,\rm{fm}$) of 2-body interactions with $\vert S \vert =1$ at $\sqrt{s_{NN}}=200~\textrm{GeV}$~\cite{STAR:2014dcy,STAR:2015kha,STAR:2018uho} and $\sqrt{s_{NN}}=3~\textrm{GeV}$~\cite{STAR:2024zvj}. The sensitivity to short-range interactions is enhanced even more in smaller collision systems ($p\,p$ $\sqrt{s_{NN}}=13~\textrm{TeV}$ and $p+\rm{Pb}$ $\sqrt{s_{NN}}=5.02~\textrm{TeV}$; source size $1-2\,\rm{fm}$), which are studied extensively by the ALICE experiment at LHC~\cite{Fabbietti:2020bfg,Fabbietti:2021cmh} for 2- and 3-body interactions for up to $\vert S \vert =3$~\cite{ALICE:2021njx,Mihaylov:2023ahn,ALICE:2019buq,ALICE:2019hdt,ALICE:2020mfd,ALICE:2022boj,ALICE:2023bny}. Notably, these studies have been used to provide independent constraints on two- and three-particle potentials~\cite{Kievsky:2023maf,Garrido:2024pwi} to study the structure of neutron stars~\cite{Vidana:2024ngv}.

\subsubsection{Theoretical Modelling}
\label{subsubsec:methods_hic_theory}
Despite tremendous efforts and advances in dynamical modelling of heavy-ion collisions during the last 40 years, a consistent description of heavy-ion collisions across all regions in the QCD phase diagram (respectively collision energies) remains challenging due to various reasons. At sufficiently high collision energies, respectively high temperature and low to moderate chemical potential, it is expected that matter becomes deconfined transitioning to a strongly interacting state of quasi-free quarks and gluons. Modelling of this QGP phase in a transport picture remains unfeasible. Therefore the QGP phase is usually modelled with relativistic (viscous) hydrodynamics supplemented by a suitable EOS (extrapolated from lattice QCD results at zero chemical potential \cite{Borsanyi:2013bia,HotQCD:2014kol}) that incorporates the change in microscopic degrees of freedom and the phase transition in a thermodynamic manner, i.e. via the pressure $P(\varepsilon)$. These calculations are commonly supplemented by a subsequent transport model calculation accounting for rescattering effects in the hadronic phase, known as the hadronic afterburner. This combined hybrid-approach has been very successfully applied in the collider energy regime \cite{Petersen:2008dd,Karpenko:2015xea,Li:2008qm,Petersen:2010cw,Huovinen:2012is,Werner:2010aa,Hirano:2010jg,Song:2011qa,Song:2010aq} and current efforts aim to push its applicability towards lower energies \cite{Goes-Hirayama:2025nls}. However, moving towards lower collision energies probing the baryon-rich region of the QCD phase diagram, the hybrid approach becomes less viable, due to the increasingly longer interpenetration time of the colliding nuclei and the subsequently longer equilibration time, while the simplified particilization process adopted from higher energies is limited in the baryon dominated energy regime \cite{Grassi:1994ng,Hung:1997du,Akkelin:2008eh,Knoll:2008sc}. One attempt to overcome these limitations is multi-fluid hydrodynamics~\cite{Katscher:1993xs,Brachmann:1997bq,Ivanov:2005yw,Batyuk:2016qmb,Cimerman:2023hjw,Werthmann:2025ueu} which treat the incoming nuclei as liquids with a friction term that creates the fireball dynamically. This approach appears as a promising tool to pin down the properties of dense nuclear matter at supra-saturation densities, respectively a phase transition, and is well suited for the collisions probed by the upcoming FAIR facility. Until these approaches are fully developed, heavy-ion collisions in the high baryon density regime remain to be mostly calculated within hadronic transport models.

In the high net-baryon density regime relevant for EOS studies and for neutron star physics the course of the collision of two impinging nuclei is typically calculated within transport models. Transport models are based on kinetic theory and describe the interaction of a (dilute) many-body systems based on microscopic degrees of freedom. Generally, transport models can be categorized in two categories: 1. (Vlasov-) Boltzmann-Uehling-Ulenbeck ((V)BUU) type models \cite{Bertsch:1984gb,Aichelin:1985zz,Kruse:1985pg,Cassing:1990dr,Danielewicz:1991dh,Ko:1987gp,Blattel:1988zz,Blaettel:1993uz,Fuchs:1995fa,Buss:2011mx,Weil:2016zrk} and 2. Quantum-Molecular-Dynamics (QMD) type models \cite{Aichelin:1991xy,Aichelin:1986wa,Hartnack:1997ez,Maruyama:1990zz,Sorge:1989dy,Bass:1998ca,Bleicher:1999xi,Graef:2014mra,Aichelin:2019tnk,Nara:2025pkg}, with some exceptions like e.g. Hadron-String-Dynamics (HSD) approaches \cite{Cassing:2008sv,Cassing:2008nn,Cassing:2009vt,Bratkovskaya:2011wp,Linnyk:2015rco,Moreau:2019vhw}. In recent years many reviews and articles have described and compared the available transport models, cf.~\cite{Bleicher:2022kcu,Reichert:2021ljd,Sorensen:2023zkk,TMEP:2022xjg,TMEP:2023ifw}. We will therefore keep the model description compact.

(V)BUU type models \cite{Bertsch:1984gb,Aichelin:1985zz,Kruse:1985pg,Cassing:1990dr,Danielewicz:1991dh,Ko:1987gp,Blattel:1988zz,Blaettel:1993uz,Fuchs:1995fa,Buss:2011mx,Weil:2016zrk} propagate the single-particle distribution function $f$ according to the relativistic Boltzmann-equation (numerics limits the models to single-particle distributions functions). The single-particle distribution function is hence subject to a collision kernel $I_{\rm coll}[f]$ and the mean-field potential $U[f]$. Because $f$ is in general not analytic, the transport equation is typically solved numerically using the test particle ansatz which represents the single-particle distribution function by $N_{\rm test}$ test particles. The cross sections for binary scatterings are treated stochastically and the mean-field is usually inferred from the ensemble-averaged phase space distribution.

In contrast to (V)BUU type models QMD type models \cite{Aichelin:1991xy,Aichelin:1986wa,Hartnack:1997ez,Maruyama:1990zz,Sorge:1989dy,Bass:1998ca,Bleicher:1999xi,Graef:2014mra,Aichelin:2019tnk,Nara:2025pkg} do not rely on test particles, but treat each of the impinging nuclei as a product state of individual nucleon wave functions. Upon assuming that the nucleon wave functions are of Gaussian shape, their centroids propagate according to Hamilton's equations of motion. The approach can be extended to additionally satisfy anti-symmetrization which is done in AMD type codes \cite{Ono:1991uz,Ono:1992uy}. QMD codes are complemented by a scattering term similar to (V)BUU type models and the EOS is implemented in QMD type models by means of a two- or n-body potential $V(\rho_B)$ (directly related to the single-particle potential $U(\rho_B)$) which enters the equations of motion through the Hamiltonian $\langle H \rangle = \langle T \rangle + \langle V \rangle$.

For both (V)BUU and QMD type transport models there exists a variety of different numerical realizations, which all rely on slightly different physical assumptions or numerical treatments. Model differences typically occur in the treatment of Pauli-blocking, the realization of relativistic invariance of the propagation and/or the mean-field, the degrees of freedom or, respectively, the hadron list, in-medium effects, the momentum dependence of the EOS, cluster production and many more. Albeit the huge success transport models have had during the last 40 years, model differences are, unfortunately, sometimes larger for given observables than the choice of the EOS and therefore make the extraction of the EOS from experimental data challenging, especially in light of the upcoming precision era.

Notably, the Transport Model Evaluation Project (TMEP)~\cite{TMEP:2022xjg} is a large scale project with the goal to evaluate existing transport codes in controlled settings (e.g. using the same initial conditions, cross sections, EOS, etc.). TMEP has been very successful in establishing transport code benchmarks, e.g. by comparing collision integrals in a box~\cite{TMEP:2017mex,TMEP:2019yci,TMEP:2021ljz} or pion production~\cite{TMEP:2023ifw}. However, at the moment this systematic code comparison is focused towards collision energies below $E_{\rm lab} \lesssim 1A$ GeV~\cite{TMEP:2023ifw}. The project thus helps to reduce and understand modelling uncertainties in the FRIB and low GSI energy regime, however, is not yet suitable for the upper GSI, RHIC-FXT and the upcoming FAIR energies. The net-baryon densities probed at FRIB and lower GSI reach $\sim 2n_{\rm sat}$ at maximum, while the GSI, RHIC and FAIR colliders will probe the high density region with $\sim 2n_{\rm sat}-6n_{\rm sat}$. An extension of the existing TMEP project or a complementary comparison of existing transport codes in the collision energy range of $E_{\rm lab} = 1 - 10A$ GeV is therefore highly desired. This is especially important for EOS sensitive observables such as flow harmonics or sub-threshold kaon production.

\subsubsection{Heavy-Ion Collision Constraints}
\label{subsubsec:methods_hic_constraints}

This motivates current and future programs such as the RHIC Beam Energy Scan, HADES at GSI, and the upcoming CBM experiment at FAIR, which focus on the high-density regime where the order of the QCD phase transition and the possible onset of new degrees of freedom remain open questions. Together, these efforts map complementary regions of the EOS: symmetric matter is best constrained by high-energy experiments, while the symmetry energy requires precise measurements at low and intermediate energies with neutron-rich systems. Although no transport model can yet reproduce all observables consistently, the growing availability of multi-differential data, coupled with systematic model–data comparisons, is steadily reducing uncertainties. Heavy-ion collisions thus remain central to constraining the EOS, both as an independent probe and as a bridge to astrophysical observations.

\begin{figure*}[!t]
    \centering
    \includegraphics[width=0.49\linewidth]{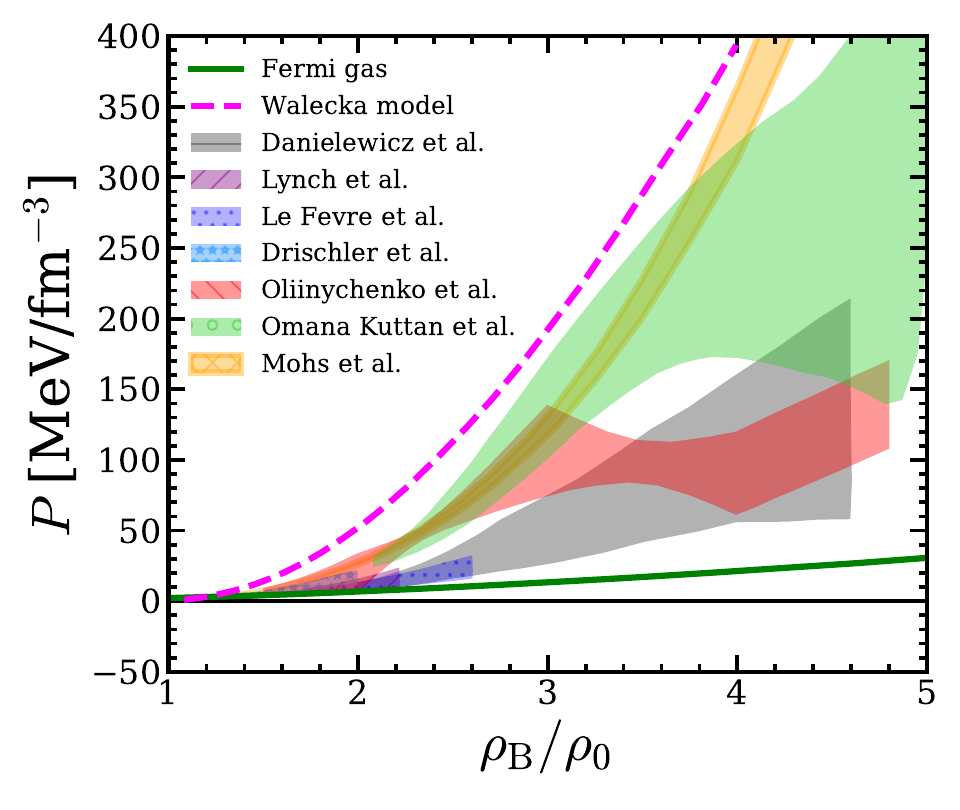}
    \includegraphics[width=0.49\linewidth]{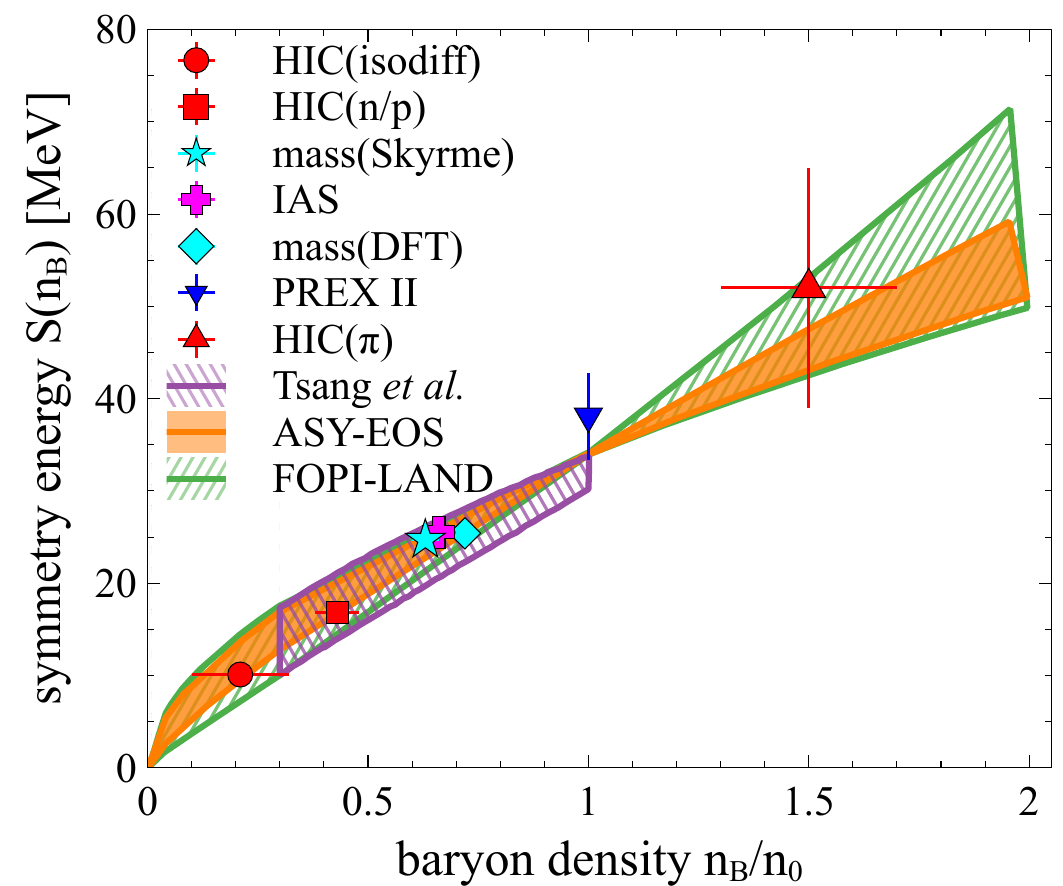}
    \caption{Selected constraints on the pressure $P(n_B)$ of symmetric matter~\cite{Danielewicz:2002pu, Fuchs:2003pc, Lynch:2009vc, LeFevre:2015paj, Drischler:2020yad, Oliinychenko:2022uvy, OmanaKuttan:2022aml, Mohs:2024gyc} (left) and the symmetry energy $S(n_B)$~\cite{Tsang:2008fd, Russotto:2011hq, Russotto:2016ucm, Lynch:2021xkq, Kortelainen:2010hv, Kortelainen:2011ft, Brown:2013mga, Danielewicz:2013upa, PREX:2021umo} (right) as functions of baryon density ($n_B/n_0$), extracted from comparisons of experimental data with hadronic transport simulations. Right figure from~\cite{Sorensen:2023zkk}.} 
    \label{fig:joint_constraints_EOS}
\end{figure*}

In practice the most common and relevant method to extract information about the dense matter EOS is by means of harmonic flow coefficients $v_n(y,p_\mathrm{T})$, first proposed in \cite{Voloshin:1994mz}, which arise from the Fourier decomposition of the azimuthal angular distribution of particles. Both the presence of initial state fluctuations as well as the (average) collision geometry create density gradients which are translated into pressure in the transverse plane. This pressure response translates into the azimuthal shape in momentum space of the final state particles and thus allows to constrain the EOS, respectively the pressure, of dense nuclear matter. Although the precise generating mechanism for the flow coefficients is rather complicated \cite{Reichert:2024ayg,Li:2022wvu}, the EOS still directly controls the magnitude of the $v_n$ coefficients. Traditionally the directed flow $v_1$ and elliptic flow $v_2$ have been studied predominantly, however, modern experiments with high-rate data recording capabilities and larger computational resources allow the study of higher harmonics and their correlations as well. Especially the investigation of triangular flow $v_3$ {\color{black}with respect to the event plane of first order} appears as a promising new tool offering more sensitivity to the EOS \cite{Hillmann:2018nmd, Hillmann:2019wlt, HADES:2020lob, HADES:2022osk, STAR:2023duf}. The direct impact of the pressure in the system onto the final state observables makes harmonic flow coefficients especially useful to constrain the pressure by means of a systematic parameter study or Bayesian inference. Complementary to harmonic flow as a tool to constrain the EOS is the investigation of charged pion production and sub-threshold Kaon production which have been extensively studied in theory \cite{Aichelin:1985rbt,Hartnack:1993bq,Maruyama:1993jb,Fang:1994cm,Li:1994xt,Fuchs:1997cp,Fuchs:2000kp,Fuchs:2005zg,Hartnack:2005tr,Ferini:2006je} but also experimentally \cite{Miskowiec:1994vj,KaoS:1999gal,KAOS:2000ekm,FOPI:2005tyo}. These efforts also highlight the importance for FAIR to continue and advance the investigation of deep sub-threshold hadron production of multi-strange and charmed observables. The left panel of fig. \ref{fig:joint_constraints_EOS} shows a compilation of pressure constraints {\color{black} from the last two decades, focusing on directed and elliptic flow measurements in the beginning, later including data from sub-threshold Kaon production, and finally obtaining results by Bayesian inference utilizing the plethora of available data} \cite{Danielewicz:2002pu, Fuchs:2003pc, Lynch:2009vc, LeFevre:2015paj, Drischler:2020yad, Oliinychenko:2022uvy, OmanaKuttan:2022aml, Mohs:2024gyc}. 
{\color{black} The low density region has been well constrained. At higher densities above $\sim 2n_\mathrm{sat}$, there is less agreement between the different model calculations. However, most observables (e.g. harmonic flow) are sensitive to the $2-3n_\mathrm{sat}$ region (even when higher densities are reached during the course of the collision \cite{LeFevre:2016vpp,Reichert:2024ayg}). Thus, the challenge for the next decades is to combine precise multi-differential data including new observables with continuous improvements of theoretical models together with Bayesian methods to constrain the high density region.}

\subsection{Multi-Messenger Astrophysics}
\label{subsec:methods_mma}

In addition to terrestrial experiments, the properties of supranuclear matter can also be studied through observations of neutron stars, particularly in environments with low temperatures and high baryonic chemical potentials, as illustrated in Fig.~\ref{fig:QCD-PD}. Various methods and channels have been employed to observe neutron stars, each providing distinct measurements. For instance, the electromagnetic signatures of individual neutron stars can be analyzed to determine their mass and radius. Conversely, in the case of binary neutron star mergers, the tidal deformabilities of the two stars are measured using the gravitational-wave signals they emit. These different observation methods explore various regimes within the EOS of neutron stars, which will be briefly summarized in this section.

\subsubsection{Isolated neutron stars}
By an isolated neutron star, we are referring to the observations that focus on the properties of a singular neutron star, while the star itself can, and most likely, be in a binary configuration.

In an edge-on binary system that contains a pulsar, the radio pulses emitted by the pulsar travel through the gravitational field of its companion star before reaching us. This interaction causes the radio signals to experience effects such as the Shapiro time delay, along with other relativistic phenomena~\cite{Demorest:2010bx}. By conducting precise timing measurements of these pulses, we can determine the mass of the companion star. Additionally, we can calculate the total mass of the binary system using the binary orbital period and velocity, which also allows us to ascertain the mass of the pulsar itself. This method has led to the mass measurement of some of the heaviest known neutron stars. For instance, the mass of PSR J0740+6620 $M$ has been determined to be $2.08 \pm 0.07 \ M_\odot$, which establishes the commonly accepted lower bound of the TOV mass to be approximately $2 \ M_\odot$.

In addition to radio signals, neutron stars can also be observed through X-ray emissions. The thermal hot spots on a neutron star, which are caused by electron-positron pair cascades~\cite{Harding:2001gm, Harding:2001at, Medin:2010tp}, lead to periodic fluctuations in X-ray emissions as the star rotates on its axis. These fluctuations depend on various factors, including the properties of the star's atmosphere, the nature of the hot spots, and, most importantly, the star's compactness and radius. The compactness affects the signal through gravitational lensing, while the radius influences the velocity and therefore the observed Doppler effect. The Neutron Star Interior Composition Explorer (NICER) main goal is to observe pulsars' X-ray emission to constrain the neutron star EOS. For PSR J0740+6620, by combining the previously measured mass using radio observation, NICER has measured its radius to be $12.49^{+1.28}_{-0.88}\ {\rm km}$~\cite{Salmi:2024aum}, and $12.92^{+2.09}_{-1.13}\ {\rm km}$~\cite{Dittmann:2024mbo}.

\subsubsection{Binary neutron stars}
In addition to studying isolated neutron stars, binary neutron star mergers offer valuable insights into the EOS of neutron stars. These mergers are multi-channel events, meaning they can be observed through gravitational waves, electromagnetic radiation, and potentially neutrinos. In this discussion, we will focus solely on the gravitational and electromagnetic channels.

The gravitational-wave signals from two binary neutron star mergers, detected by Advanced LIGO~\cite{LIGOScientific:2014pky} and Advanced Virgo~\cite{VIRGO:2014yos}, namely, GW170817~\cite{LIGOScientific:2017vwq} and GW190425~\cite{LIGOScientific:2020aai}, have initiated the effort to constrain the EOS of neutron stars using gravitational-wave observations. During the inspiral of the two neutron stars, energy and angular momentum are lost due to gravitational radiation, causing the stars to spiral into each other and deform under the tidal field of their companion. This deformation leaves an imprint on the observed signal, which is determined by the tidal deformabilities of the individual stars. Since the tidal deformability is heavily dependent on the neutron star's equation of state, this relationship enables us to constrain the EOS through these observations. Based on GW170817, the tidal deformability of a $1.4M_\odot$ neutron star is projected to be $190^{+390}_{-120}$~\cite{LIGOScientific:2018cki}.

GW170817 was linked to a variety of electromagnetic signals, including the kilonova AT2017gfo~\cite{LIGOScientific:2017pwl, Andreoni:2017ppd, Coulter:2017wya, Lipunov:2017dwd, Shappee:2017zly, Tanvir:2017pws, J-GEM:2017tyx}, a short gamma-ray burst (GRB) known as GRB170817A~\cite{LIGOScientific:2017zic, Goldstein:2017mmi, Savchenko:2017ffs}, and its afterglow~\cite{Hallinan:2017woc, Alexander:2018dcl, Margutti:2018xqd, Ghirlanda:2018uyx, Troja:2017nqp, DAvanzo:2018zyz}. The kilonova, a type of pseudo-blackbody radiation, is produced by the radiative decay of heavy elements synthesized in the material ejected during the merger of binary neutron stars via the r-process. This kilonova was first detected approximately 11 hours after the merger and lasted for about two weeks. The GRB, associated with the launch of a relativistic jet, was observed around 1.7 seconds after the merger. The afterglow of the GRB has been observed in radio and X-ray channels more than a week following the merger and has continued to be monitored for several years~\cite{Wang:2022msf}.

In contrast to isolated neutron stars or the gravitational waves produced by binary neutron star mergers, electromagnetic signals are directly related to the behavior of matter at the moment of merger. For example, the amount of mass ejected during the merger is correlated with the maximum mass that a given EOS can support~\cite{Dietrich:2020efo}. As a result, these signals are expected to probe densities that exceed those found in the centers of individual neutron stars. This offers valuable information that cannot be obtained from neutron stars in equilibrium.

\subsubsection{Astrophysical constraints}
\begin{figure*}
    \centering
    \includegraphics[width=0.48\linewidth]{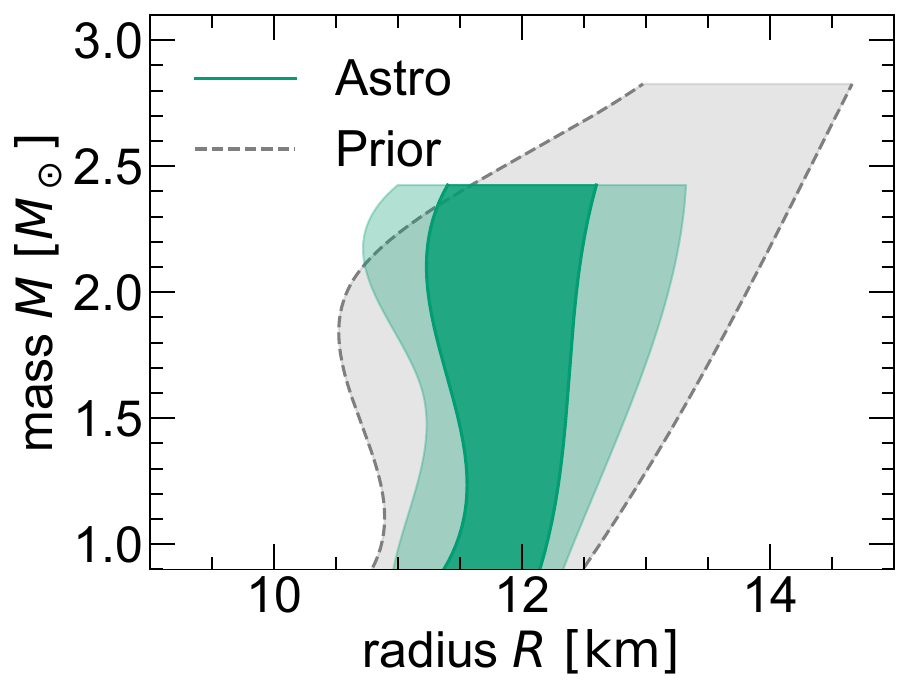}
    \includegraphics[width=0.51\linewidth]{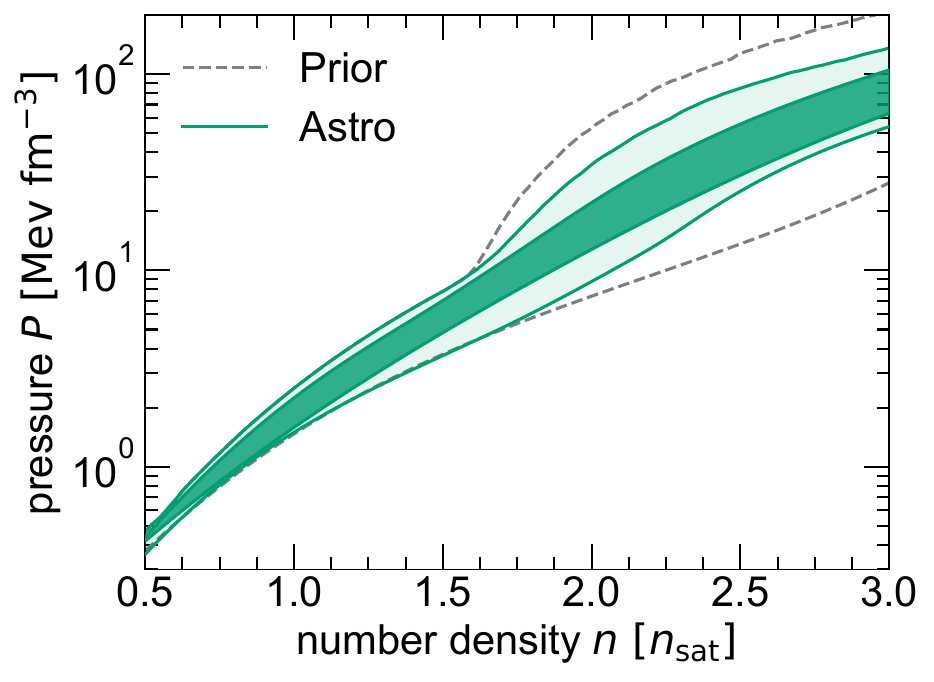}
    \caption{Constraints on the neutron star EOS derived from a combination of multiple astrophysical observations, including electromagnetic and gravitational channels. Posterior on the neutron star EOS presented in the mass-radius (left) and pressure-density (right) plane. The shading corresponds to the $95\%$ (light green) and $68\%$ credible intervals (dark green), the $95\%$ prior bound is shown for comparison (grey dashed lines). The images are sourced from Ref.~\cite{Huth:2021bsp}.}
    \label{fig:astro_only_constraint}
\end{figure*}

By combining the previously mentioned astrophysical observations on neutron stars with various aspects, robust constraints on the neutron star EOS are obtained. Such constraints can be visualized in Fig.~\ref{fig:astro_only_constraint}. While the constraints on densities below $1.5n_{\rm sat}$ are not pronounced, the constraints on densities above are significant and showcase the unique value from these neutron star observations.

\section{Frameworks to Combine Multiple Information Sources}
\label{sec:frameworks}

\subsection{Nuclear Multi-Messenger Astronomy (NMMA) framework}
\label{subsec:nmma}

The Nuclear Multi-Messenger Astronomy (NMMA) framework~\cite{Pang:2022rzc}\footnote{\url{https://github.com/nuclear-multimessenger-astronomy/nmma}} is an open-source Python package designed for conducting Bayesian multi-messenger parameter estimation, with a focus on joint analyses of gravitational-wave and electromagnetic data. While originally developed for binary neutron star mergers to analyze the gravitational-wave signal, kilonova, and GRB afterglow simultaneously, its scope has expanded to analyzing and classifying multi-component astronomical transients~\cite{Hussenot-Desenonges:2023oll, Kann:2023ulv, Kunert:2023vqd}.

NMMA's flexible interface enables integration of diverse astrophysical observations on neutron stars, including radio mass measurements and NICER's mass-radius measurements, as well as theoretical and experimental nuclear constraints. This comprehensive approach has yielded some of the strongest constraints on dense nuclear matter~\cite{Koehn:2024set}, as shown in Fig.~\ref{fig:nmma_results}.

The multi-channel observation of GW170817 demonstrates this multi-messenger approach, combining the gravitational-wave signal~\cite{LIGOScientific:2017vwq}, the kilonova AT2017gfo~\cite{LIGOScientific:2017pwl, Andreoni:2017ppd, Coulter:2017wya, Lipunov:2017dwd, Shappee:2017zly, Tanvir:2017pws, J-GEM:2017tyx}, the gamma-ray burst GRB170817A~\cite{LIGOScientific:2017zic, Goldstein:2017mmi, Savchenko:2017ffs}, and its afterglow~\cite{Hallinan:2017woc, Alexander:2018dcl, Margutti:2018xqd, Ghirlanda:2018uyx, Troja:2017nqp, DAvanzo:2018zyz}. This milestone event has provided significant constraints on the neutron star equation of state (EOS)~\cite{Dietrich:2020efo,Huth:2021bsp,Essick:2021kjb,Legred:2021hdx,Pang:2021jta,Pang:2022rzc,Essick:2023fso,Pang:2023dqj,Tsang:2023vhh,Koehn:2024set,Legred:2025aar,Finch:2025bao}.

Building on these astrophysical successes, the multi-messenger concept has expanded to incorporate theoretical nuclear physics calculations, such as chiral effective field theory~\cite{Tews:2012fj} and perturbative quantum chromodynamics (QCD)\cite{Freedman:1976ub, Kurkela:2009gj}, along with experimental results from PREX-II\cite{PREX:2021umo}, CREX~\cite{CREX:2022kgg}, and heavy-ion collision experiments like FOPI~\cite{LeFevre:2015paj} and ASY-EOS~\cite{Russotto:2016ucm}.

The NMMA framework addresses the need for a robust and flexible Bayesian analysis platform capable of handling current and future multi-messenger constraints through its modular design, which facilitates the seamless integration of additional observational and theoretical inputs.

{\color{black}
\subsubsection{EOS prior construction}
In our latest iteration of the NMMA framework, we construct a set of 100,000 EOS candidates using a procedure similar to that described in Ref.~\cite{Dietrich:2020efo}. At low densities, matter is assumed to be purely nucleonic, while a model-agnostic scheme is used at high densities to accommodate possible exotic phases. At the lowest densities, neutron star matter forms a solid crust described by the model of Ref.~\cite{Douchin:2001sv}, which is held fixed for all EOSs. This crust EOS is applied up to the crust-core transition density of $0.076~\mathrm{fm}^{-3}$.

Above the crust, purely nucleonic matter in beta equilibrium is assumed up to a breakpoint density $n_\mathrm{break}$, drawn uniformly from $[1,2]\,n_\mathrm{sat}$. The EOS in this regime is modelled using the metamodel (MM) of Refs.~\cite{Margueron:2017eqc, Margueron:2017lup}, a density functional approach in which the energy per particle in symmetric matter and the symmetry energy are expanded in Taylor series about saturation density, with coefficients given by nuclear empirical parameters (NEPs). These NEPs are sampled uniformly over the ranges as shown in Table 1 in Ref.~\cite{Koehn:2024set}. The crust and core EOSs are joined via a cubic spline in the sound speed between $0.076$ and $0.12~\mathrm{fm}^{-3}$.

Above $n_\mathrm{break}$, we use a speed-of-sound construction following Ref.~\cite{Tews:2018iwm}. The squared sound speed is sampled uniformly in $[0, c^2]$ at nine density grid points between $n_\mathrm{break}$ and $25\,n_\mathrm{sat}$, with linear interpolation in between. The resulting $c_s^2(n)$ profile is integrated to yield the full EOS. First-order phase transitions are not included explicitly, but the sampling scheme allows for arbitrarily soft EOSs that closely mimic them.

This EOS set defines a natural prior on derived quantities such as the canonical radius $R_{1.4}$ and the TOV mass $M_\mathrm{TOV}$. For constraints that admit a direct likelihood evaluation, posteriors are obtained by reweighting the candidates by $\mathcal{L}(\mathrm{EOS}|d)$. For the gravitational-wave analysis and the joint analysis of gravitational-wave, kilonova, and GRB afterglow data, the candidate set instead serves as the prior for nested sampling using \textsc{dynesty}~\cite{Speagle:2019ivv}.
}
\subsubsection{Gravitational-wave likelihood}
Given a gravitational wave data strain $d$, the likelihood for the specified parameters $\vec{\theta}$, denoted as $\mathcal{L}(\vec{\theta})$, is defined as
\begin{equation}
    \label{eq:likelihood}
    \mathcal{L}(\vec{\theta}) \propto \exp\left(-\frac{1}{2}\langle d - h(\vec{\theta})  \vert d - h(\vec{\theta})\rangle\right),
\end{equation}
where $h(\vec{\theta})$ is the gravitational-wave waveform generated using a particular waveform model for parameters $\vec{\theta}$ and $\langle a \vert b \rangle$ is the noise-weighted inner product, defined as
\begin{equation}
    \langle a \vert b \rangle = 4\Re\int^{f_{\rm max}}_{f_{\rm min}}df \frac{\tilde{a}\tilde{b}^*(f)}{S_n(f)},
\end{equation}
where tilde denotes the Fourier transform and $S_n(f)$ is the one-sided power spectral density of the noise. \textcolor{black}{It is worth noting here that Eq.~\eqref{eq:likelihood} assumes stationary Gaussian noise.} The frequency range $[f_{\rm min}, f_{\rm max}]$ is selected to ensure that the entire signal in the band is captured. For a binary neutron star signal, this is typically $[20{\rm Hz}, 2048 {\rm Hz}]$.

For a standard parameter estimation of a binary neutron star signal, the parameters include the component masses $m_{1,2}$, component spins $\vec{s}_{1,2}$, and their tidal deformabilities $\Lambda_{1,2}$. These 10 parameters are referred to as intrinsic parameters. In addition, there are 6 extrinsic parameters, which include luminosity distance $d_{\rm L}$, inclination $\iota$, sky location $(\alpha, \delta)$, merger time $t_{\rm c}$, and merger phase $\phi_{\rm c}$.

For the waveform model, NMMA can make use of any models available in the \texttt{LALSimulation} library, one of the typical choices, which is also one of the state-of-the-art models, is the \texttt{IMRPhenomXP\_NRTidalv3} waveform model. The model combines the IMRPhenomXP model~\cite{Pratten:2020ceb} with the newly developed NRTidalv3 model, which accounts for tidal effects~\cite{Abac:2023ujg}. IMRPhenomXP describes the (2,2) mode of a precessing circular binary of point masses based on a phenomenological approach. Its advantage over previous models lies in a more refined description of the inspiral phase and its calibration to a broader set of merger simulations in numerical relativity. For the modeling of binary neutron star systems, NRTidalv3 adds tidal phase contributions to the IMRPhenomXP waveform. Unlike its predecessors, this model utilizes a larger set of numerical relativity simulations that encompass systems with high mass ratios and a wide range of EOSs. NRTidalv3 also incorporates dynamical tides, taking into account that tidal deformability is not adiabatic but rather a function of the gravitational wave frequency.

\subsubsection{Electromagnetic-wave likelihood}
Given a set of photometry data in AB magnitude $m(t_j, d)$ with a statistical error $\sigma_{{\rm stat}}(t_j)$, at time $t_j$ and wavelength $f$, the likelihood of the photometry data is given by
\begin{equation}
\mathcal{L}(\vec{\theta}) \propto \exp\left(-\frac{1}{2} \sum_{f,j} \frac{(m^{\rm est}(t_j, \vec{\theta}) - m(t_j,d))^2}{(\sigma_{\rm sys})^2 + (\sigma_{{\rm stat}}(t_j))^2}\right),
\end{equation}
where the expected AB magnitudes $m^{\rm est}(t_j, \vec{\theta})$ with parameters $\vec{\theta}$.

An auxiliary systematic uncertainty, denoted as $\sigma_{\rm sys}$, is introduced to account for systematic errors in the light curve models. Various approaches have been employed for this purpose. Some methods set $\sigma_{\rm sys}$ to fixed values~\cite{Dietrich:2020efo,Heinzel:2020qlt,Pang:2022rzc}, while others sample it in order to marginalize the uncertainty~\cite{Hussenot-Desenonges:2023oll,Kann:2023ulv,Kunert:2023vqd}. In the latest developments, $\sigma_{\rm sys}$ is sampled as a function of both time and frequency~\cite{Jhawar:2024ezm}.

As the likelihood suggests, one requires a model that links physical system parameters, such as the energy of the central engine, ejecta mass, and velocity, to the emitted light curves. Several different models for the kilonova emission are available in the literature, e.g., Refs.~\cite{Kasen:2017sxr, Dietrich:2020efo, Anand:2023jbz}, and as well as for the GRB afterglow emission, e.g., Refs~\cite{Ryan:2023pzk, Nedora:2024vrv, Wang:2024wbt}.

To simulate the lightcurves emitted by a kilonova, one can rely on semi-analytical models, e.g., Refs~\cite{Kasen:2017sxr,Nicholl:2021rcr} or numerical simulations with the so-called radiative transfer code, one of which is \texttt{POSSIS}. \texttt{POSSIS}~\cite{Bulla:2019muo, Anand:2023jbz, Koehn:2025zzb} is a three-dimensional Monte Carlo radiative transfer code. The material from the ejecta undergoes homologous expansion, and the emitted photon packages are calculated based on temperature and opacity distributions. The latest model from \texttt{POSSIS} utilizes five intrinsic parameters, the masses and velocities of both the dynamical and wind ejecta, as well as the average electron fraction of the dynamical ejecta. For extrinsic parameters, other than the trivially dependent luminosity distance, the view angle also plays a major role in the observed light curves. In comparison to its predecessor, the new model benefits from enhanced prescriptions of heating rates, thermalization efficiencies, and opacities in \texttt{POSSIS}. For further details, we refer readers to Refs.~\citep{Bulla:2022mwo, Anand:2023jbz, Koehn:2025zzb}.

The GRB afterglow is typically modeled using semi-analytical frameworks~\cite{Ryan:2023pzk, Nedora:2024vrv, Wang:2024wbt} for efficient parameter estimation. In these models, the observed emission arises from relativistic electrons spiraling around magnetic field lines, which are amplified by collisionless shock processes. The electrons are accelerated via diffusive shock acceleration, and their distribution is described by the power law $dn/d\gamma \propto \gamma^{-p}$, where $\gamma$ is the electron Lorentz factor and $p$ is the spectral index. The microphysics is encapsulated by the equipartition parameters $\epsilon_e$ and $\epsilon_B$, representing the fractions of post-shock energy contained in electrons and magnetic fields, respectively. These parameters, along with $p$, determine the characteristic synchrotron spectrum of the afterglow, with synchrotron self-absorption typically neglected at the frequencies and timescales of interest.

The jet structure is discretized into axisymmetric conical layers, each defined by its initial mass, velocity, and polar angle. The angular dependence of the kinetic energy can be prescribed using various profiles; here we adopt a Gaussian distribution, $E(\theta) \propto E_0 \exp(-\frac{1}{2}\theta^2/\theta_c^2)$, truncated at $\theta_w$, where $\theta_c$ is the core width. The dynamical evolution of each layer is computed semi-analytically using the thin-shell approximation, which recasts energy conservation and shock-jump conditions into evolution equations for the blast-wave Lorentz factor and radius. Lateral expansion, driven by the pressure gradient perpendicular to the shock front, is included and becomes significant at late times when the jet decelerates. The observed flux density is then obtained by integrating over equal-arrival-time surfaces,
\begin{equation}
F_\nu = \frac{1}{4\pi d_L^2} \int d\theta\, d\phi\, R^2 \sin\theta\, \epsilon_\nu \alpha_\nu (1 - e^{-\tau}),
\end{equation}
where $\epsilon_\nu$ and $\alpha_\nu$ are the synchrotron emissivity and absorption coefficients, $\tau$ is the optical depth, and $d_L$ is the luminosity distance. This integration accounts for relativistic beaming and time-delay effects, ensuring that the total observed $F_\nu$ properly includes emission from regions with different comoving times and Doppler factors.

While the light curves of GRB afterglows can be evaluated quickly enough for likelihood assessments, this is not the case for \texttt{POSSIS}. The extensive computation time required for generating a single \texttt{POSSIS} light curve can take several hours, making it impractical to conduct the numerous likelihood evaluations needed during sampling. To address this issue, we use a feed-forward neural network to interpolate the light curves for any given point in the parameter space, based on a fixed grid of \texttt{POSSIS} simulations~\cite{Anand:2023jbz, Pang:2022rzc}. Recent advancements have led to the development of surrogate models that estimate the spectral flux across a wide range of frequencies. These models make use of simple feed-forward neural networks and conditional variational autoencoders (cVAEs)~\cite{Koehn:2025zzb}. Similar techniques are also employed to accelerate the evaluation of GRB afterglow light curves~\cite{Koehn:2025zzb}.

\subsubsection{Joint EOS inference}
To connect the observed properties of the kilonovae, and GRB afterglows to the characteristics of their binary systems, we rely on phenomenological relations, which are fits derived from numerical-relativity simulations. We will showcase the fits presented in Ref.~\cite{Kruger:2020gig} and Ref.~\cite{Dietrich:2020efo} as examples. However, it is important to note that various alternative fitting formulas can also be found in the literature~\cite{Dietrich:2016fpt,Coughlin:2018miv,Radice:2018pdn,Coughlin:2018fis,Nedora:2020qtd}. 

In NMMA, the dynamical ejecta mass $m^{\rm ej}_{\rm dyn}$ is connected to the binary properties through the phenomenological relation~\cite{Kruger:2020gig}
\begin{equation}
    \frac{m^{\rm ej}_{\rm dyn, fit}}{10^{-3}M_{\odot}} = \left(\frac{a}{C_1} + b\left(\frac{m_2}{m_1}\right)^n + cC_1\right) + (1 \leftrightarrow 2),
\end{equation}
where $m_i$ and $C_i$ represent the masses and compactness of the two components of the binary system. The best-fit coefficients are given as $a = -9.3335$, $b = 114.17$, $c = -337.56$, and $n = 1.5465$. This relationship enables a precise estimation of the ejecta mass, which is associated with a well-approximated error characterized by a zero-mean Gaussian distribution. The standard deviation of this distribution is $0.004M_{\odot}$~\cite{Kruger:2020gig}. Consequently, the dynamical ejecta mass can be estimated as follows,
\begin{equation}
    m^{\rm ej}_{\rm dyn} = m^{\rm ej}_{\rm dyn, fit} + \alpha,
\end{equation}
where $\alpha\sim\mathcal{N}(\mu=0, \sigma=0.004M_{\odot})$.

To determine the disk mass $m_{\rm disk}$, we follow the description of Ref.~\cite{Dietrich:2020efo},
\begin{align}
    \log_{10}\left(\frac{m_{\rm disk}}{M_{\odot}}\right) = \textrm{max}\left(-3, a\left(1 + b\tanh\left(\frac{c - (m_1+m_2) M_{\rm threshold}^{-1}}{d}\right)\right)\right), \label{eq:Mdisk_fit}
\end{align}
with $a$ and $b$ given by
\begin{equation}
\begin{aligned}
    a &= a_o + \frac{\delta a}{2}\tanh\left(\beta \left(q-q_{\rm trans}\right)\right)\,\\
    b &= b_o + \frac{\delta b}{2}\tanh\left(\beta \left(q-q_{\rm trans}\right)\right)\,,
\end{aligned}
\end{equation}
where $a_o$, $b_o$, $\delta a$, $\delta b$, $c$, $d$, $\beta$, $q_{\rm trans}$ are free parameters with $q \equiv m_2/m_1 \leq 1$ is the mass ratio. The best-fit model parameters are $a_o=-1.581$, $\delta a=-2.439$, $b_o=-0.538$, $\delta b=-0.406$, $c =0.953$, $d=0.0417$, $\beta=3.910$, $q_{\rm trans}=0.900$.

The threshold mass $M_{\rm threshold}$ for a given EOS is estimated as~\cite{Agathos:2019sah}
\begin{equation}
    M_{\rm threshold} = \left(2.38 - 3.606 \frac{M_{\rm TOV}}{R_{1.6}}\right) M_{\rm TOV},
\end{equation}
where $M_{\rm TOV}$ and $R_{1.6}$ are the maximum mass of a neutron star and the radius of a $1.6M_{\odot}$ neutron star.

We note that we assume that the disk-wind ejecta component is proportional to the disk mass, i.e., $m^{\rm ej}_{\rm wind} = \xi m_{\rm disk}$. And the energy of the central engine for the GRB afterglow is proportional to the leftover disk mass, i.e., $E_0 = \epsilon(1 - \xi) m_{\rm disk}$.

\subsubsection{Integration of nuclear theoretical calculation and experimental constraints}
In the above-mentioned analysis, one needs to provide a set of candidate EOSs for the analysis, which is also a natural place for NMMA to take into account additional constraints or information. To integrate the nuclear information into the astrophysical observations, NMMA allows for mainly two channels to do so, i) implicitly included in the EOS set used for analysis, and ii) explicitly as the weighting for each EOS in the EOS set.

The former approach has been taken to account for the chiral effective field theory (CEFT)~\cite{Hebeler:2013nza,Tews:2012fj,Lynn:2016piq,Drischler:2017wtt} in Ref.~\cite{Dietrich:2020efo}. In that study, the set of candidate EOSs is generated using CEFT calculations up to $1.5n_{\rm sat}$, after which a generic speed-of-sound extension is used to explore the potential new degrees of freedom realized inside a neutron star. This approach, which involves implicit inclusion of prior constraints, is simpler to implement for various theoretical and experimental conditions. It only requires the prediction of some representations of the neutron star EOS within their framework. However, a major disadvantage is the lack of flexibility, all subsequent analyses and related outcomes will depend on these prior constraints, making comparisons with other approaches more challenging.

\begin{figure*}[!t]
    \centering
    \includegraphics[width=0.9\linewidth]{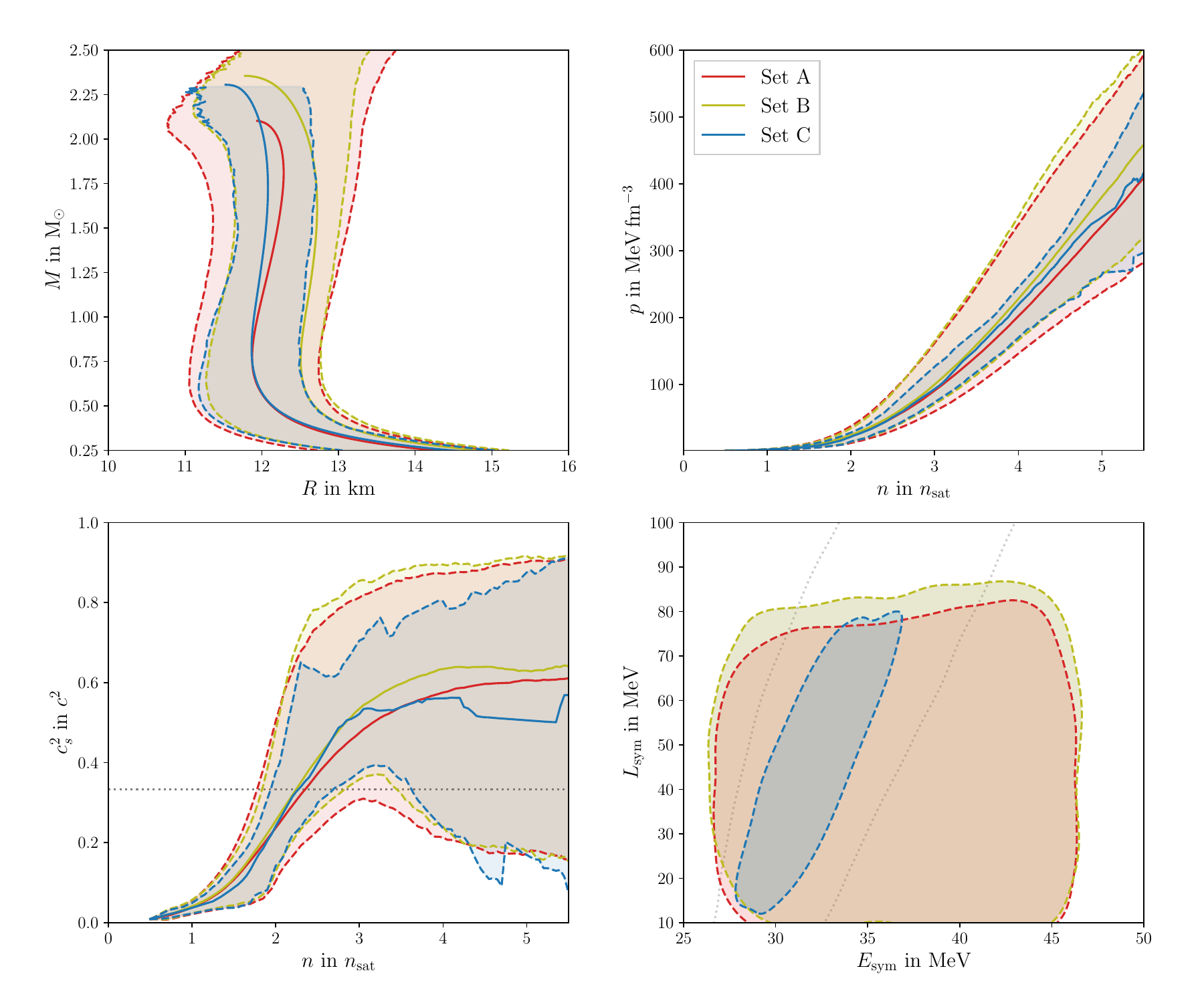}
    \label{fig:nmma_results}
    \caption{The final posterior estimates for the mass-radius ($M$-$R$) and pressure-number density ($p$-$n$) relationships, the speed of sound, and symmetry-energy parameters are presented based on three different sets of constraints: A (red), B (yellow), and C (blue).  In the top left panel, the solid lines represent the $M$-$R$ curves with the highest posterior likelihood, while the dashed lines indicate the 95\% credibility intervals for radius at each given mass. The top-right panel similarly shows the 95\% credible intervals for pressure as a function of number density, with the solid line indicating the median. The bottom left panel illustrates the same quantities for the speed of sound as a function of number density, again with the median represented by solid lines. The gray dotted line in this panel denotes the conformal limit, where $c^2_s = c^2/3$. Finally, the bottom right panel displays the 95\% credible regions for the nuclear symmetry parameters $E_{\rm sym}$ and $L_{\rm sym}$. The gray dotted line represents the 95\% credible region for these parameters based on the combined results of PREX-II and CREX.} 
\end{figure*}

In the latter approach, a generic candidate EOS set is used, with the information provided in different forms of likelihoods. Such an approach is taken in Ref.~\cite{Koehn:2024set}, where likelihoods of CEFT, perturbative QCD~\cite{Gorda:2018gpy,Gorda:2021znl,Gorda:2023mkk}, PREX-II~\cite{PREX:2021umo}, CREX~\cite{CREX:2022kgg} and heavy-ion collision~\cite{Huth:2021bsp,LeFevre:2015paj,Russotto:2016ucm} are used. The major advantage of such an approach is the flexibility compared to the previous one, as you can now easily include or exclude different constraints as one wishes. However, this method also requires the most modelling as the likelihood function for these experiments or theoretical calculations is not trivial. We will give a brief overview of the choice of likelihood functions for each of the experiments considered in Ref.~\cite{Koehn:2024set} as follows.

\vspace{3mm}
{\color{black}
\noindent\textbf{Chiral effective field theory}\\
CEFT provides a systematic expansion in nucleon momenta over a breakdown scale $\Lambda_b$, with NN, 3N, and higher multi-nucleon interactions expressed through pion exchanges and short-range contact terms. Truncating at a given order yields a nuclear Hamiltonian from which $E/A$, and hence the EOS, can be computed, with truncation uncertainties estimated from order-by-order convergence~\cite{Epelbaum:2014efa,Drischler:2020yad}. Such bands have been computed using a variety of many-body methods and CEFT interactions~\cite{Tews:2012fj,Lynn:2015jua,Drischler:2017wtt,Tews:2018kmu,Keller:2022crb}, yielding broadly consistent results~\cite{Huth:2020ozf}.

Rather than treating the CEFT band as a hard boundary, we construct a score function $f(p,n)$ that is unity inside the band and decays exponentially outside,
\begin{align}
    f(p, n) = \begin{cases} 1 & \text{if } p \in [p_{-}, p_{+}],\\
    \exp\left(-\beta \frac{|p - p_{\pm}|}{p_{+} - p_{-}}\right) & \text{otherwise,}
    \end{cases}
\label{eq:score_function}
\end{align}
where $p_{\pm}$ are the density-dependent band boundaries, and $\beta$ is chosen to match the $j/(j+1)$ credibility of the truncation error at order $j$~\cite{Furnstahl:2015rha}. In the chiral counting at N$^2$LO one has $j = 3$, so $j/(j +1)=3/4 = 75\%$; the score function integrates to a fraction $\beta/(\beta + 2)$ inside the band, and equating this to $3/4$ gives $\beta =6$. The total EOS likelihood is then~\cite{Brandes:2023hma}  

\begin{equation}
    \mathcal{L}(\text{EOS}|\textrm{CEFT}) \propto \exp\left(\int_{0.75\,n_{\rm sat}}^{n_{\text{break}}}\ \frac{\log f(p(n, \text{EOS}), n)}{n_{\text{break}} - 0.75\,n_{\rm sat}}\ dn \right),
   \label{eq:chiEFT_likelihood}
\end{equation}
integrated over the density range where the nucleonic metamodel applies. Since $n_\mathrm{break}$ is EOS-dependent, individual EOSs may deviate strongly from the CEFT prediction above their own breakdown density.

\vspace{3mm}
\noindent\textbf{PREX-II and CREX}\\
Much of the EOS uncertainty at nuclear densities stems from poorly constrained nuclear interactions at large isospin asymmetry, which maps onto uncertainty in the symmetry energy slope parameter $L_\mathrm{sym}$. This parameter is directly related to the pressure of pure neutron matter at saturation density,
\begin{align}
    p_\mathrm{pnm}(n_\mathrm{sat}) = \frac{1}{3} n_\mathrm{sat} L_\mathrm{sym}\,,
\end{align}
and correlates strongly with the neutron-skin thickness $R_\mathrm{skin} \equiv R_n - R_p$ of atomic nuclei~\cite{Horowitz:2000xj,Carriere:2002bx,Klupfel:2008af}, motivating its experimental measurement.

The PREX-II and CREX experiments measured the parity-violating asymmetry of scattered electrons off $^{208}$Pb and $^{48}$Ca, respectively, yielding neutron-skin thicknesses of $R_\mathrm{skin} = 0.28 \pm 0.14$~fm~\cite{PREX:2021umo} and $0.12 \pm 0.07$~fm~\cite{CREX:2022kgg}. To translate these to constraints on $E_\mathrm{sym}$ and $L_\mathrm{sym}$, one can perform a linear fit to a representative collection of covariant and non-relativistic energy density functionals~\cite{Chen:2014sca,Reed:2021nqk}, i.e.,
\begin{align}
    R^{(\mathrm{fit},\,208)}_\mathrm{skin}(L_\mathrm{sym}) &= 1.69\times10^{-3}\,L_\mathrm{sym} + 0.09\ \mathrm{fm}\,, \label{eq:Rskin208}\\
    R^{(\mathrm{fit},\,48)}_\mathrm{skin}(L_\mathrm{sym}) &= 8.90\times10^{-4}\,L_\mathrm{sym} + 0.13\ \mathrm{fm}\,. \label{eq:Rskin48}
\end{align}
The associated likelihood function for PREX-II is therefore given by,
\begin{align}
   &\log \mathcal{L}(E_\mathrm{sym}, L_\mathrm{sym}|\text{PREX-II}) \nonumber\\
   &= -\frac{1}{2} \Biggl(\biggl(\frac{R_\mathrm{skin} - R^{(\mathrm{fit},\,208)}_\mathrm{skin}(L_\mathrm{sym})}{\sigma_\mathrm{expt}}\biggr)^2 + \biggl(\frac{E_\mathrm{sym} - E^{(\mathrm{fit})}_\mathrm{sym}(L_\mathrm{sym})}{\sigma_\mathrm{fit}(L_\mathrm{sym})}\biggr)^2\Biggr)\,,
\label{eq:sym_sampling_likelihood}
\end{align}
with an analogous expression for CREX. The $E_\mathrm{sym}$ term accounts for the energy-density-functional-derived correlation between $E_\mathrm{sym}$ and $L_\mathrm{sym}$, parameterized as
\begin{align}
    E^{(\mathrm{fit})}_\mathrm{sym}(L_\mathrm{sym}) = a L_\mathrm{sym}^3 + b L_\mathrm{sym}^2 + c L_\mathrm{sym} + d\,,
\end{align}
with $a = 3.54\times10^{-6}$, $b = -1.81\times10^{-4}$, $c = 0.063$, $d = 29.01$, and $\sigma_\mathrm{fit}(L_\mathrm{sym})$ denoting the 68\% fitting error~\cite{Koehn:2024set}.

\vspace{3mm}
\noindent\textbf{Heavy-ion collisions}\\
In heavy-ion collision (HIC) experiments, relativistic collisions of heavy nuclei compress nuclear matter, probing the EOS in the density range $1$--$2\,n_\mathrm{sat}$~\cite{Danielewicz:2002pu,LeFevre:2015paj,Russotto:2016ucm,Tsang:2019mlz}. The key constraints come from the FOPI~\cite{LeFevre:2015paj} and ASY-EOS~\cite{Russotto:2016ucm} experiments at GSI, as well as the symmetric nuclear matter results of Ref.~\cite{Danielewicz:2002pu}, and have been used to constrain the NS EOS in Refs.~\cite{Huth:2021bsp,Koehn:2024set}. The S$\pi$rit experiment~\cite{SpiRIT:2021gtq}, which probes the symmetry energy via charged pion spectra in tin isotope collisions, yields results consistent with ASY-EOS but with similarly large uncertainties and has not been included in these analyses.

The key observable is the elliptic flow $v_2$, the second Fourier coefficient of the azimuthal particle distribution relative to the reaction plane,
\begin{equation}
    \frac{d\sigma(y,p_t)}{d\Phi} = C\left[1 + 2v_1\cos(\Phi-\Phi_\mathrm{RP}) + 2v_2\cos(2(\Phi-\Phi_\mathrm{RP})) + \dots\right],
\end{equation}
where $y = \frac{1}{2}\ln\!\left(\frac{E+p_z}{E-p_z}\right)$ is the longitudinal rapidity. The neutron-to-proton flow ratio $v_2^\mathrm{np} = v_2^\mathrm{n}/v_2^\mathrm{p}$ is particularly sensitive to the symmetry energy~\cite{Russotto:2011hq}, and has been analyzed using UrQMD transport simulations~\cite{Russotto:2016ucm}, with consistency verified across other transport models~\cite{Huth:2021bsp}.

The EOS functional used in the transport simulations is
\begin{equation}
    \frac{E}{A}(n,\delta) \approx e_\mathrm{sat}(n) + e_\mathrm{sym}(n)\,\delta^2 + \dots\,,
    \label{eq:expandE}
\end{equation}
where $\delta = 1 - 2n_p/n$ is the isospin asymmetry. The symmetric matter term is parametrized as~\cite{LeFevre:2015paj}
\begin{equation}
    e_\mathrm{sat}(n) = \frac{3}{5}\left(\frac{n}{n_\mathrm{sat}}\right)^{2/3}E_F + \frac{\alpha}{2}\frac{n}{n_\mathrm{sat}} + \frac{\beta}{\gamma+1}\left(\frac{n}{n_\mathrm{sat}}\right)^\gamma\,,
    \label{eq:asy_ESNM}
\end{equation}
with $E_F = 37$~MeV and $\alpha$, $\beta$, $\gamma$ fitted to $E_\mathrm{sat} = -16$~MeV, zero pressure at saturation, and a varying $K_\mathrm{sat}$. The symmetry energy is parametrized as
\begin{equation}
    e_\mathrm{sym}(n) = E_\mathrm{kin,0}\left(\frac{n}{n_\mathrm{sat}}\right)^{2/3} + E_\mathrm{pot,0}\left(\frac{n}{n_\mathrm{sat}}\right)^{\gamma_\mathrm{asy}}\,,
    \label{eq:Esym_QMD}
\end{equation}
with $E_\mathrm{kin,0} = 12$~MeV and $E_\mathrm{pot,0} = E_\mathrm{sym} - E_\mathrm{kin,0}$. Fitting to ASY-EOS data yields $\gamma_\mathrm{asy} = 0.68 \pm 0.19$ for $E_\mathrm{sym} = 31$~MeV and $\gamma_\mathrm{asy} = 0.72 \pm 0.19$ for $E_\mathrm{sym} = 34$~MeV, with linear interpolation for intermediate values. Following Ref.~\cite{Huth:2021bsp}, $E_\mathrm{sym}$ is varied uniformly over $[31, 34]$~MeV and $K_\mathrm{sat}$ is drawn from a Gaussian with mean $200$~MeV and standard deviation $25$~MeV.

For each density, this yields a pressure distribution $P(p, n|\mathrm{HIC})$, and the EOS likelihood is
\begin{align}
    \mathcal{L}(\text{EOS}|\text{HIC}) = \int dn\ P(p(n, \text{EOS}), n|\text{HIC})\ C(n)\,,
    \label{eq:HIC_likelihood}
\end{align}
where $C(n)$ is the experimental sensitivity curve of the ASY-EOS flow ratio~\cite{Russotto:2016ucm}.

\vspace{3mm}
\noindent\textbf{Perturbative QCD}\\
At $n \approx 40\,n_\mathrm{sat}$, matter is expected to be in a quark-matter phase, where the EOS can be computed from perturbative QCD (pQCD)~\cite{Gorda:2021znl,Kurkela:2009gj,Freedman:1976ub}. Although this density is far beyond any terrestrial or astrophysical laboratory, the pQCD EOS can be used to exclude certain pressure regions at lower densities by checking whether a low-density point $(\mu_L, n_L, p_L)$ can be connected to the pQCD point $(\mu_H, n_H, p_H)$ via a causal and mechanically stable interpolation~\cite{Komoltsev:2021jzg,Gorda:2022jvk}. Such a connection exists if and only if
\begin{align}
  \frac{\mu_H^2 - \mu_L^2}{2\mu_L}n_L\, \leq p_H - p_L \leq \frac{\mu_H^2 - \mu_L^2}{2\mu_H}n_H\,.
  \label{eq:pQCD_condition}
\end{align}
Following Refs.~\cite{Gorda:2022jvk, Gorda:2023usm}, the posterior for a given EOS point is
\begin{align}
    P(\epsilon_L, p_L |n_L, \mu_H, n_H, p_H) = \begin{cases} 
        1 & \text{if Eq.~(\ref{eq:pQCD_condition}) is fulfilled,} \\
        0 & \text{otherwise.}
    \end{cases}
    \label{eq:pQCD_stepfunction}
\end{align}
Ref.~\cite{Gorda:2023usm} accounts for uncertainties in $p_H$ and $n_H$ from missing higher-order (MHO) contributions, estimated via the Bayesian algorithm MiHO~\cite{Duhr:2021mfd}, and from the renormalization scale $\bar{\Lambda}$. The latter enters through the dimensionless parameter $X = \frac{3}{2}\bar{\Lambda}/\mu_H$, over which we marginalize log-uniformly between $1/2$ and $2$. The full likelihood is~\cite{Gorda:2023usm}
\begin{align}
    &\mathcal{L}(\text{EOS}|\text{pQCD}) = \nonumber\\
    &\int dX\, dp_H\, dn_H\,P(\epsilon_L, p_L|n_L, \mu_H, n_H, p_H)\,P_{\text{MHO}}(p_H, n_H|\vec{p}^{(j)}, \vec{n}^{(j)})\,P_{\text{SM}}(X|\vec{p}^{(j)})\,,
\end{align}
where $P_\mathrm{SM}$ is the marginalized likelihood of $X$ within the MiHO statistical model, and $\mu_H = 2.6~\mathrm{GeV}$ corresponds to $\approx 40\,n_\mathrm{sat}$. MiHO~\cite{Duhr:2021mfd} places a Bayesian prior on the size of the next-omitted pQCD coefficient and conditions on the computed series $\vec{p}^{(j)}, \vec{n}^{(j)}$ to yield a posterior distribution $P_{\text{MHO}}(p_H, n_H)$ for the all-orders values; Ref.~\cite{Gorda:2023usm} applies this construction at $\mu_H = 2.6~\mathrm{GeV}$.

For a causal and stable EOS, it suffices to check Eq.~\eqref{eq:pQCD_condition} only at the highest density point, and the constraint becomes less restrictive at lower matching densities. Two approaches have been commonly used to connect the pQCD prediction to the NS density regime. In the more conservative approach~\cite{Somasundaram:2022ztm}, the EOS is matched at $n_\mathrm{TOV}$, and the unstable branch above it is discarded. A known limitation is that this asymmetrically permits strong phase transitions immediately above $n_\mathrm{TOV}$~\cite{Gorda:2022jvk}. More stringent constraints can be obtained either by extending the EOS to a fixed higher density~\cite{Gorda:2022jvk}, or by constructing the pQCD likelihood from an ensemble of Gaussian-process-generated EOS segments conditioned on the pQCD speed-of-sound band between $25$ and $40\,n_\mathrm{sat}$~\cite{Komoltsev:2023zor}, though both come at the cost of additional model dependence. A comparison of these approaches in Ref.~\cite{Koehn:2024set} shows that differences are most pronounced for very massive NSs, highlighting that the constraining power of pQCD depends sensitively on how its high-density prediction is propagated down to densities of a few $n_\mathrm{sat}$.
}

By a comprehensive Bayesian analysis combining diverse constraints—from nuclear theory, nuclear experiments, and neutron star observations, Ref.~\cite{Koehn:2024set} presented one of the strongest constraints of ultra-dense matter across all density regimes, as shown in Fig.~\ref{fig:nmma_results}. In which, different sets of constraints are imposed, from set A as the conservative set and towards set C as the aggressive set, depending on the systematics involved. Please see Ref.~\cite{Koehn:2024set} for details.

\subsection{Modular Unified Solver of the Equation of State (MUSES)}
\label{subsec:MUSES}


A recent effort to unify the study of the dense matter EOS has been undertaken by the Modular Unified Solver of the Equation of State (MUSES) collaboration, bringing both low- and high-energy nuclear physicists, astrophysicists and computer scientists together \cite{MUSES_WEBSITE}. 
Unlike approaches such as NMMA and BAND, which primarily focus on providing tools for statistical synthesis of information from distinct domains (discussed in Sect. \ref{subsec:nmma} and Sect. \ref{sec:band}, respectively), MUSES aims to build a fully modular, extensible computational framework that allows theorists and experimentalists to incorporate first-principle-based calculations, microscopic models and effective theories, within a single consistent environment.

\subsubsection{Overview}
\label{subsubsec:MUSES_overview}

\indent Neutron stars and heavy-ion collisions probe different but complementary regimes of the QCD phase diagram, yet both ultimately describe the same underlying strongly interacting matter. 
In neutron stars, supra-saturation densities govern global properties such as masses, radii, and tidal deformability, while in heavy-ion collisions, laboratory experiments recreate matter at extreme temperature and finite baryon density, offering insight into the quark–gluon plasma, the nature of the phase transition to deconfined matter, and the possible existence of a critical point. 
Although each domain employs dedicated modelling strategies tailored to its conditions, the EOS that emerges in one regime must remain consistent with those in other regimes, while fulfilling different constraints for each regime. 
Bridging these approaches within a single framework is therefore essential: it allows us to confront the EOS with astrophysical and terrestrial constraints simultaneously when possible, test its robustness across disparate conditions, and ultimately move toward a coherent description of dense matter valid throughout the phase diagram.

The philosophy of MUSES is to establish a platform where different theoretical ingredients can be interchanged, benchmarked, and systematically combined. 
This modularity enables researchers to test how sensitive EOS predictions are to underlying assumptions, and to trace the impact of microscopic physics all the way to macroscopic observables, such as neutron star masses and radii for instance.
This can be done either by using observable modules already embedded in the MUSES cyberinfrastructure, or by using the EOSs generated by MUSES as inputs to simulations tools, such as the ones developed to describe neutron star mergers or relativistic heavy-ion collisions (discussed in Section \ref{subsubsec:methods_hic_theory}).
More importantly, MUSES provides tools to merge different EOSs with partially overlapping, and at the same time complementary regimes, offering end-products with a broad coverage of the QCD phase diagram, which are needed to describe the evolution of realistic complex systems in simulations.

Finally, MUSES functions as a community-driven cyberinfrastructure: anyone can not only change parametrizations and other options of pre-existing modules, but also download and modify the codes, or even contribute by adding new modules with different approaches to compute the nuclear matter EOS and related observables.
All codes implemented in the MUSES cyberinfrastructure are open-source, ensuring transparency, reproducibility, and providing a common platform for community-wide development.


\subsubsection{The MUSES framework}
\label{subsubsec:MUSES_framework}

The MUSES framework provides a standardized specification for packaging calculation codes and their dependencies, defining programming interfaces, and documenting their functionality. 
By enforcing these common requirements, it enables otherwise-independent EOS calculation codes and related physics libraries to interoperate seamlessly within larger applications and data analysis workflows. 
The establishment of such a well-defined environment thus facilitates its extension via the integration of new software packages, called ``modules" \cite{Manning:2025MUSES_CE}.

\begin{figure*}[!tb]
    \centering
    \includegraphics[width=0.9\linewidth]{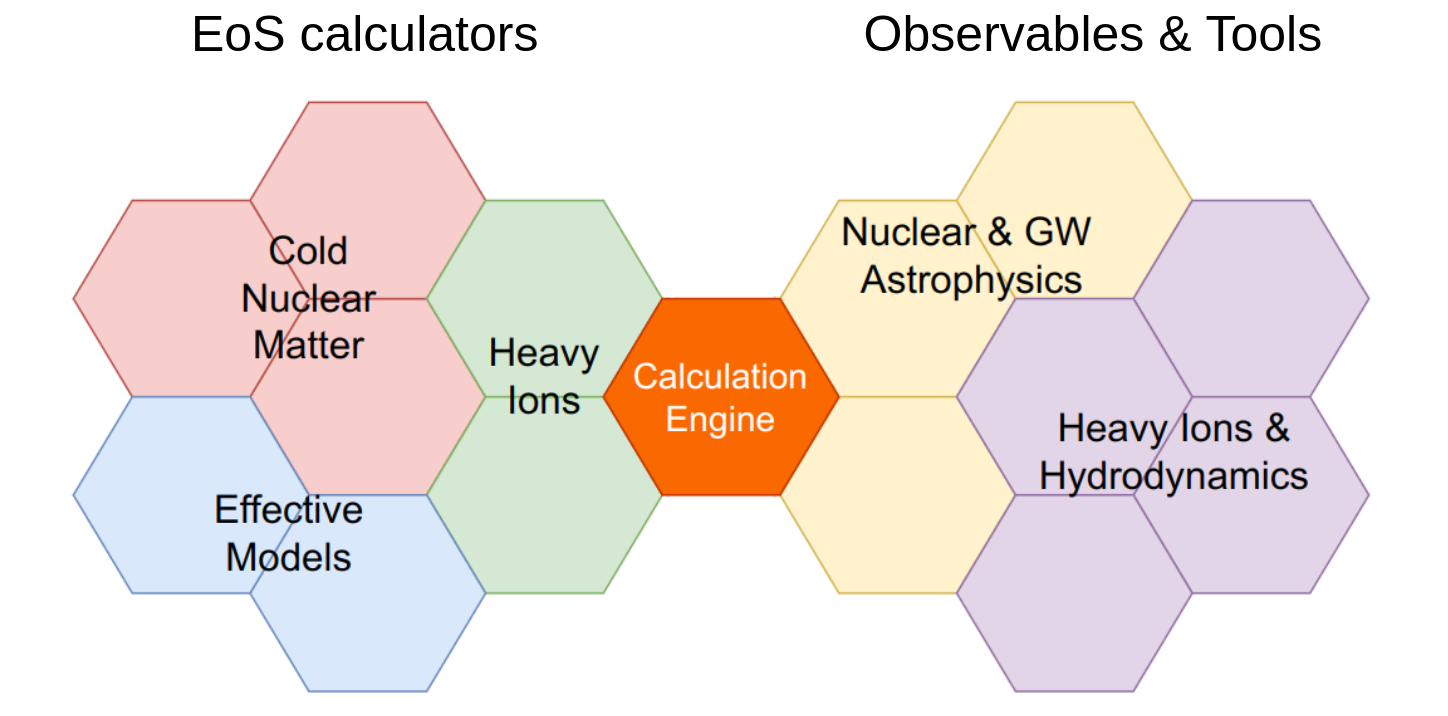}
    \caption{
    Graphical representation of the different modules for nuclear EOS computation and for related observables computation and tools, in the different relevant regimes and fields of application, integrated to the MUSES Calculation Engine. Figure from Ref. \cite{Manning:2025MUSES_CE}.
    } 
    \label{fig:muses_CE}
\end{figure*}

The central element of the MUSES framework is its Calculation Engine (CE), linking all the different MUSES modules together as illustrated in Fig. \ref{fig:muses_CE}, the first version of which was released early 2025 \cite{manning_2025_14721912}. 
It provides a web interface to a workflow management system, that enables any registered user to build and execute a specific sequence of calculation using one or several MUSES modules, defined through a directed acyclic graph (DAG). 
This workflow is then executed on a computation cluster as a job, first entering a Celery distributed task queue system, then executed asynchronously by a pool of worker processes. This allows to take advantage of the workflow’s DAG structure to parallelize tasks where possible, making effective use of available computing resources. For complex workflows involving the sequential execution of several modules, files can be transferred between modules, and all produced files can be downloaded for local use \cite{ReinkePelicer:2025vuh}.
Currently, the Kubernetes cluster running the jobs submitted through the MUSES CE is composed of 10 worker nodes, each including 16 cores for a total of 160 cores and 640 GB of memory, and a limit of 12 concurrently-running tasks per node \cite{Manning:2025MUSES_CE}. 
Each user has a limit of maximum 10 jobs running simultaneously, made out of up to 20 processes per single workflow \cite{MUSES_CE}.

All modules registered on the MUSES CE are defined as a software package containerized to be executed with Docker \cite{Merkel:2014docker} by the CE.
This feature gives the possibility to run any module's code inside a container on their own local computer or cluster, thus without having to worry about installing the required dependencies. 
The module inputs and outputs are defined in a set of machine-level specifications, based on the OpenAPI standard format, to which they are validated against during the module's execution. 
After defining some unit tests that allow to automatically verify the functioning of the module alone, they are finally integrated to the CE by declaring their information into the CE manifest.

Upon integration, workflows including registered modules can then be constructed once defined in the CE. 
While most modules have at least one so-called ``singleton" workflow defined (\textit{i.e.} where the module is executed as a stand-alone), more complex workflows involving several modules executed in a sequential order are also possible.
Some examples will be discussed in Sect. \ref{subsubsec:MUSES_modules&workflows}.

%
%

\subsubsection{Calculating EOS and related observables with MUSES}
\label{subsubsec:MUSES_modules&workflows}

Following up from the discussion of the motivations behind MUSES in Sect.\ref{subsubsec:MUSES_overview}, and the characteristics of the MUSES cyber-infrastructure and framework in Sect. \ref{subsubsec:MUSES_framework}, we discuss at last the physics cases that can be studied through the available capabilities of the current version of the MUSES CE.

\paragraph{Equations of state for neutron stars}

The MUSES CE provides a way to build a complete neutron star EOS, thanks to the current available models for nuclear EOS, spanning over different regimes of nuclear density and charge fraction:

\begin{itemize}
    \item \textbf{Chiral Mean Field model (CMF++)} \cite{cruz_camacho_2025_14860593} describes dense nuclear matter, suited to model neutron star's both inner and outer core, based either on light and strange quarks for the densest phase, or based on full baryon octet and decuplet to describe hadronic phase
    \textcolor{black}{; in its current implementation, the CMF coupling constants correspond to the original parametrizations of Refs.} 
    \cite{Dexheimer:2008ax, Dexheimer:2009hi, Cruz-Camacho:2024odu}
    \textcolor{black}{and are not recalibrated within MUSES using external constraints: as such, the model should be regarded as a phenomenological input rather than a globally constrained EOS};
    
    \item \textbf{Chiral Effective-Field Theory ($\chi$-EFT)} \cite{david_friedenberg_2024_14611275} offers \textit{ab-initio} calculation 
    \textcolor{black}{based on chiral interactions, currently implemented using many-body perturbation theory approaches,} 
    of a purely nucleonic EOS for densities ranging from $\sim0.5$ $n_\text{sat}$ to $\sim1.5$ $n_\text{sat}$ \cite{Machleidt:2011zz, Drischler:2021kxf},
    \textcolor{black}{representative of state-of-the-art $\chi$EFT constraints on neutron-rich matter};

    \item \textbf{Crust Density-Functional Theory (Crust-DFT)} \cite{steiner_2025_14714273} allows to calculate the EOS
    \textcolor{black}{of nuclear matter in the crust regime using density-functional theory, based on Skyrme-type energy density functionals (e.g. SLy and BSk families), describing nuclei in equilibrium with dripped nucleons and reproducing the nuclear liquid–gas phase transition}
    \cite{Du:2018vyp, Du:2021rhq}.
\end{itemize}

\begin{figure}[!t]
    \centering
    \includegraphics[width=0.5\linewidth]{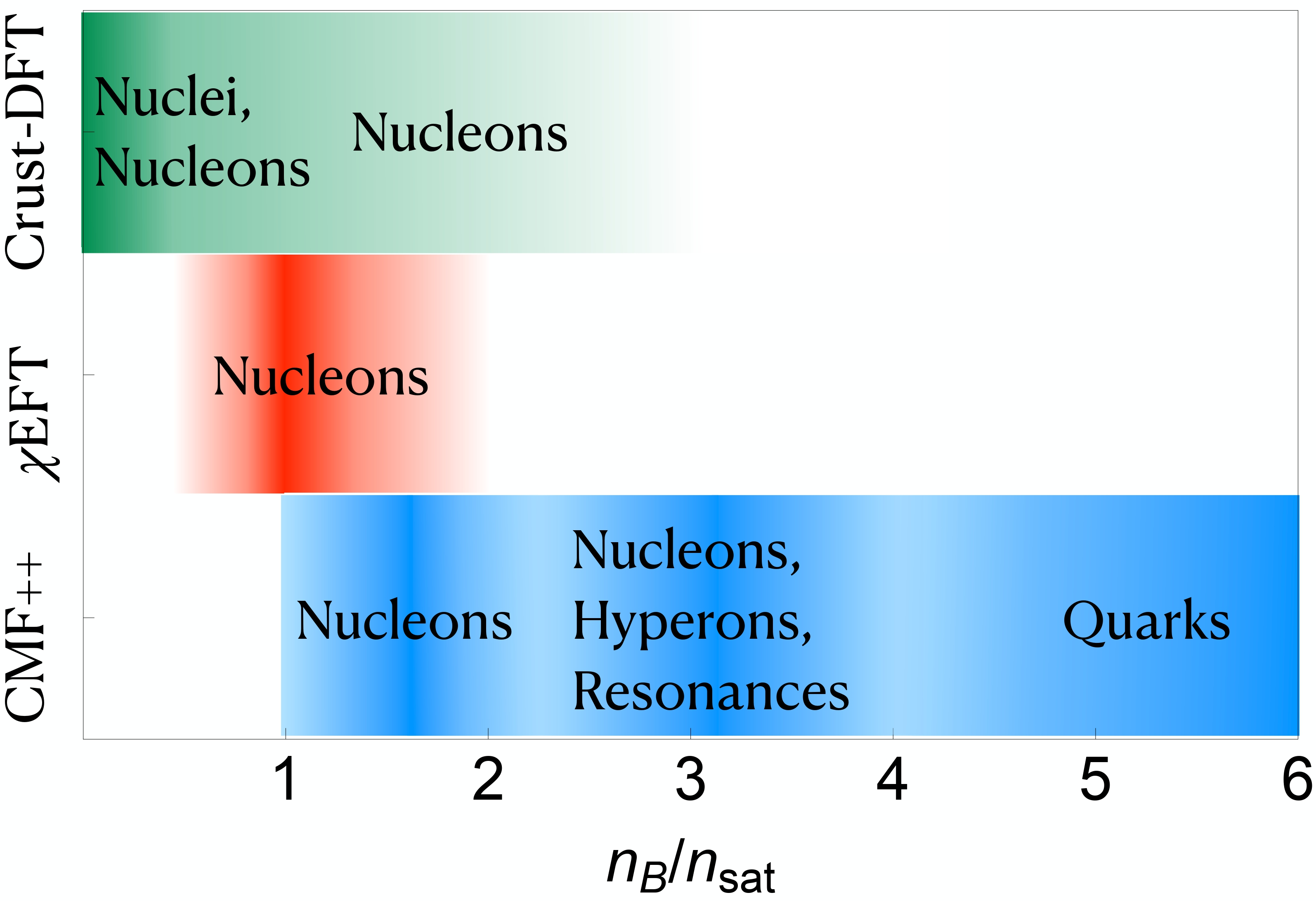}
    \hfill
    \includegraphics[width=0.45\linewidth]{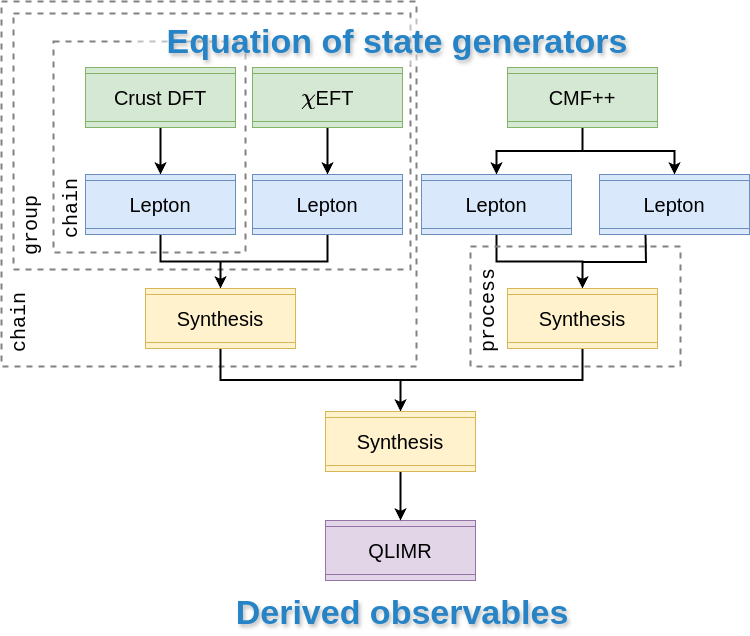}
    \caption{
    \textbf{Left:} baryon density coverage of different neutron star MUSES EOS modules at $T \sim 0$, with associated degrees of freedom.
    \textbf{Right:} sketch of a multi-module workflow composed of several chains of modules, used to generate a hybrid $\beta$-equilibrated and charge-neutral nuclear EOS, coupled to an observable module for physics analysis.
    Figures taken from Ref. \cite{ReinkePelicer:2025vuh}.}
    \label{fig:MUSES_NS_composition}
\end{figure}

Together, these modules span density ranges going from what is expected in the outer crust to the core (as illustrated in Fig. \ref{fig:MUSES_NS_composition}, left), ensuring all relevant degrees of freedom can be described by one way or another, although only for $T \sim 0$ MeV at the moment.
MUSES offers thus a large panel of possibilities, together with another leptonic EOS module \cite{pelicer_2025_14654137} based on a free Fermi gas description of electrons, muons and tauons, that can be used to calculate a charge-neutral nuclear EOS and calculate the $\beta$-equilibrium condition, with the option to include neutrinos as well \cite{ReinkePelicer:2025vuh}.

While it is of course possible to generate EOS tables from a single one of these  modules, the strength of the MUSES framework resides in the possibility to merge them together or with any external EOS, and to perform observable calculation from the resulting EOS, through elaborated module workflows.
The right-side of Fig. \ref{fig:MUSES_NS_composition} illustrates one of the possible advanced workflows that can be built through the MUSES CE, representing a complete process of generating a realistic nuclear EOS. 
In the example displayed here, the crust EOS is generated with the Crust-DFT module, the outer core EOS with the $\chi$-EFT module, and two EOS are generated through CMF++ (one baryonic, and one with quarks for the most inner core). 
All these EOSs are then individually $\beta$-equilibrated and turned neutral by using the Lepton module, before being merged together two-by-two, into a final single table, which can eventually be fed into the so-called ``observable" modules to calculate physical properties of the modelled neutron star.

\begin{itemize}
    \item \textbf{Synthesis module} \cite{pelicer_2025_14654584} allows to merge two EOSs into a single one, given that some of their thermodynamic quantities overlap in a common region of applicability.  
    \textcolor{black}{When a consistent overlap exists, the module} offers several smooth matching prescriptions (e.g., via pressure, energy density, or speed of sound) or first-order transition constructions (Maxwell or Gibbs), depending on the provided tables and user's choice \cite{ReinkePelicer:2025vuh}
    \footnote{\textcolor{black}{A thermodynamically consistent matching is not guaranteed for arbitrary pairs of EOSs. In particular, inconsistencies may arise if the two EOSs do not share a common domain in thermodynamic space (e.g. in density or pressure), or if their underlying behavior leads to incompatible slopes (e.g. negative compressibility or discontinuities in the chemical potential). In such cases, MUSES does not enforce a matching and instead requires the user to adjust the chosen models or matching region.}}.

    \item The \textbf{QLIMR module} \cite{conde_ocazionez_2025_14525356} (for Quadrupole, Love number, Inertia, Mass, Radius) is designed to compute stellar observables of slowly-rotating neutron stars, such as mass–radius curves, tidal deformability, moment of inertia or quadrupole moment, based on the Hartle-Thorne approximation and quasi-universal relations of neutron stars \cite{Hartle:1968si, Flanagan:2007ix, Yagi:2013awa}.

    \item The \textbf{Flavor Equilibration module} \cite{alford_2025_14537518} evaluates weak-interaction relaxation rate, bulk viscosities and isothermal incompressibility of neutron stars that are driven out of $\beta$-equilibrium, by oscillations or merger dynamics \cite{Alford:2021ogv, Alford:2023gxq}. This capability allows systematic studies of dissipation in neutron star interiors, crucial for interpreting gravitational-wave post-merger signals.
\end{itemize}
    
\begin{figure*}[ht]
    \centering
    \includegraphics[width=0.585\linewidth]{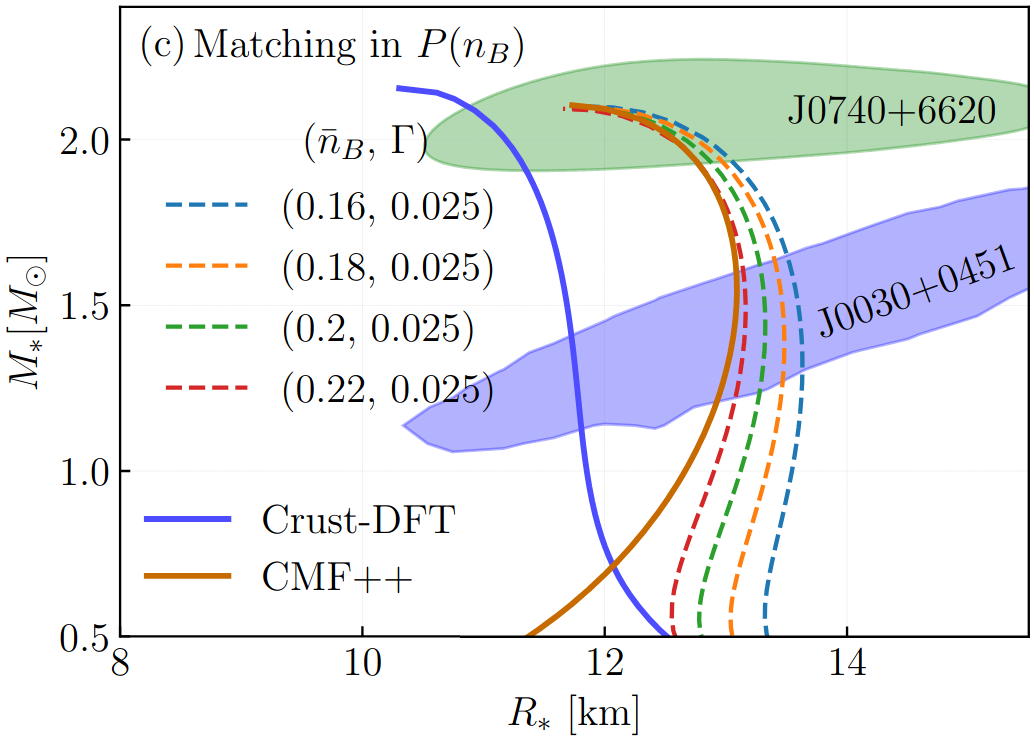}
    \hfill
    \includegraphics[width=0.39\linewidth]{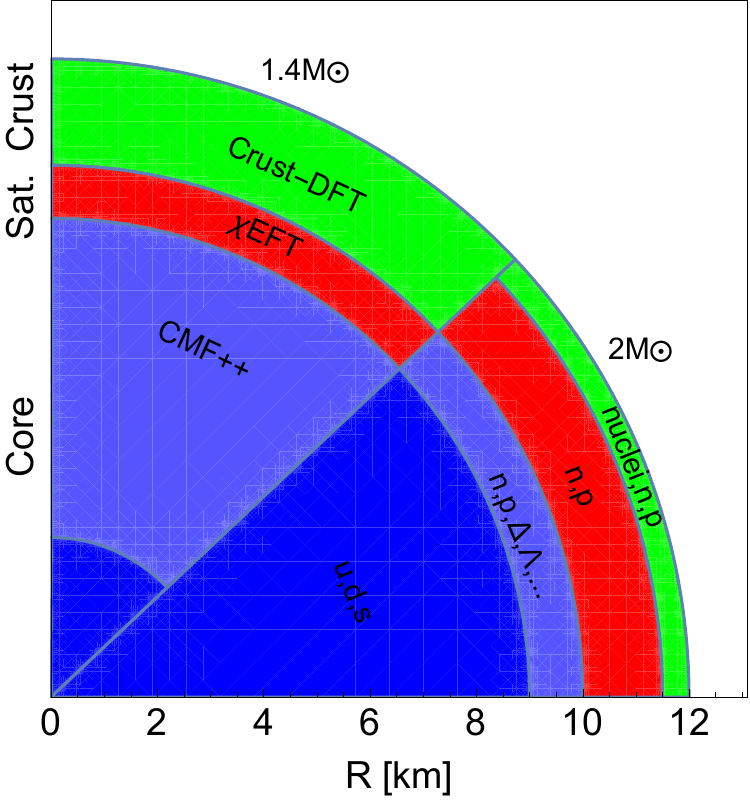}
    \caption{
    \textbf{Left:} mass-radius diagrams obtained with the QLIMR module. Colored dashed lines correspond to The numbers in the labels indicate the midpoint and width of the matching. The shaded areas identify the 2sigma confidence mass-radius measurements from NICER: (i) the green area represents J0740+6620; (ii) the purple area represents pulsar J0030+0451.
    \textbf{Right:} schematic representation of the structure of two neutron stars of different masses, with matter phases described by CMF++ in blue, region around saturation density described by $\chi$-EFT in red, and the crust described by Crust-DFT in green (displayed numbers are rough estimations).
    Figures taken from Ref. \cite{ReinkePelicer:2025vuh}.}
    \label{fig:MUSES_NSresults}
\end{figure*}

To perform the merging of two different 1D EOS dubbed $I$ and $II$, using a given thermodynamic function $Y$ depending on another variable $x$, the Synthesis module employs the following equation: 
\begin{equation}
    Y(x) =  Y^I(x) f_- (x)+ Y^{II}(x) f_+(x)\,,
\end{equation}
where the interpolating functions $f_\pm(x)$ are hyperbolic tangent-based functions

\begin{equation}
    f_\pm (x) = \frac{1}{2} \left( 1 \pm \tanh{\left[ \frac{x - \bar x }{\Gamma} \right]} \right) \,,
\end{equation}

\noindent relying on an interpolation midpoint value $\bar{x}$ and width $\Gamma$.
In the context of \cite{ReinkePelicer:2025vuh}, the quantity used to merge the EOS for a crossover-type transition, as well as the merging parameters were varied.
\textcolor{black}{We stress that the matching of different EOSs within MUSES does not, by itself, constitute a propagation of microscopic constraints across density regimes. In particular, combining a low-density EOS constrained by $\chi$EFT with a high-density phenomenological model (such as CMF) does not enforce consistency of the latter with these constraints beyond the matching region.}
Then coupled to the QLIMR module, it led to confirm that matching procedures only weakly affect global properties (mass and radius vary by a few percent), while quasi-universal I-Love-Q relations remain intact.
One can thus observe on Fig. \ref{fig:MUSES_NSresults} (left) how different interpolating parameters for a merging in $P(n_B)$ affect the resulting $M-R$ curves (in dashed lines), as compared to a purely nuleonic \& nuclear EOS from Crust-DFT (in thick blue), or to a quark and baryon EOS from CMF++ (in thick orange).
Additionally, the right panel of Fig.~\ref{fig:MUSES_NSresults} illustrates the difference of the interior of two neutron stars from the same sequence (\textit{i.e.} from the same line on a $M-R$ diagram), only based on their masses.
It shows that even for one EOS, stellar composition varies, with more massive stars having much larger portion of high-density matter.

Using the same set of variously-merged EOSs with the Flavor Equilibration module, results show viscosities on the order of $10^{24}-10^{27}$ MeV$^3$, depending on the microscopic model used as an input ($\chi$-EFT vs. CMF).

At last, it is worth mentioning that the EOS construction discussed here is purely illustrative and far from being the only option given by the MUSES CE.
From merging Crust-DFT with CMF++ (since they overlap in density, as shown in Fig. \ref{fig:MUSES_NS_composition}), to use only one of the two phases from CMF++ (including strangeness or not), through different choices and parametrisation of transition types, and even the use of external EOSs... 
Many other workflow configurations are possible, making MUSES a scientific sandbox for studies of the EOS of neutron stars and its impacts on their macroscopic properties.

%

\paragraph{Equations of state for heavy-ion collisions}

Through (ultra-)relativistic HICs, one can probe the opposite side of the phase diagram, namely the high-temperature / sub-saturation density regime. 
In the current version of the MUSES CE, several EOSs can be computed to be used in transport codes, in order to describe HICs at LHC (low to vanishing $\mu_B$ values) down to lower energies at RHIC and SPS (up to $\mu_B \sim 800$ MeV). 
MUSES provides first of all two 4D EOS modules based on first-principle lattice QCD calculations, both obtained with series expansions to reach finite values of $\mu_{B,Q,S}$, chemical potentials of baryon number ($B$), electric charge ($Q$) and strangeness ($S$) respectively. These expansions are needed to circumvent the limitations of lattice QCD restricted to simulations at $\mu_i = 0$, due to the sign problem.

\begin{itemize}
    \item \textbf{4D Taylor-expanded lattice QCD (or BQS EOS)} \cite{jahan_2025_14639786} is based on a direct Taylor expansion of the pressure in powers of $\hat{\mu}_i = \mu_i/T$, with $i=B,Q,S$, at constant $T$:
    \begin{equation}
        \frac{p(T,\hat{\mu}_B,\hat{\mu}_Q,\hat{\mu}_S)}{T^4} = \sum_{i,j,k} \frac{1}{i!j!k!} \chi_{ijk}^{BQS}(T) \hat{\mu}_B^i \hat{\mu}_Q^j \hat{\mu}_S^k \, ,
    \end{equation}
    where $i+j+k = 2n$ with $n \in \mathbb{N}$, up to next-to-next-to-leading-order (NNLO), \textit{i.e.} for ${i+j+k\leq4}$ \cite{Noronha-Hostler:2019ayj}.
    \item \textbf{4D lattice QCD $T^\prime$-Expansion Scheme (4D-TExS)} \cite{jahan_2025_16749257} consists of an extrapolation along lines of constant generalised charge density $X_1$:
    \begin{equation}
        X_1(T, \hat{\mu}_B,\hat{\mu}_Q,\hat{\mu}_S) =
        \frac{\overline{X}_1(\hat{\mu}_B,\hat{\mu}_Q,\hat{\mu}_S)}{\overline{X}_2(\hat{\mu}=0)}
        X_2\left(T^{\, \prime}(T,\hat{\mu}_B,\hat{\mu}_Q,\hat{\mu}_S)\right)
        \, ,
    \end{equation}
    where $X_2$ is the generalised second-order susceptibility (linear combination of second-order susceptibilities for $B$, $Q$ and $S$), and $\overline{X}_1$, $\overline{X}_2$ are Stefan-Boltzmann limits of associated quantities (non-interacting gas limit when $T\to\infty$). Here, both the expanded temperature, $T^\prime (T,\hat{\mu}_B,\hat{\mu}_Q,\hat{\mu}_S) = T \left( 1 + \lambda_{2}(T) (\hat{\mu}_B^2 + \hat{\mu}_Q^2 + \hat{\mu}_S^2)\right)$, as well as $\overline{X}_1$ carry the finite chemical-potential dependence \cite{Abuali:2025tbd}.
\end{itemize}

\begin{figure}[!t]
    \centering
    \includegraphics[width=0.9\linewidth]{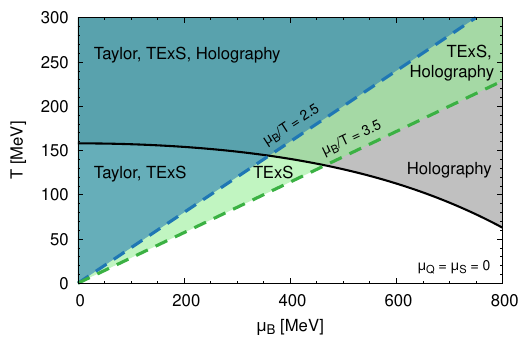}
    \caption{
    \textcolor{black}{Comparison of the coverage of validity of different EOS types from MUSES heavy-ion modules in the finite-$\mu_B$ phase diagram. 
    Figure taken from Ref.} \cite{ReinkePelicer:2025vuh}.
    }
    \label{fig:MUSES_HI}
\end{figure}

The Taylor expansion construction of the BQS EOS module is built upon parametrisations of susceptibilities $\chi_{ijk}^{BQS}(T)$ up to the fourth order, based on lattice QCD simulations with $N_\tau = 12$ timeslices \cite{Borsanyi:2018grb}, and is usually trusted for $\hat{\mu}_i \lesssim 2.5$ \cite{Noronha-Hostler:2019ayj}.
In comparison, the 4D-TExS EOS module is using a resummation of the Taylor expansion, and the second-order expansion coefficient $\lambda_{2}(T)$ (at $\mu_{BQS} = 0$) is constructed with a more recent set of \textit{continuum-estimated} lattice QCD susceptibilities up to the fourth order \cite{jahan_2025_16749145}.
For matching order of $\chi_{ijk}^{BQS}$, 4D-TExS offers a broader coverage of the phase diagram, since it is trusted up to $\hat{\mu}_B$ of 3.5, as displayed in Fig. \ref{fig:MUSES_HI} (a reach that can vary in other finite-$\mu_i$ directions).
In both cases, the susceptibilities, which are the base ingredients of the expansions, are merged from lattice QCD, and ideal-HRG \cite{Dashen:1969ep, Venugopalan:1992hy} at low temperature (for $T\lesssim120$ MeV), to ensure an optimum description of the QCD EOS across a large range of temperature, in both QGP and hadronic phases. Two other effective-model-based EOS modules are also available, namely:

\begin{itemize}
    \item the \textbf{Ising 2D $T^\prime$-Expansion Scheme (Ising-2DTExS)} \cite{kahangirwe_2025_14637802}, based on a 2D version of the TExS of lattice QCD at finite $\mu_B$, embedding the contribution of a critical point and associated first-order transition line from the 3D Ising model universality class \cite{Kahangirwe:2024cny};
    \item the \textbf{Holographic model} \cite{yang_2025_14695243}, where an finite $(T, \mu_B)$ EOS is constructed using the AdS/QCD correspondence from an Einstein-Maxwell dilaton field theory, describing a crossover at low $\mu_B$ which turns into a first-order phase transition at higher density, starting from the critical point \cite{Critelli:2017oub, Grefa:2021qvt, Hippert:2023bel}.
\end{itemize}

While the second one predicts the existence of a critical point when matched with lattice QCD results at $\mu_B = 0$ \cite{Hippert:2023bel}, but can only describe strongly-coupled deconfined matter, the first one offers a temperature ranging from hadronic gas to QGP regime (as explained previously), but does not predict the existence of a QCD critical point. 
It is designed as a tool for the user to choose freely where to set a critical point, and to play with the Ising model $\leftrightarrow$ QCD mapping, varying thus it strength and the size of the critical region \cite{Kahangirwe:2024cny}.
In a similar manner, the Holographic EOS can be changed by modifying the parametrisation of the black hole model it is constructed from, although more indirectly since there is no straightforward connection with the location of the critical point then obtained.
This feature makes both models (of which the coverage are displayed in Fig. \ref{fig:MUSES_HI}) very good candidates for Bayesian studies of the location of the critical point.
\\

At the moment, the different modules available in the MUSES CE offer two distinctive regimes where the nuclear EOS can be generated, which are not connected so far.
At $T \sim 0$ and up to high densities, for the modelling of neutron stars, and at finite $T$ but from 0 to moderate baryonic densities, more dedicated to the simulations of ultra-relativistic heavy-ion collisions.
However, upgrades of the  MUSES CE are already planned in order to bridge that gap. 
The inclusion of a module to compute the quantum van-der-Waals hadron resonance gas (QvdW-HRG) EOS based on the \texttt{Thermal-FIST} package \cite{Vovchenko:2019pjl}, describing the whole hadronic phase region up to the nuclear liquid-gas phase transition, will fill a large missing piece of the phase diagram coverage of MUSES. 
\textcolor{black}{
On the other side of the transition, an additional module relying on an improved formulation of the Nambu-Jona Lasinio model \cite{Gholami:2024diy} will cover the color-superconducting phase.
Finally, in order to help connecting to the very large baryon densities reached in NS mergers, a module containing pQCD EOS calculation are also planned to be added soon to the MUSES CE, based on the scheme developed in Ref.~\cite{Danhoni:2025qpn} at $T=0$ MeV, using data from Ref.~\cite{Graf:2015tda}.} 
All of these will come with extensions of pre-existing modules for neutron stars EOS to finite-$T$ \cite{Mroczek:2024sfp, Wellenhofer:2014hya}, accompanied by a $n$-dimensional method for the merging of different equations of state \textcolor{black}{\cite{Yang:2026brr}}, thus extending the possibilities of workflows from the production of 1D to 2/3/4D EOS tables.
These new awaited theoretical tools will open the gate to connect the experimental domains of neutron stars and heavy-ion collisions, offering novel ways to infer properties of nuclear matter while imposing their respective constraints on one another.
\textcolor{black}{Moreover, the MUSES CE being an extendable framework by essence, other models and approaches complementary to the ones currently available (such as alternative approaches to similar models, or even new approaches not yet available) can easily be added in the future through the integration of new modules, upon request from interested authors.
A complete list of currently available EOS modules and their underlying implementations is provided in Ref.~\cite{MUSES_Doc} and the MUSES online repository.
A fully self-consistent calibration of phenomenological models such as CMF, to simultaneously reproduce $\chi$EFT constraints at low density and pQCD limits at high density, could also be considered for future developments.
}

%


\subsection{Bayesian Analysis of Nuclear Dynamics (BAND)} \label{sec:band}

In this section, we provide an overview of the BAND collaboration's efforts at the intersection of Bayesian methods and nuclear physics. We discuss their cyberinfrastructure framework, give detailed insights to the tools available, and highlight BAND's contributions to the field of dense matter and the EOS.


\subsubsection{Overview}

Nuclear physics is progressing into the ``precision era", where knowing both experimental and theoretical uncertainties is critical for ongoing advances in the field. As this occurs, several overarching questions have intrigued the community:
\begin{itemize}
    \item How do we determine model uncertainties?
    \item How do we calculate the uncertainties of complex models with large numbers of variable parameters?
    \item What do we do when there are several models to describe one physical phenomenon, but it is unclear which, if any, are correct?
    \item How do we establish sustainable software that is both comprehensible and extendable by the nuclear physics community?
\end{itemize}
In answer, the BAND collaboration has developed a modular, community-friendly cyberinfrastructure framework that enables nuclear physicists to capitalise on state-of-the-art computational, theoretical, and statistical methods to yield rigorous Bayesian uncertainties for their problems~\cite{Phillips:2020dmw, bandframework}. This framework directly provides solutions to these questions through:
\begin{itemize}
    \item \textbf{Model calibration.} The BAND collaboration has collected and designed tools to help users with a given theoretical model perform uncertainty quantification (UQ) through Bayesian calibration of the model parameters. This produces full probability distributions that can be propagated to observables computed using the model in question.
    \item \textbf{Emulation.} For models with extensive numbers of parameters, it can be difficult to compute reliable UQ without very large computational times. Using a surrogate model, i.e., an emulator, can speed up these calculations and avoid computational bottlenecks. This is especially useful for large frameworks, such as those used by astrophysical collaborations, e.g., LIGO-Virgo-KAGRA.
    \item \textbf{Model mixing.} When there is more than one theoretical model to describe a physical process, and it is not clear that the true model is even in the set of those available, Bayesian model mixing (BMM) is a trustworthy option to combine these models into an optimal, \textit{mixed} model. This resulting model possesses the best attributes of each individual component, which is implemented via local weighting, which will be described in more detail in the following sections. BMM is a frontier field, with much needed development still to come as it matures. In an effort to introduce this technique to nuclear physicists and other scientists who may need it, BAND has provided the community with \texttt{Taweret}, a Python package that enables users of any field to implement different techniques of BMM according to their needs (see Sect.~\ref{sec:bandframework}).
    \item \textbf{Modular, sustainable software.} The (primarily Python-based) cyberinfrastructure is meant for both static use and extension by the nuclear physics community, and has been written with versatility and longevity in mind. BAND has also provided tutorials and case studies using the software in the framework so that new researchers will be able to take up the tools presented there and use them effectively in their own research. Many of the packages contained in the framework are designed to be applicable to a wide range of physics problems, eliminating difficulties with generalising code.
\end{itemize}

The BAND framework possesses both the computational tools and user-friendly tutorials to help newcomers understand and incorporate Bayesian methods into their own work. In the next section, we discuss the framework's components and some of the available tools in detail.


\subsubsection{The BAND framework} \label{sec:bandframework}

\begin{figure*}[t]
    \centering
    \includegraphics[width=\linewidth]{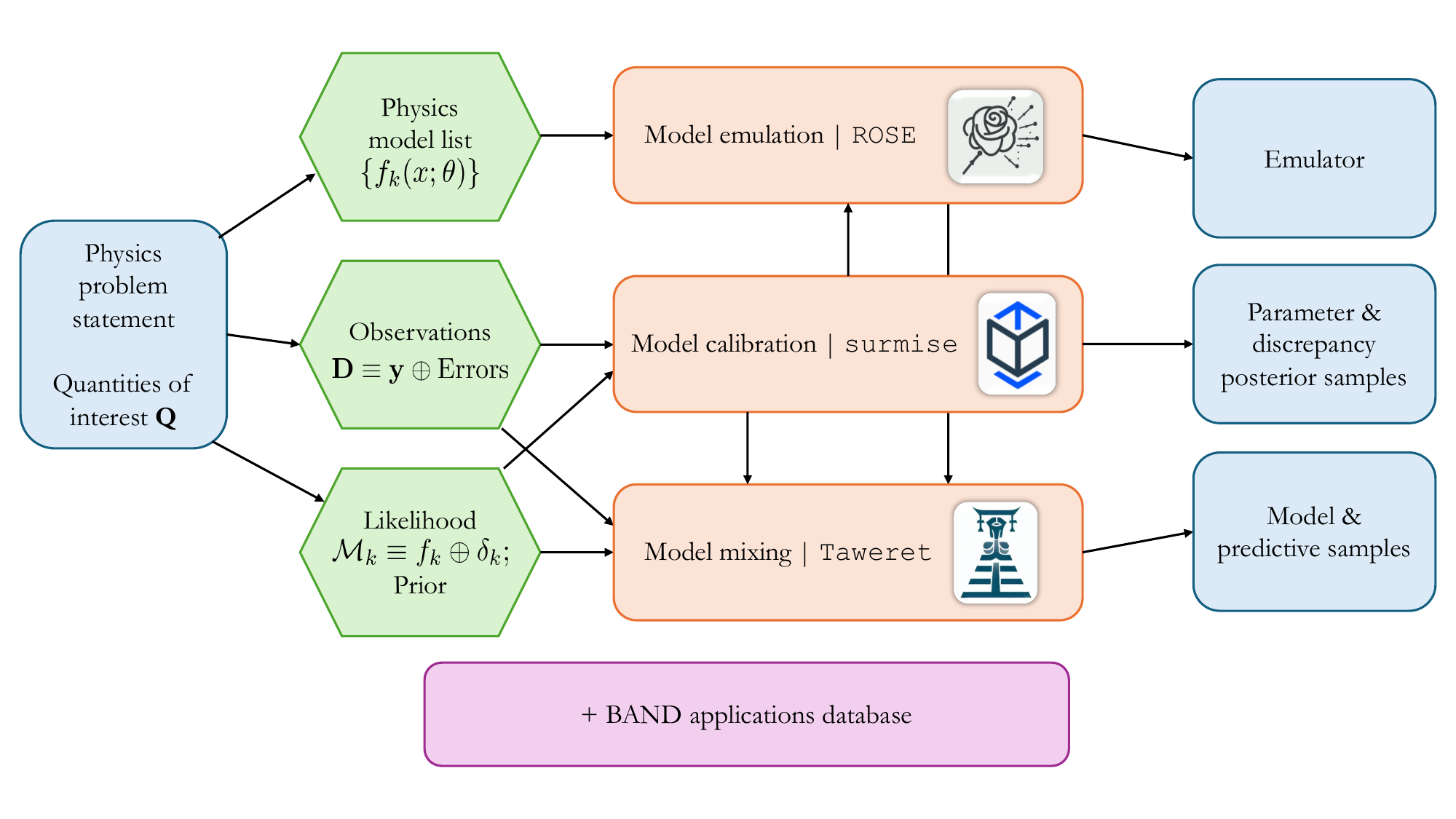}
    \caption{The BAND framework, updated to reflect the current status of each potential branch taken by a user with a physics problem of interest (here, the dense matter EOS). The BAND applications database consists of GitHub repositories and code packages that deal with specific physics problems, whereas the branches of the framework are generic and able to be used for a wide range of fields, including some outside of physics.} 
    \label{fig:band_framework}
\end{figure*}

The BAND framework's general structure is shown in Fig.~\ref{fig:band_framework}. To use this framework, a researcher supplies the physics problem of interest, here the dense matter EOS, and is able to use one (or more) of the three major branches in the framework---model emulation, calibration, or model mixing. These produce, respectively, an emulator of the EOS, posterior samples of the EOS or resulting observables, or a model of the EOS and corresponding predictive samples. Each of these branches has been realised in an individual Python package, built for wide application in nuclear physics. Model calibration can be achieved using the package \texttt{surmise}~\cite{surmise2021} and model mixing is accessible in \texttt{Taweret}~\cite{Ingles:2023nha, Taweret}. Model emulation can be achieved by both \texttt{surmise} and \texttt{ROSE}~\cite{Odell:2023cun}. \texttt{ROSE} is designed for nuclear scattering problems and is not as generic;  however, the mathematical and computational tools used in \texttt{ROSE} are adaptable to other fields. BAND users also have access to a comprehensive database of physics applications where BAND tools have been developed and applied to a wide range of problems, e.g., the dense matter EOS~\cite{Semposki:2024vnp, Semposki:2025etb}, the saturation point of nuclear matter~\cite{Drischler:2024ebw}, and relativistic heavy-ion collision calculations~\cite{Liyanage:2023nds}. Regarding the dense matter EOS specifically, BAND possesses the tools to achieve calibration of parameters in an EOS (see Sect.~\ref{sec:outlook} on potential future uses of these modules to this purpose). \texttt{surmise} could be used in this context; it focuses on calibrating the parameters of a model via emulating the model first and calibrating the model parameters by calling the emulated model, or calibrating the original model to experimental data directly. \texttt{surmise} also accepts user-built calibration methods, making it a very flexible tool. It is already being implemented in analysis of data significant to dense matter studies by physicists that are not in the BAND collaboration~\cite{Andronic:2025ylc}.

The most progress to date with respect to dense nuclear matter in the BAND collaboration has been in terms of BMM. BMM generally involves selecting a method to weight models such that each one is locally dominant in its respective region, allowing models to supersede others where they are more precise, or, with the proper interpolation techniques, allowing the mixed model to cross between models that do not overlap in the chosen input space. Mathematically, this translates to weighting the \textit{posterior predictive distributions} of each individual model $\mathcal{M}_k$, which are given by
\begin{equation}
    \label{eq:singlePPD}
   p(\bm{\tilde {y}} | \bm{y}, \bm{x}, \mathcal{M}_{k}) = \int p(\bm{\tilde y} | \bm{y}, \bm{x}, \bm{\theta}) p(\bm{\theta} | \bm{y}, \bm{x}) d\bm{\theta},
\end{equation}
where $\bm{y}$ and $\bm{x}$ are the observations at points in the input space, and $\bm{\tilde y}$ are new predicted observations. The right-hand side of the equation denotes the distribution of these predicted observations given the known observations and parameter values $\bm{\theta}$, and the posterior of these parameter values given $\bm{y}$ and $\bm{x}$. The parameter values are marginalized over via the integral. The parameters relate to the individual models through 
 \begin{equation}                   \label{eq:stand_model}
   \mathcal{M}_k : y_i = f_k(x_i; \bm{\theta}) + \varepsilon_{i,k},
 \end{equation}
where $\varepsilon_{i,k}$ indicates the discrepancy term on each model from theoretical and experimental uncertainties. 

The BMM weighting process can generally be defined as
\begin{equation}     \label{eq:PPD_weighting}
     p(\bm{\tilde y} | \bm{y}, \bm{x}) = \sum_{k=1}^{K} \hat w_k(x) p (\bm{\tilde y} | \bm{y}, \bm{x}, \mathcal{M}_k),
 \end{equation}
where we denote $\bm{\hat{w}}_{k}(x)$ as the location-dependent weights on the individual posterior predictive distributions $p (\bm{\tilde y} | \bm{y}, \bm{x}, \mathcal{M}_k)$ corresponding to each model  $\mathcal{M}_k$. Models can either be weighted as full probability distributions across the input space, or weights can instead be applied to the means and variances of the models. The output mixed model will then either be the full mixed distribution or the mean and variance, from which a full distribution can be reconstructed if it is assumed to be a Gaussian form.  

\begin{figure*}[!t]
    \centering
    \includegraphics[width=\linewidth]{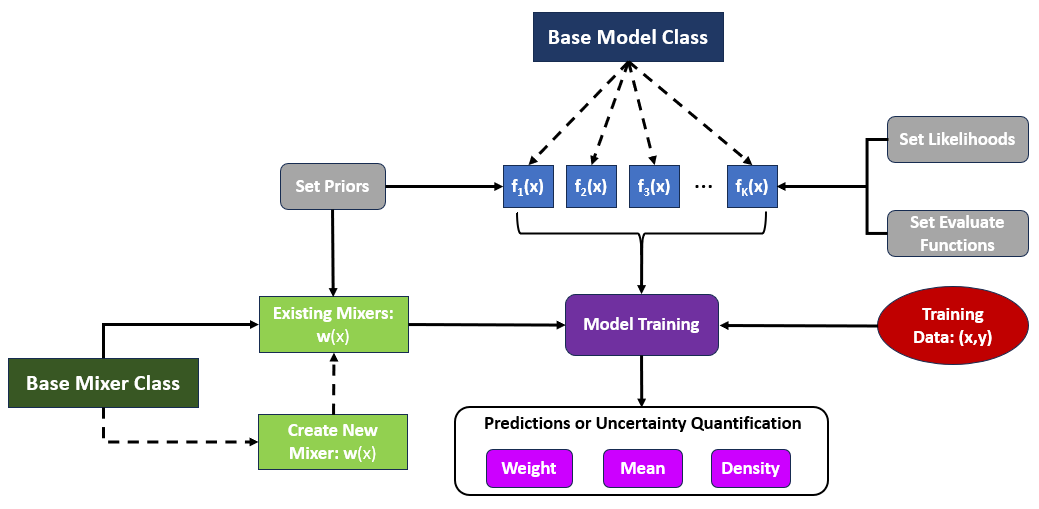}
    \caption{The structure of the workflow in the \texttt{Taweret} Python package. Users are able to use a mixing class located in \texttt{Taweret} to combine their models of interest, or include their own module for model mixing in the package for their use as well as others in the community. Figure taken from Ref.~\cite{Ingles:2023nha}.}
    \label{fig:taweret_diagram}
\end{figure*}

The BAND BMM-specific package, \texttt{Taweret}, takes in several theoretical models, in the form of model likelihoods and priors (see Fig.~\ref{fig:taweret_diagram}) into a base model class that gets subsequently fed into the package, where the model mixing method can be chosen by the user to perform the BMM step (see Ref.~\cite{Ingles:2023nha} for explicit details of each available model mixing technique). Then the routine will return a mixed model to the user that possesses rigorous uncertainties, the form of which will depend on the chosen mixing method. Currently this package is only able to accept theoretical models that have fixed parameters, called \textit{calibrated} models; future extensions plan to include the option of calibration alongside model mixing~\cite{Ingles:2023nha}, which could be very advantageous for forming the theoretical prior distribution of the EOS.


\subsubsection{BAND and the dense matter EOS} \label{sec:bandeoswork}

BAND efforts have been widespread in nuclear physics to date, but a significant number of works have been published that help to constrain the dense matter EOS. The primary focus has been to develop BMM tools for the zero-temperature EOS spanning from chiral EFT ($\chi$EFT) to perturbative QCD~\cite{Semposki:2024vnp, Semposki:2025etb}, but there has also been notable work to constrain the saturation point of nuclear matter~\cite{Drischler:2024ebw}. Below we discuss both advances, the computational tools for which can be found in corresponding open source repositories~\cite{bandframework, Taweret, EOS_BMM_ANM}.

\textbf{BMM for the dense matter EOS.} The BAND collaboration focused its efforts early on in developing model mixing techniques for general physics applications, with the dense matter EOS in mind as a possible case study of BAND tools for both model mixing and UQ~\cite{Phillips:2020dmw}. Specifically, a Gaussian process (GP) approach was tested to bridge the gap between two theories at extreme values of the input space~\cite{Semposki:2022gcp}. GPs have been used extensively in UQ for the EOS in recent years, and their development and exploration in the field is still ongoing~\cite{Essick:2020flb, Drischler:2020yad, Drischler:2020hwi, Finch:2025bao}. Within the BAND collaboration, they have been used to bridge the gap between the EOS from $\chi$EFT to pQCD, all the while preserving uncertainties of both theories within their respective density regions~\cite{Semposki:2024vnp}. 
\textcolor{black}{Here we discuss their use in this application, including their limitations.}

GPs are constructed from a mean function $m(x)$ and a covariance function, or \textit{kernel} $\kappa(x,x')$, and are denoted by~\cite{rasmussen2006gaussian}
\begin{equation}
    f(x) \sim \mathcal{GP}[m(x), \kappa(x,x')], \quad x \in \mathbb{R}^{d},
\end{equation}
where $x$ is some $d$-dimensional input space. The user's ability to choose a desired mean function or kernel to help capture the trends in the data from their scientific problem makes them very flexible objects. The kernel chosen may also have intrinsic parameters, called \textit{hyperparameters}, that need to be optimized. These parameters can be fixed by a user, or priors (called \textit{hyperpriors}) can be placed on them so that the GP produces realistic, physical results. This can be especially useful for situations where a quantity must remain positive, or can only vary between fixed values. For a more detailed introduction to GPs, we refer the reader to Refs.~\cite{Melendez:2019izc, duvenaud_PhD_2014, rasmussen2006gaussian}, and for an exposition of possible GP kernel and hyperprior choices we recommend  Ref.~\cite{Semposki:2025etb}.

\begin{figure*}
    \centering
    \includegraphics[width=\linewidth]{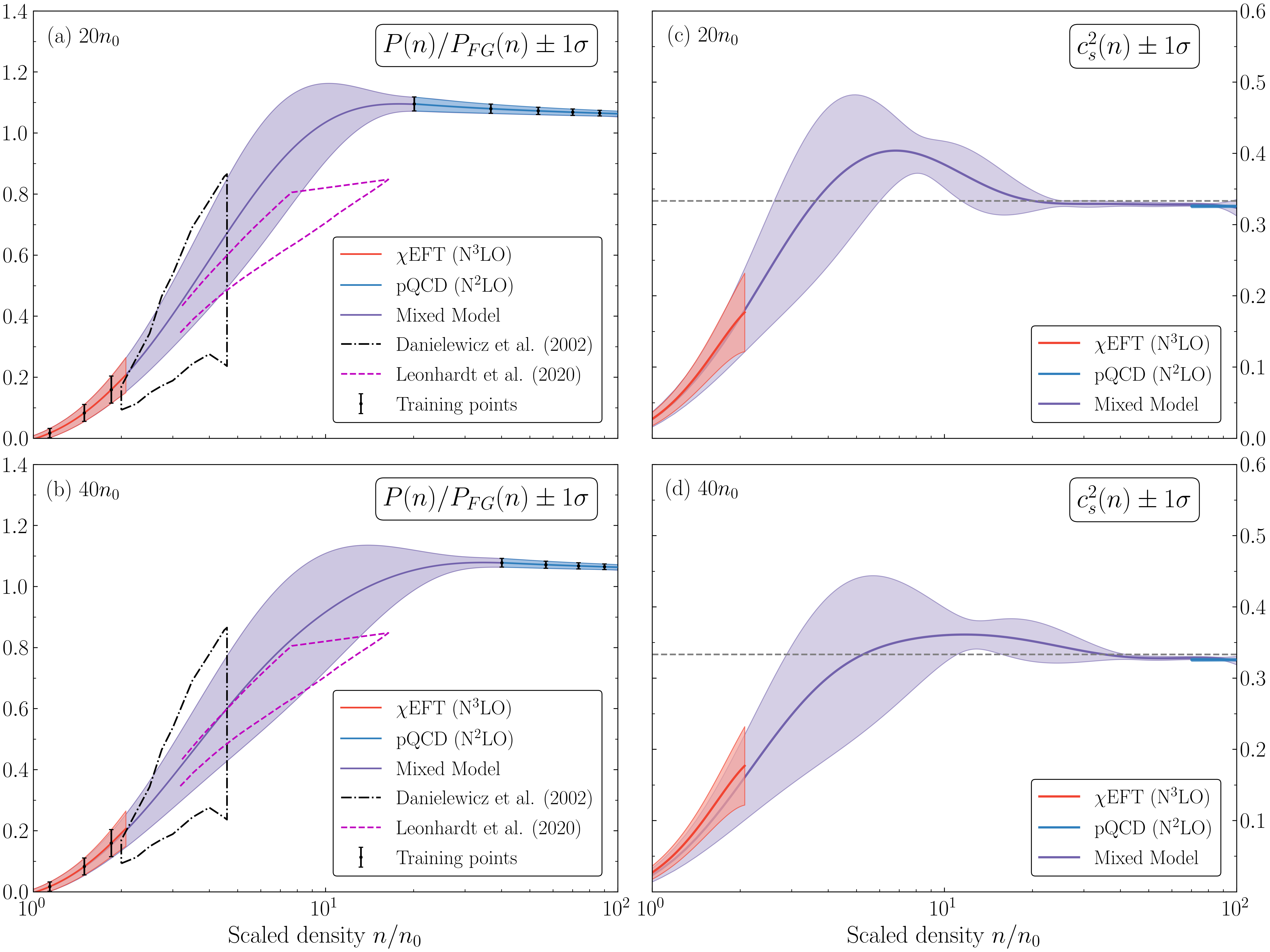}
    \caption{The results for the mixed model EOS using physics-informed hyperpriors on the squared-exponential RBF kernel hyperparameters. [(a),(b)] The pressure $P(n_B)$ of the EOS, scaled by the Fermi gas pressure $P_{FG}(n_B)$. [(c),(d)] The  corresponding speed of sound squared curves as a function of density. The results reproduce the speed of sound (including uncertainties) obtained in pQCD at asymptotically high densities, and overlap quite well the speed-of-sound uncertainty band of $\chi$EFT at low densities. Reprinted with permission from A. C. Semposki, C. Drischler, R. J. Furnstahl, J. A. Melendez, and D. R. Phillips, Phys. Rev. C. \textbf{111}, 035804 (2025)~\cite{Semposki:2024vnp}. Copyright 2025 by the American Physical Society.
    } 
    \label{fig:band_results}
\end{figure*}

To perform BMM on the symmetric matter EOS from $\chi$EFT and pQCD, rigorous UQ was first performed for the individual models using the BUQEYE truncation error model in Ref.~\cite{Melendez:2019izc, Drischler:2020yad, Drischler:2020hwi}, and a GP was designed to be trained on means and covariance matrices of data\footnote{Here we use the term ``data'' to refer to mean values and covariance matrices from the theoretical model, not from any experimental or observational data.} drawn from the two EOSs. The full model-mixed EOS was then extracted from the trained GP. This result can be seen in Fig.~\ref{fig:band_results}, where the purple bands indicate the GP result for the EOS in terms of scaled pressure and scaled number density. The kernel to construct this model-mixed GP is given by the stationary, squared-exponential radial basis function kernel with a hyperparameter called the marginal variance $\bar{c}^{2}$,\footnote{This parameter controls the variation of the GP in fitting the data; see Refs.~\cite{Melendez:2019izc, Semposki:2025etb} for more details.}
\begin{equation}
    \kappa_{\textrm{RBF}}(x,x') = \bar{c}^{2}\exp\left( \frac{-(x-x')^{2}}{2\ell^{2}} \right).
\end{equation}
The hyperparameters of this kernel were constrained using a hyperprior formed from a truncated normal distribution. 
The resulting smooth, causal EOS could be used in heavy-ion calculations, similar to Ref.~\cite{Yao:2023yda}. Users of the software produced (see Ref.~\cite{EOS_BMM_ANM}) can easily update these predictions with more heavy-ion data and further constrain the intermediate region between $\chi$EFT and pQCD. An extension to neutron star matter and heavy-ion data is shown explicitly in the following case.

\begin{figure*}
    \centering
    \includegraphics[width=\linewidth]{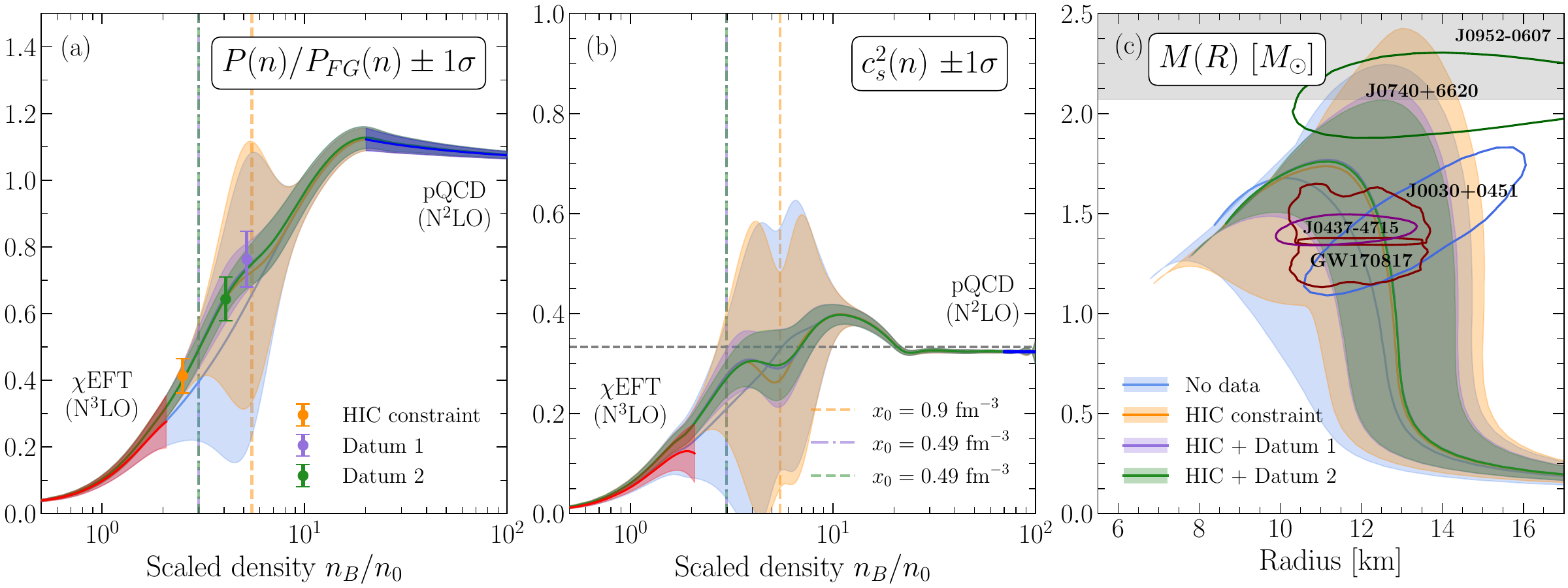}
    \caption{(a) The resulting asymmetric matter, mixed-model EOS in terms of scaled pressure $P(n_B)/P_{FG}(n_B)$ with respect to the scaled baryon number density $n_B/n_0$. Here a changepoint kernel with a sigmoid mixing function and a cubic spline mean function was applied as a GP prior. Each data point corresponds to an iteratively added constraint from heavy-ion data (orange) and two possible model constraints (purple and green). The resulting EOS shrinks in the intermediate region as these constraints are included. (b) The corresponding speed of sound squared, $c_{s}^{2}(n_B)$, for each added constraint. The vertical dashed and dash-dot lines show the locations of the changepoint $x_{0}$ for each iterative training step. (c) The resulting 90\% mass-radius uncertainties for each mixed-model EOS as the constraints are included. Notable NICER~\cite{Choudhury:2024xbk, Riley:2019yda, Riley:2021pdl, Miller:2021qha, Miller:2019cac, Romani:2022jhd} and LIGO-Virgo~\cite{LIGOScientific:2018cki} measurements are included as contours overlaying the mixed-model uncertainty bands. Figure modified from Ref.~\cite{Semposki:2025etb}.}
    \label{fig:band_nsm_results}
\end{figure*}

To incorporate constraints from experimental or other theoretical model data, e.g., from possible intermediate models between $\chi$EFT and pQCD, the symmetric nuclear matter code was extended to $\beta$-equilibrated matter. The GP model mixing framework itself was also extended to include more flexible kernels and hyperpriors. To calculate the model-mixed EOS of $\beta$-equilibrated, zero-temperature dense matter, a non-stationary GP kernel was implemented, i.e., one that varies directly with number density, given by
\begin{align}
    \label{eq:changepointkernel}
    \kappa_{f}(x,x'; \ell_{1}, \ell_{2}, \bar{c}_{1}^{2}, \bar{c}_{2}^{2}, \bm{\xi}) &= [1 - \alpha(x, \bm{\xi})] 
    \times \kappa_{1}(x,x'; \bar{c}_{1}^{2}, \ell_{1}) [1 - \alpha(x', \bm{\xi})] \nonumber \\
    &\quad+ \alpha(x, \bm{\xi}) \kappa_{2}(x, x'; \bar{c}_{2}^{2}, \ell_{2}) \alpha(x', \bm{\xi}),
\end{align}
where $\alpha$ is a mixing function (here, a sigmoid function) with parameters $\bm{\xi}$. This formalism combines the two stationary radial basis function kernels describing $\chi$EFT and pQCD, $\kappa_{1}$ and $\kappa_{2}$, respectively. A truncated normal hyperprior is placed on the parameters of the mixing function during the optimization process of the GP. This kernel also allows for simple inclusion of measurements from heavy-ion collision experiments~\cite{Huth:2021bsp} and pseudo-data from potential theoretical models that would cover some of the intermediate densities~\cite{Semposki:2025etb}. These additions were compared as each point was included in the calculation of the model-mixed EOS, as shown in Fig.~\ref{fig:band_nsm_results}a. As each additional point is added, the model-mixed EOS, and its subsequent speed of sound result in  Fig.~\ref{fig:band_nsm_results}b, becomes more and more constrained in both the intermediate and $\chi$EFT density regions, exhibiting what may happen in the $n_B - P(n_B)$ and $n_B - c_{s}^{2}(n_B)$ planes if more precise theoretical models were devised in the intermediate regime. 

In Fig.~\ref{fig:band_nsm_results}c, the model-mixed EOSs are shown overlaid with the 90\% contours of several LIGO-Virgo and NICER mass-radius posteriors~\cite{Choudhury:2024xbk, Riley:2019yda, Riley:2021pdl, Miller:2021qha, Miller:2019cac, Romani:2022jhd, LIGOScientific:2018cki}. The shrinking of the uncertainties in the $M-R$ region, and the slight shift of the mean curve as data is added iteratively to the model-mixed EOS, indicates the significant influence of data in the intermediate regime (i.e., $\geq 2n_0$) on the EOS and its observables. 

These proof-of-principle results for the dense matter EOS can be further extended to arbitrary temperature, proton fraction, and chemical potential in future versions of this BMM framework, or combined with other EOS codes to provide a theory prior on which the full EOS posterior can be built, including astrophysical and heavy-ion data directly into the prediction, as in the frameworks discussed in Refs.~\cite{Koehn:2024set, Cartaxo:2025jpi}. The BAND \texttt{neutron-rich-bmm} repository contains the GP mixing research described here~\cite{EOS_BMM_ANM}; in future BAND releases, this package will be generalized and included in \texttt{Taweret} for widespread use.

\textcolor{black}{At this point, it is important to note the limitations of this technique. The model-mixed EOS posteriors discussed above are built from a specific choice of GP kernel, hyperprior(s), and mean function. As such, this GP is somewhat limited by these decisions; it therefore holds some bias from these initial priors, which influences the size of the uncertainty bands in the intermediate region, regardless of how flexible the prior is chosen to be. This effect is especially prevalent when no data is included at these intermediate densities. Hence, these intermediate uncertainties are not fully thermodynamically informed---they are influenced by these initial modelling choices. We note here that the authors of Refs.~\cite{Legred:2025aar, Essick:2020flb, Essick:2021kjb, Essick:2023fso} attempt to mitigate this type of bias by sampling from thousands of GPs with different length scales. More recently, the authors of Ref.~\cite{Gorda:2025aiu} uses a Brownian ridge framework to reduce the bias of the initial choice of prior at intermediate densities, and is also able to include constraints from thermodynamic integrals, causality, and stability. In future, it would be interesting to explore how these other types of frameworks could be used in the unification workflow discussed later in Sect.~\ref{sec:unifiedframework}.}

\begin{figure*}[t]
    \includegraphics[width=\linewidth]{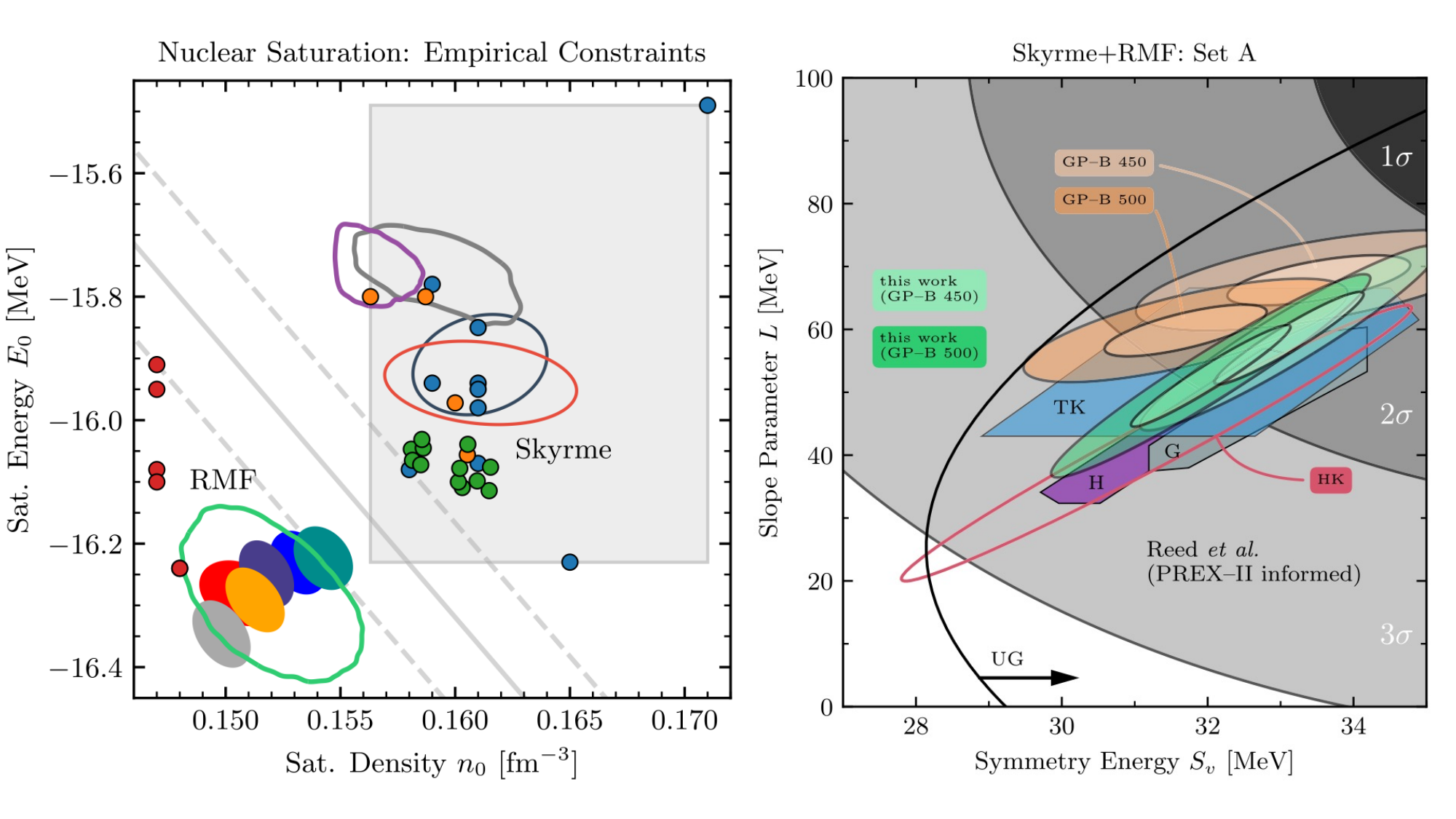}
    \caption{(a): Empirical constraints from multiple DFT calculations using relativistic mean field (RMF) and Skyrme models for the nuclear saturation point $(n_0,E_0)$. The contours represent 95\% credible intervals. The grey box represents the constraint from Ref.~\cite{Drischler:2015eba}. The light green contour corresponds to the kernel density estimation of Giuliani et al.~\cite{Giuliani:2022yna}, which was used in the analysis of Ref.~\cite{Drischler:2024ebw}. A support vector machine was used to calculate the light grey dashed margins and maximum margin separating hyperplane (light solid grey line) that show the clear discrepancy between the RMF and Skyrme results. See Fig. 1 of Ref.~\cite{Drischler:2024ebw} for the authors and papers corresponding to each calculation shown. (b): A comparison of several calculations of the symmetry energy $S_{v}$ and slope parameter $L$: the unitary gas (UG) in black; microscopic constraints from: Hebeler et al.~\cite{Hebeler:2010jx} (H) in purple, Gandolfi et al.~\cite{Gandolfi:2011xu} (G) in grey, Tews and Kr\"uger et al.~\cite{Tews:2012fj} (TK) in blue, and Holt and Kaiser~\cite{Holt:2016pjb} (HK) in red, and the PREX-II informed constraint from Reed et al.~\cite{Reed:2021nqk} in grey in the background of the figure. The ellipses in the foreground represent the BUQEYE collaboration constraint from Refs.~\cite{Drischler:2020hwi, Drischler:2020yad}, and the green ellipses represent the work of Ref.~\cite{Drischler:2024ebw} to further constrain this result via UQ on the saturation point as described in the text. The ellipses represent two different confidence regions: the lighter is $1 \sigma$ (39\%) and the darker is $2 \sigma$ (86\%). The mixture model used to obtain these results is using DFT predictions with quantified uncertainties in the mixture model. Figure taken and modified with permission from Ref.~\cite{Drischler:2024ebw}.}
    \label{fig:nsat_constraints}
\end{figure*}

\textbf{Constraining the nuclear saturation point.} Collaborations with BAND have also included constraining the EOS of dense matter, specifically at the nuclear saturation point, which can be denoted as the pair of values corresponding to the saturation density, $n_0$, and the energy per particle minimum in SNM, $E_0$~\cite{Drischler:2024ebw}. When comparing the numerical value of this pair derived from two classes of density functional theory (DFT) approaches, Skyrme and Relativistic Mean Field, there are discrepancies between the results (see the left panel of Fig.~\ref{fig:nsat_constraints} and Fig. 1 of Ref.~\cite{Drischler:2024ebw}). To remedy this issue, a more general BMM approach, here called a Bayesian mixture model, was devised and applied to these DFT models to better determine a reliable uncertainty band for the $(n_0, E_0)$ pair (for full details of the calculation itself see Ref.~\cite{Drischler:2024ebw}). To summarize, each DFT prediction of $(n_0, E_0)$ was taken as an individual draw from an underlying unknown distribution, thereby including the potential for these draws to be correlated with one another. From here, a hierarchical model was devised which was able to incorporate knowledge of the uncertainties on the individual DFT models producing these $(n_0, E_0)$ pairs. This knowledge, along with informative prior distributions, produced a posterior for the saturation point with reliable estimates of uncertainties~\cite{Drischler:2024ebw}.

The uncertainty band produced in this investigation puts a rigorous error bar on the saturation point, and therefore allows the propagation of the uncertainties in this quantity to observables computed in nuclear matter, e.g., the symmetry energy $S_{v}$ and its slope parameter $L$. When including the uncertainties of the saturation point in the predictive distribution of $S_{v}$ and $L$, the mean value of the $S_{v}-L$ predicted contour is shifted to smaller values of $L$ and larger values of $S_{v}$ than when they are not included (see Fig.~\ref{fig:nsat_constraints}). These studies are paramount in helping to eliminate detrimental effects in nuclear matter calculations stemming from discrepant theoretical models. This type of analysis is also useful for improving our progress in solving the \textit{inverse problem}, i.e., using constraints in $S_{v}-L$ and other observables in dense matter to narrow the field of possible EOSs that are able to generate them.

\section{Outlook and Opportunities}
\label{sec:outlook}

In the past decade, the advent of multi-messenger astrophysics and the development of analysis frameworks with input from both nuclear theory and experiments have paved the way for a more unified understanding of the dense nuclear matter EOS. This comes at an important moment, as the next generation of nuclear experiments and astrophysical observations will provide far more precise data, marking a transition of EOS studies toward the regime of ``precision physics" (see tab.~\ref{tab:eos_datafuture}). 

\begin{table}[!b]
\centering
\renewcommand{\arraystretch}{1.35}
\resizebox{\textwidth}{!}{%
\begin{tabular}{@{}ccccccccccccccccccc@{}}
\toprule
 &
   &
  22 &
  23 &
  24 &
  25 &
  26 &
  27 &
  28 &
  29 &
  30 &
  31 &
  32 &
  33 &
  34 &
  35 &
  36 &
  37 &
  38 \\ \midrule \midrule
\multirow{10}{*}{\begin{tabular}[c]{@{}c@{}}HIC\end{tabular}} &
 STAR    & \multicolumn{4}{c}{\cellcolor{orange!50}}             &  &  &   &   &   &   &   &  &  &       &       &   &   \\
 & HADES    & \multicolumn{4}{c}{\cellcolor{orange!50}}             &  &  &   &   &   &   &   &  &  &       &       & &     \\
 & NA61    & \multicolumn{13}{c}{\cellcolor{orange!50}}             &  &  &        \\
 & ALICE    & \multicolumn{17}{c}{\cellcolor{orange!50}}                 \\
 & ASY-EOS  &     &     & \multicolumn{2}{c}{\cellcolor{orange!50}} &  &  &   &   &   &   &   &  &  &       &       &      \\
 & NICA      &     &     & \multicolumn{10}{c}{\cellcolor{orange!50}}                             &  &       &       &      \\
 & FRIB     &     &     &           & \multicolumn{14}{c}{\cellcolor{orange!50}}                                           \\
 & HIAF      &     &     &           &          \multicolumn{14}{c}{\cellcolor{orange!50}}                       \\ 
 & CBM      &     &     &           &          &  &  &   & \multicolumn{10}{c}{\cellcolor{orange!50}}                       \\ 
 & NA60+      &     &     &           &          &  &  & &  & \multicolumn{9}{c}{\cellcolor{orange!50}}                       \\ 
\midrule
\multirow{3}{*}{\begin{tabular}[c]{@{}c@{}}BNS\\ GW\end{tabular}} &
  LIGO (O5) &
  \multicolumn{10}{c}{\cellcolor{blue!50}} &
   &
   &
   &
   &
   &
   \\
 & ET       &     &     &           &          &  &  &   &   &   &   &   &  &  & \multicolumn{4}{c}{\cellcolor{blue!50}} \\
 & CE       &     &     &           &          &  &  &   &   &   &   &   &  &  & \multicolumn{4}{c}{\cellcolor{blue!50}} \\
 & LISA     &     &     &           &          &  &  &   &   &   &   &   &  &  &       &       \multicolumn{3}{c}{\cellcolor{blue!50}}     \\ \midrule
\multirow{4}{*}{\begin{tabular}[c]{@{}c@{}}(B)NS\\ EM\end{tabular}} &
  NICER (AO8) &
  \multicolumn{6}{c}{\cellcolor{green!50}} &
   &
   &
   &
   &
   &
   &
   &
   &
   
   &
   \\
 & VRO  &   &   &   & \multicolumn{10}{c}{\cellcolor{green!50}} & &   &       \\
 & eXTP     &     & &   &     &           &         &  &  & \multicolumn{9}{c}{\cellcolor{green!50}}            \\
 & NewATHENA   &     &     &           &          &  &  &   &   &   &   &   &  &  & & & \multicolumn{2}{c}{\cellcolor{green!50}} \\
  & THESEUS &     &     &    & & & & &       &          &  &  &     &   &   & & \multicolumn{2}{c}{\cellcolor{green!50}}           \\ \bottomrule
\end{tabular}
}
\caption{Compilation of the schedules timelines of various running and planned heavy-ion collision (HIC) facilities and programmes, along with the observational runs for astrophysical programmes (gravitational waves from binary neutron star mergers (BNS-GW) and electromagnetic observations (EM)) providing complementary constraints on the nuclear high density EOS. Information about the individual sources is detailed in the text.}
\label{tab:eos_datafuture}
\end{table}

\subsection{Relativistic Heavy-Ion Collisions}

\vspace{6pt}
{\textbf{Frameworks: Consistent Transport Codes and Hybrid Approaches}}
\vspace{3pt}

The existing and upcoming experimental programmes will only reach their full potential if accompanied by transport and hydrodynamic frameworks whose systematic uncertainties are under quantitative control. The Transport Model Evaluation Project (TMEP) has taken an essential step toward establishing benchmarking frameworks that identify systematic differences across different codes. So far, these efforts have focused on $E_{\rm lab} \lesssim 1A$~GeV, where densities up to $\sim 2\,n_0$ are reached. Extending such systematic comparisons into the $E_{\rm lab}=1$–$10A$~GeV regime relevant for upper GSI, FAIR, NICA and RHIC-FXT energies is therefore a high-priority step toward precision EOS extraction at $n_B \gtrsim 2\,n_0$. Ultimately, such an endeavour has to provide a quantitative systematic uncertainty assessment of i) the chosen method of cluster production, ii) the implemented hadron list, their properties and all binary cross sections, iii) the treatment of in-medium cross sections, iv) the description and propagation of resonances, v) the choice of the EOS (density-, isospin-, momentum- and particle-ID-dependent) as well as the actual implementation of the EOS, vi) the choice of the initial state (shape deformations, high momentum tail, neutron skin, ...), vii) a variety of code-specific technical parameters such as e.g. the number of test particles, widths used for Gaussian smearing etc.

In parallel, hybrid descriptions, in which a pre-equilibrium transport stage is followed by viscous hydrodynamics and a hadronic afterburner, are being pushed to lower beam energies~\cite{Petersen:2008dd,Karpenko:2015xea,Werner:2010aa,Song:2011qa,Goes-Hirayama:2025nls}. Multi-fluid approaches, which treat projectile, target and produced matter as distinct interacting fluids~\cite{Brachmann:1997bq,Ivanov:2005yw,Batyuk:2016qmb,Cimerman:2023hjw,Werthmann:2025ueu}, offer a complementary route to describe finite stopping and incomplete equilibration at intermediate energies. But also all hydrodynamic event generators need to be benchmarked providing a quantitative handle on their systematic uncertainties. Ultimately, a consistent description across all event generators, whose systematics are quantitatively under control, will be required to exploit the increasing precision of flow and particle production data of upcoming experiments. 

A second key development is the integration of Bayesian methods and emulators into heavy-ion analyses. First studies combining transport calculations with Bayesian parameter estimation have already demonstrated that simultaneous constraints on the EOS, effective masses and in-medium cross sections are generally feasible~\cite{Huth:2021bsp,OmanaKuttan:2022aml,Oliinychenko:2022uvy,Mohs:2024gyc}. Thus, building a flexible Bayesian framework around modern transport codes (i.e. with quantitative uncertainty control), supported by fast Gaussian-process emulators for computationally expensive simulations, will be essential to turn the wealth of forthcoming data into quantitative statements about the pressure of dense matter.

\vspace{6pt}
{\textbf{Symmetric Nuclear Matter - High-Statistics Measurements}}
\vspace{3pt}

Several heavy-ion collision experiments have traditionally spanned the energies corresponding to supranuclear densities albeit sparsely. However, systematic, precision scans both in collision energies and systems for experiments with large rapidity acceptance have been largely missing. Over the next decade, this will be addressed by several experiments, all with the aim of doing multi-differential studies of the EOS-sensitive observables.

\begin{itemize}
\setlength{\itemsep}{3pt}
    
    \item \underline{In the density regime $1-2.5n_{0}$} -- The HADES experiment at SIS-18 has already successfully demonstrated the impact of precision and wide acceptance measurements by measuring collective flow coefficients up to $v_6$ in Au+Au collisions at $\snn = 2.42\ \rm{GeV}$, enabling the reconstruction of the full three-dimensional emission pattern in momentum space~\cite{HADES:2020lob,HADES:2022osk}. In addition, HADES also allows the extraction of fireball temperatures via dielectron reconstruction~\cite{HADES:2019auv}, thus enabling the search for other observables sensitive to potential deconfinement signatures. This has triggered a systematic beam-energy scan by HADES across the SIS-18 energy range ($\snn = 1.96$--$2.23\ \rm{GeV}$ Au+Au collisions) in 2024-25, therefore, soon providing high-precision measurements of the aforementioned EOS-sensitive observables~\cite{HADES:QM25Plenary}. 
    
    \item \underline{In the density regime $>2.5n_{0}$} -- The recently concluded STAR Fixed-Target (FXT) program of the Beam Energy Scan (BES) at RHIC ($\snn = 3$--$13.7\ \rm{GeV}$ Au+Au collisions)~\cite{STAR:QM25Plenary} has paved the way for several upcoming experiments at new facilities with high-intensity heavy-ion beams~\cite{Musa:2024euo,Musa:2025yhp}. These include the Facility for Antiproton and Ion Research (FAIR) in Darmstadt~\cite{Durante:2019hzd, Aumann:2024unk}, the Nuclotron-based Ion Collider fAcility (NICA) in Dubna~\cite{Blaschke:2016NICA,Kekelidze:2017ghu}, the High Intensity heavy-ion Accelerator Facility (HIAF) in Huizhou~\cite{Yang:2013yeb}, and the Japan Proton Accelerator Research Complex - Heavy Ion Project (JPARC-HI) in Tokai~\cite{Sako:2019hzh,Miake:2021rgy}. Notably, the Compressed Baryonic Matter (CBM) experiment at FAIR is a general-purpose fixed-target experiment that will operate at unprecedentedly high beam-target interaction rates of up to 10~MHz to measure both hadronic and electromagnetic observables ($\snn = 2.9$--$4.9\ \rm{GeV}$ Au+Au collisions)~\cite{Friman:2011zz,CBM:2016kpk,Agarwal:2022ydl,Agarwal:2023otg}. Moreover, the Multi-Purpose Detector (MPD) at NICA is a $4\pi$ spectrometer, which is also a general-purpose experiment in collider configuration ($\snn = 4$--$11\ \rm{GeV}$ Au+Au collisions)~\cite{MPD:2022qhn,MPD:AllTDRs}. These efforts will be further complemented by experiments at existing facilities, such as the NA61/SHINE and the proposed NA60+ experiment at the SPS ($\snn = 6.3$--$17.3\ \rm{GeV}$ Pb+Pb collisions)~\cite{NA61:2014lfx, NA60:2022sze, Arnaldi:2025ikz} and the future relocation of the HADES experiment from GSI to FAIR for collisions with moderate particle multiplicities~\cite{CBM:2016kpk,HADES:QM25Plenary}. Altogether, these endeavours are pivotal to probe the QCD phase structure and the possible onset of new degrees of freedom in the high-density regime.

\end{itemize}

\vspace{6pt}
{\textbf{Symmetry Energy: Breaching the $2n_{0}$ Frontier}}
\vspace{3pt}

In contrast to symmetric matter, the density dependence of the symmetry energy is currently well constrained only around and below saturation. Heavy-ion collisions with neutron-rich beams offer a path to extend these constraints into the supra-saturation regime. The FOPI-LAND and ASY-EOS measurements of the neutron-to-proton elliptic-flow ratio in Au+Au at SIS18 energies~\cite{Russotto:2011hq,Russotto:2016ucm}, combined with transport simulations, already limit the slope parameter $L$ and place non-trivial bounds on $S(n_B)$ up to $\sim 1.5\,n_0$~\cite{Lynch:2021xkq}. A follow-up ASY-EOS-2 programme at SIS18 \cite{Russotto:2021mpu} aims to improve the statistics and systematics of these measurements and to extend the set of isospin-sensitive observables. Moreover, studying the sub-threshold production of neutron-to-proton-like particle ratios, such as $\Sigma^-/\Sigma^+$ at the higher energies accessible with SIS-100 can further probe high-density nuclear symmetry energy~\cite{Yong:2022pyb}. Therefore, CBM's silicon vertexing and tracking detectors~\cite{Heuser:2013nft,Klaus:246516}, in conjunction with novel track reconstruction methods~\cite{Kisel:2018vkh}, enable the high-statistics measurement of $\Sigma$ hyperons, which are short-lived ($c\tau \sim 2$~cm) and  decay with at least one neutral daughter particle.

At lower beam energies but with extreme neutron-rich systems, the S$\pi$RIT experiment at RIKEN has systematically studied pion ratios for radioactive Sn+Sn systems at $\snn\simeq 2.01$~GeV~\cite{SpiRIT:2020sfn,SpiRIT:2021gtq}. In the longer term, FRIB and its planned FRIB400 upgrade will provide high-intensity rare-isotope beams with isospin asymmetries significantly beyond stable nuclei~\cite{FRIB400,Brown:2024rml}. Combined with large-acceptance neutron detectors and time-projection chambers, this will enable precision measurements of flows, isospin diffusion, and pion and light-cluster production in collisions that reach densities up to, and eventually beyond, $2\,n_0$. Complementary efforts are also foreseen at FAIR with the Super-FRagment Separator (Super-FRS) to produce high-intensity radioactive ion beams in the energy range up to about 2 GeV/nucleon~\cite{Aumann:2024unk}. 

When interpreted with systematically benchmarked transport models, such data will be crucial to reduce the present spread in the symmetry-energy band of Fig.~\ref{fig:joint_constraints_EOS} at supra-saturation densities and to connect heavy-ion constraints more tightly to neutron-star radii and tidal deformabilities.

\vspace{6pt}
{\textbf{Emerging Observables for Symmetric Matter and Symmetry Energy}}
\vspace{3pt}

While flow and sub-threshold kaon production remain cornerstone observables, new observables have increasing relevance due to the unrivalled statistics modern experiments can deliver. With the unprecedented high collision rate that FAIR will deliver, precision measurements of deep sub-threshold probes will become viable. This stretches multi-strange hadrons ($\Xi$, $\Omega$) \cite{Steinheimer:2015sha,Song:2020clw}, but also open and hidden charm ($J/\Psi$, $D$, $\bar{D}$, $\Lambda_c$) which are expected to be produced copiously and to probe the EOS \cite{Steinheimer:2016jjk,Steinheimer:2025trr}. Novel observables that tie to the traditional flow coefficients have been proposed via flow correlations \cite{Reichert:2022gqe,Reichert:2023eev}, 3D flow decomposition \cite{Reichert:2022yxq}, the measurement of di-lepton production and flow \cite{Seck:2020qbx,Savchuk:2022aev,Reichert:2023eev,Goes-Hirayama:2024aqz,Nishimura:2023not}. Femtoscopy, through the measurement of $R_{\rm out}$, $R_{\rm side}$, and $R_{\rm long}$, provides direct sensitivity to emission timescales $\Delta\tau$ and thus to the softness or stiffness of the EOS and signals of the phase transition \cite{Li:2022iil}. Continuously improving treatments of cluster production in HIC also receive more relevance for extracting features of the EOS \cite{Kireyeu:2024hjo,Zhou:2025zgn,Bratkovskaya:2025oys}. Moreover, recent ideas to measure the neutron skin in HIC at ultra-relativistic collision energies are also interesting for CBM~\cite{Hammelmann:2019vwd,Jia:2022qgl,Giacalone:2023cet,Pihan:2025pep} and tie to measurements of symmetry energy. These complement the still relevant traditional measurement of flow coefficients~\cite{Hillmann:2019wlt,Oliinychenko:2022uvy,Steinheimer:2022gqb,Tarasovicova:2024isp,Mohs:2024gyc}. 

The STAR collaboration has already initiated femtoscopic analyses at low beam energies \cite{STAR:2024zvj}, with forthcoming high-statistics measurements expected to deliver tighter constraints on the EOS as well as scattering lengths, while HADES is actively investigating flow correlations as well as di-lepton production and flow. At HL-LHC, the ALICE 3 experiment can probe this even further with unique measurements of three-body correlations and $A = 4$ or $A = 5$ hyper-nuclei across collision systems~\cite{ALICE:2022wwr}. On the symmetry energy side, pion yield ratios in radioactive-beam collisions (e.g., the S$\pi$RIT experiment at RIKEN) have proven powerful, and new opportunities may arise from isotopic collisions at the LHC, where asymmetric systems could probe isospin effects at higher energies. These novel observables promise to expand the EOS toolbox well beyond traditional flow measurements.

\vspace{12pt}
Taken together, these developments outline a clear roadmap for the next decade of EOS studies. High-statistics datasets will transform long-established observables into quantitative probes, while transport and hydrodynamic models evolve to meet the precision frontier. Complementary observables such as femtoscopy, isotopic collisions, and multi-strangeness production will open previously inaccessible windows on dense matter. On the symmetry energy side, FRIB and FRIB400, coupled with next-generation detectors and Bayesian multi-parameter inference, promise transformative advances in probing isospin effects at supra-saturation densities. Finally, the CBM physics programme will provide the experimental breadth needed to explore densities beyond $2.5n_{0}$ and search for signs of phase transitions. These opportunities promise not only improved constraints on the EOS of symmetric matter and symmetry energy, but also the possibility of uncovering new states of QCD matter at high density.

\subsection{Multi-Messenger Astrophysics}
In the coming decade, numerous generational leaps are expected in the astrophysical observations of neutron stars across different directions.

In the gravitational-wave side of the story, the third-generation ground-based gravitational-wave detectors, e.g., the European-driven Einstein Telescope (ET)~\cite{Punturo:2010zz, Hild:2010id} and American-driven Cosmic Explorer (CE)~\cite{Reitze:2019iox, Evans:2021gyd, Evans:2023euw}, promised to have an order of magnitude improvement in sensitivity across the frequency band. For the observation of binary neutron star mergers, such an improvement can advance the study of supranuclear matter in two directions. The first direction is the great increase in observations and a largely decreasing associated uncertainty. In the current generation, we have observed two binary neutron star mergers, i.e., GW170817~\cite{LIGOScientific:2017vwq} and GW190425~\cite{LIGOScientific:2020aai}\footnote{There is also a candidate detection of GW231109\_235456 by external effort~\cite{Niu:2025nha}.}, in the third-generation detector era, it is expected to observe $10^4-10^5$ mergers per year, with $10^3$ of them with less than $10\%$ uncertainty on the tidal deformabilities measured~\cite{Branchesi:2023mws}. Such a wealth of detections with unprecedentedly low uncertainties can yield a well-constrained mass-tidal deformability relation, with radius uncertainty down to $0.1{\rm km}$ and further inform us about the cold supranuclear equation of state~\cite{ET:2025xjr}. In addition to the usual inspiral of binary neutron star mergers, third-generation detectors also have the potential to detect gravitational waves from a different phase of the merger, namely the post-merger phase. Other than promptly collapsing into a black hole, the merger can result in a dense remnant that can withstand gravitational collapse with timescales ranging from several milliseconds to a few minutes after the merger~\cite{Rosswog:2001fh,Shibata:2006nm}. During such a phase, the temperature can reach $50$MeV, with different transport
coefficients leaving significant imprints on the gravitational-wave waveform~\cite{Hammond:2021vtv,Raithel:2021hye,Rosswog:2001fh}. Therefore, it allows us to explore a new dimension in the QCD phase diagram and potentially exotic nuclear phenomena~\cite{ET:2025xjr}.

There are also exciting advancements in electromagnetic observations. Two key sources for studying neutron stars are the signals from binary neutron star mergers, including kilonovae and gamma-ray burst afterglows, and those from rotating neutron stars, which can be analyzed through X-ray pulsar profiles. The Vera Rubin Observatory~\cite{LSST:2008ijt} and the Wide-field Spectroscopic Telescope (WST)~\cite{WST:2024rai} are well-equipped to detect kilonovae from events localized to within a few tens of square degrees, with redshifts less than $0.5$~\cite{Loffredo:2024gmx}. On the other end of the spectrum, the X-ray signals from the gamma-ray burst afterglow can be potentially captured by next-generation X-ray telescopes, e.g., NewATHENA~\cite{Nandra:2013jka,Cruise:2024mgo} and  THESEUS~\cite{THESEUS:2017wvz,Rosati:2021yjd}. For X-ray pulse profiling, the enhanced X-ray Timing and Polarimetry mission (eXTP)~\cite{Zhang:2025iae}, is expected to be $\sim3\times$ more sensitive than NICER and can measure the mass and radius of a neutron star with $\pm0.1M_\odot$ and $\pm0.1{\rm km}$, respectively~\cite{Li:2025uaw}. \textcolor{black}{Additionally, the anticipated measurement of the moment of inertia of PSR J0737-3039A~\cite{Bejger:2005jy, Greif:2020pju, Wang:2022cpi} would provide the first precision measurement of this observable, allowing us to compare model predictions and determine if this measurement produces tighter constraints on the validity of EOS predictive posteriors.}

\subsection{Community Tools for Unified Frameworks} \label{subsec:communitytools}

Beyond the three major frameworks previously discussed, there are several tools that have been developed with unification of the dense matter EOS in mind. Below we review some of these and briefly discuss their potential inclusion or use alongside the frameworks in Sect.~\ref{sec:frameworks}. 

\begin{itemize}
\setlength{\itemsep}{3pt}

    \item \textbf{\texttt{nucleardatapy}.} The \texttt{nucleardatapy} toolkit~\cite{Margueron:2025ugs} has an incredibly broad inclusion of experimental, theoretical, and observational data and analysis tools to construct the EOS. It is constructed as a Python package intended for continuous updating by the community as new data is taken and publicly released. This package could be used by the frameworks discussed (i.e., BAND, MUSES, and NMMA) to include state-of-the-art measurements and updated theoretical models, alleviating the difficulties of collecting and individually implementing various theoretical models and experimental datasets. This would make computation of the EOS much simpler and friendlier to new members of the dense matter community.

    \item \textbf{\texttt{CompactObject}.} The \texttt{CompactObject} collaboration is centred around a software framework built to allow users to perform Bayesian inference on the neutron star EOS~\cite{Huang:2024rfg, Cartaxo:2025jpi}. It accomplishes three major goals: supplying the community with a straightforward TOV solver; computing the neutron star EOS using a relativistic mean field approach, and outputting EOS posteriors given likelihoods based on experimental and astrophysical observations, as well as theoretical models. The software is also open source and meant to be extended by the community, just like BAND and \texttt{nucleardatapy}, enabling researchers to use it in conjunction with BAND or MUSES software to perform full Bayesian inference on the EOS.
    
    \item \textbf{\texttt{CompOSE}}.
    The \textbf{\texttt{CompOSE}} (Compact Star Online Supernovae Equations of State) database \cite{CompOSECoreTeam:2022ddl, Typel:2013rza} provides a standardized repository of tabulated equations of state for dense matter relevant to neutron stars, supernovae, and related astrophysical environments. 
    It offers access to a wide range of EOS tables derived from different theoretical models, all formatted consistently to facilitate comparison, interpolation, and use in simulations. 
    By centralizing and standardizing EOS data, CompOSE enables reproducibility and interoperability across the nuclear astrophysics community, and can thus be fed by MUSES thanks to its potential to generate a wide variety of EOSs.

    \item \textbf{Emulators.} Recent advances in model emulation for low-energy nuclear physics have ignited efforts to develop emulators for the field of dense matter physics. An emulator's purpose is to act as a surrogate model to assume the place of the original, computationally prohibitive model, often in a larger, intensive codebase. In nuclear physics, emulators have been used in the past few years largely for nuclear scattering~\cite{Drischler:2022ipa, Bonilla:2022rph, Giuliani:2022yna, Garcia:2023slj, Odell:2023cun, Maldonado:2025ftg, Gnech:2025gsy}, \textcolor{black}{but significant advances have also been made for many-body calculations of EFT-based potentials for infinite matter}~\cite{Jiang:2022oba}, and there has been recent effort in emulating the TOV equations for large Bayesian inference frameworks as well. The TOV equations can be a bottleneck for large-scale Bayesian computations, and both Refs.~\cite{Reed:2024urq} and~\cite{Lalit:2024vmu} devise sophisticated emulators to solve this problem.\footnote{In addition to emulators for the TOV equations, hardware acceleration using GPUs has also been explored and shows promising results for resolving bottlenecks in Bayesian inference~\cite{Wouters:2025zju}, also offer the functionality of auto-differentiation.} Since emulators can be adapted to fit a vast array of problems, other computationally costly scenarios in dense matter, e.g., model calibration of EOSs with large numbers of parameters, will certainly be addressed by emulators in the near future.
\end{itemize}


\subsection{Steps toward a unified framework} 
\label{sec:unifiedframework}

\begin{figure}[!b]
    \centering
    \includegraphics[width=\linewidth]{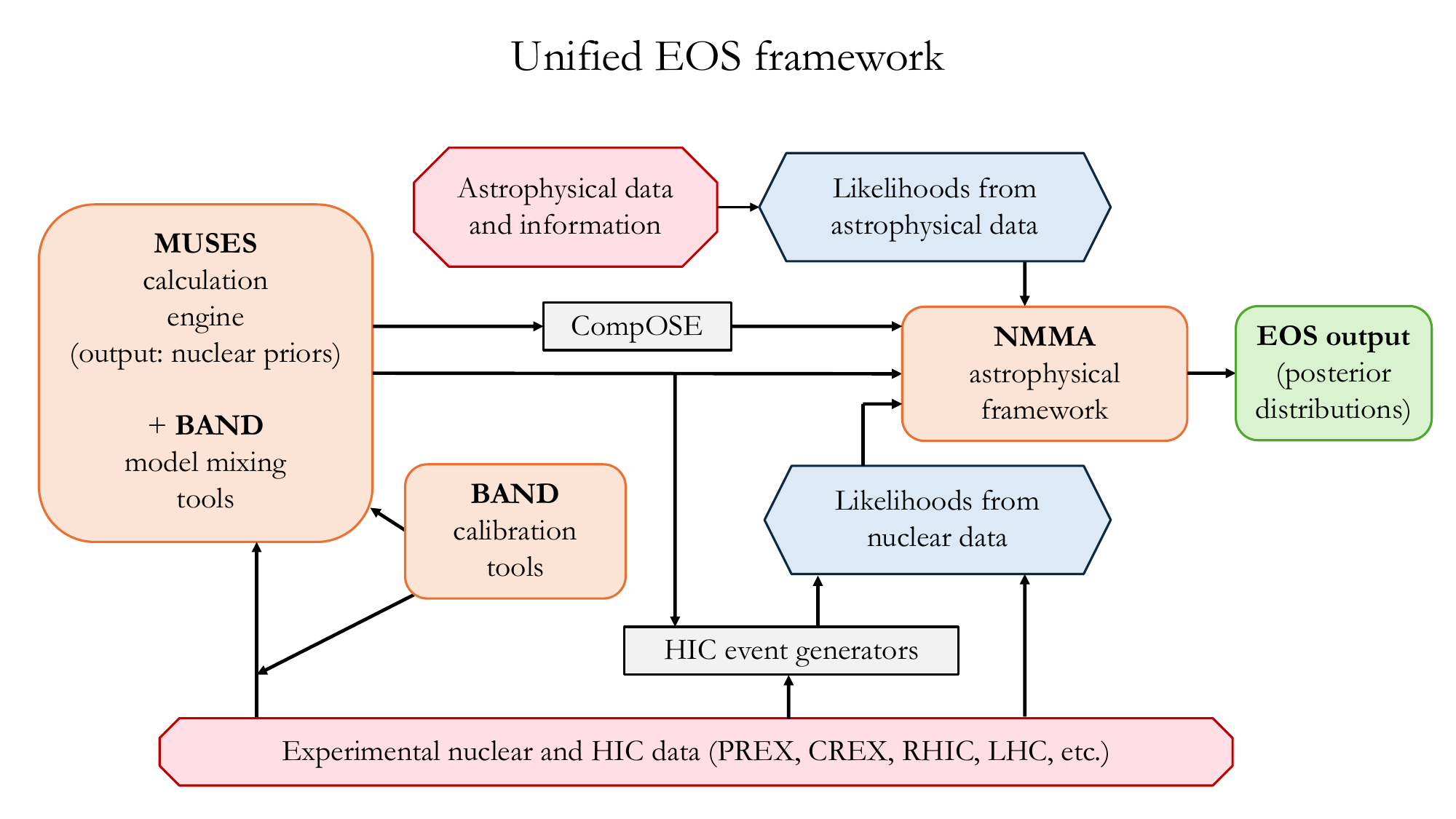}
    \caption{\textcolor{black}{A schematic of a possible unified framework for the EOS using all three overarching cyberinfrastructure frameworks discussed in Sect.~\ref{sec:frameworks}, and including various astrophysical and experimental data sources. The orange boxes are where the three main frameworks are used, the red octagons highlight where the data is fed into the workflow, and the blue hexagons represent the likelihoods formed from this data. One workflow would be, e.g., the production of the nuclear theory prior by the MUSES calculation engine, which would then be fed into the full workflow along with data-driven likelihoods to produce, via the NMMA framework, an EOS output that may consist of posterior draws for various macroscopic quantities. Other possible configurations of this framework, including the use of emulators at several steps of the workflow, are briefly discussed in the main text.}}
    \label{fig:unifiedframeworkdiagram}
\end{figure}

Moving forward in the field of dense nuclear matter requires us to consider how best to capitalize on the work already achieved, and how to advance further by using that as a foundation. 
To aid the reader in this effort, we have put together a diagram, displayed in Fig.~\ref{fig:unifiedframeworkdiagram}, in an attempt to summarize the critical development in terms of available frameworks and methods employed in both theory and experiment, and to provide a potential workflow to tie them all into one unified analysis framework to calculate the nuclear matter EOS.

In Fig.~\ref{fig:unifiedframeworkdiagram}, experimental nuclear data, such as that from relativistic HICs and parity-violating electron scattering experiments (e.g., PREX-II~\cite{PREX:2021umo, Reed:2021nqk} and CREX~\cite{CREX:2022kgg}) is treated as input to HIC event generators and nuclear data likelihoods, to be used for the Bayesian inference of the EOS posterior. Together with BAND framework's calibration (\texttt{surmise}) and model mixing (\texttt{Taweret}) tools, this data can be used to the fix free parameters of the different theoretical models available in the MUSES CE.
The integration of BAND tools into the MUSES framework would be the ideal use of the rigorous Bayesian statistics that the BAND collaboration has enabled the community to take advantage of, and would allow for full UQ of the theoretical models to propagate through this EOS framework.
While MUSES can provide EOSs both for HIC EGs as well as EOS priors for NMMA, the \texttt{CompOSE} database~\cite{CompOSECoreTeam:2022ddl} could also be used alternatively in the latter case to provide EOS tables from models not yet available through MUSES, depending on the desire of the user.

On the upper portion of the diagram, astrophysical data from, e.g., pulsar profile modelling, gravitational wave measurements, and radio astronomy is included in the astrophysical likelihoods that are input to the NMMA framework; \textcolor{black}{these could also incorporate the pQCD likelihood from, e.g., the KoKu method.} Once the EOS prior and likelihoods are constructed, the NMMA framework combines this information in a rigorous Bayesian way to produce the final product, the EOS posterior and any relevant quantities that a user may request. In producing the EOS posterior, it may be useful to employ emulators to speed up this process; emulators would also be helpful in other places on the diagram, including calibrating the EOS prior within the MUSES framework or to fasten simulations in EGs.
As the development of emulators progresses, integrating them into this unified framework will greatly assist in making this workflow efficient without compromising too strongly on precision or accuracy.

The exact determination of how this unified workflow will look will be heavily impacted by the available tools and requested inputs/outputs by the user. However, the most crucial step now is attaining the goal of these frameworks and databases being able to interface efficiently and effectively with one another. 
We stress that this is the optimal time for the field of dense nuclear matter
to multiply efforts in that direction, given all of the tools we now possess with these cyberinfrastructure frameworks, and the many smaller databases and tools that are currently being developed and expanded (see Sect.~\ref{subsec:communitytools}). Future experimental and astrophysical campaigns are expected to produce more data, and it is important that those of us whose goal is to 
exploit it with state-of-the-art theoretical models and Bayesian UQ methods are able to anticipate this influx of information and plan accordingly, to make sure we are able to capitalize on it when it arrives.


\section*{Acknowledgements}
 All authors express their sincere gratitude to Subhasis Chattopadhyay and Peter Senger for inviting us to contribute this article in the special volume of the European Physics Journal Special Topics (EPJ ST) titled “High Density Nuclear Matter”. This article has benefited from talks and discussions at the workshop -- Dense Nuclear Matter Equation of State from Theory and Experiments -- organised by the International Research Laboratory for Nuclear Physics and Nuclear Astrophysics (IRL NPA) at the Facility for Rare Isotope Beams (FRIB) from October 28 to November 1, 2024.
 
 K.A. thanks Hans-Rudolf Schmidt, Peter Senger, and Agnieszka Sorensen for insightful discussions and Laura Fabbietti for helpful comments in Sect.~\ref{subsubsec:methods_hic_exp}. A.C.S. thanks Christian Drischler for helpful suggestions pertaining to Sect.~\ref{sec:bandeoswork}, and Sudhanva Lalit for reviewing Secs.~\ref{sec:band} and~\ref{sec:unifiedframework}. J.J. thanks Veronica Dexheimer and Claudia Ratti for reviewing Sect. \ref{subsec:MUSES}. 

 
 J.J. acknowledges support by the MUSES collaboration under National Science Foundation (NSF) grant number OAC-2103680. B.K. gratefully acknowledges support from the Helmholtz Forschungsakademie HFHF and the GSI F\&E program. P.T.H.P. is supported by the research program of the Netherlands Organization for Scientific Research (NWO) under grant number VI.Veni.232.021. T.R. thanks Agnieszka Sorensen for fruitful discussions and Steffen Bass for the kind hospitality at Duke University. T.R. gratefully acknowledges financial support by the Fulbright U.S. Scholar Program, which is sponsored by the U.S. Department of State and the German-American Fulbright Commission. This article’s contents are solely the responsibility of the authors and do not necessarily represent the official views of the Fulbright Program, the Government of the United States, or the German-American Fulbright Commission. T.R. gratefully acknowledges support from The Branco Weiss Fellowship - Society in Science, administered by the ETH Z\"urich. A.C.S. warmly acknowledges the Facility for Rare Isotope Beams for its hospitality during the completion of this work.

\bibliography{main}



\end{document}